\documentclass[11pt]{iopart} 

\usepackage{hyperref}
\usepackage{graphicx}   
\usepackage{latexsym}   %
\usepackage{color}     	%
\usepackage{dcolumn}    
\usepackage{bm}        	
\usepackage{amssymb}    
\usepackage{graphics}
\usepackage{epsfig}
\usepackage{caption}
\usepackage{subfigure}
\usepackage{nicefrac}

\begin{document}

\title[Longitudinal and transverse structure functions in MHD turbulence]{Longitudinal and transverse structure functions \\
in high Reynolds-number magneto-hydrodynamic turbulence}

\author{J Friedrich$^1$, H Homann$^2$, T Sch\"afer$^{3,4}$ and R Grauer$^1$}
\address{$^1$ Theoretische Physik I, Ruhr-Universit\"{a}t Bochum, Universit\"{a}tsstr. 150,
D-44780 Bochum, Germany}
\address{$^2$ Universit\'e de Nice-Sophia, CNRS, Observatoire de la C\^ote d'Azur,
         CS 34229, 06304 Nice Cedex 4, France}
\address{$^3$ Department of Mathematics, College of Staten Island, CUNY, USA}
\address{$^4$ Physics Program at the CUNY Graduate Center, 365
5th Ave, New York, NY 10016, USA}

\begin{abstract}
We investigate the scaling behavior of longitudinal and
transverse structure functions in homogeneous and
isotropic magneto-hydrodynamic (MHD) turbulence by means of an exact hierarchy of
structure function equations as well as by direct
numerical simulations of two- and three-dimensional MHD turbulence.
In particular, rescaling relations between longitudinal
and transverse structure functions are derived and utilized in order
to compare different scaling behavior in the inertial range.
It is found that there are no substantial differences between
longitudinal and transverse structure functions in MHD turbulence.
This finding stands in contrast to the case of hydrodynamic turbulence
which shows persistent differences even at high Reynolds numbers.
We propose a physical picture that is based on an effective reduction of
pressure contributions due to local regions
of same magnitude and alignment of velocity and magnetic field fluctuations.
Finally, our findings underline the importance of the pressure term
for the actually observed scaling differences in hydrodynamic turbulence.
\end{abstract}



\noindent{\it high Reynolds-number turbulence, \/ magneto-hydrodynamics, \/ structure functions, \/ direct numerical simulations}

\maketitle

\section{Introduction}\label{sec:intro}
The question whether  longitudinal and transverse structure functions
posses different scaling behavior in highly turbulent flows is still
an open and unsolved problem.  Symmetry considerations
\cite{biferale-procaccia:2005} of the underlying Navier-Stokes
equation suggest no difference in scaling behavior. However,
experimental data
\cite{vanderwater-herweijer:1999,zhou-antonia:2000,shen-warhaft:2002}
and high resolution numerical simulations
\cite{boratav-pelz:1997,gotoh-fukayama-etal:2002,
ishihara-gotoh-etal:2009,benzi-biferale-etal:2010,grauer-homann-pinton:2012}
show a consistent difference between longitudinal and transverse
structure functions. So far it is absolutely not clear what the cause
of this observation is and possible explanations include a remaining
small-scale anisotropy of the flow
\cite{biferale-lanotte-etal:2008,romano-antonia:2001} or a finite
Reynolds number effect \cite{hill:2001}.
Thus a natural question and
approach is to ask whether these differences between longitudinal and transverse
structure functions exist or vanish in other turbulent systems in order to better
understand and identify a possible cause.
For instance,
it was shown in the context of two-dimensional
electron magneto-hydrodynamics \cite{germaschewski-grauer:1999} that
the differences between longitudinal and transverse structure functions
in the direct cascade range
decreased with increasing Reynolds number.

In this paper we investigate the behavior of
longitudinal and transverse
structure functions in the context of magneto-hydrodynamic (MHD) turbulence.
The motivation for looking at MHD turbulence arose from the known
fact that current and vortex sheets play a key role for the
understanding of intermittency in this particular flow.
Since sheets represent quite
anisotropic structures compared to vortex filaments as their hydrodynamic
counterparts, the hope was to attribute the different scaling of
longitudinal and transverse structure functions to the nature of the
dissipative structures. It turned out that this statement is correct
but in a way that was not anticipated and which is
counterintuitive at first glance. The result of this paper is
that there is no substantial difference
longitudinal and transverse structure functions in
the inertial range of high Reynolds number MHD flows.
Moreover, this observation can be attributed to regions of preferential alignment of
the magnetic and the velocity field which result in an effective depletion
of pressure contributions. In reverse, these results also open up the way
for a better understanding of the problem in hydrodynamic turbulence.
The outline of the paper is as follows: First-of-all,
we derive a hierarchy of structure function relations from the basic MHD
equations, a rather technical part in section \ref{sec:hie}
which will be accompanied by \ref{app:increment_evo} to \ref{app:viscous}.
These relations are then used in section \ref{sec:rescale}, in order to derive
rescaling relations between transverse and longitudinal
structure functions along the lines of
Grauer, Homann and Pinton \cite{grauer-homann-pinton:2012}.
The scaling behavior of longitudinal and transverse structure functions
in direct numerical
simulations of MHD turbulence will then be investigated with the help of these
rescaling relations in section \ref{sec:num}. The paper concludes
with a simple examination of local regions
of same magnitude and alignment of velocity and magnetic field
contributions and their depleting
effect on the total pressure.

\section{Hierarchy of structure functions in MHD turbulence}
\label{sec:hie}
The aim of this section is to establish relations between longitudinal and transverse
structure functions in MHD turbulence similar to the ones that have been derived by
Hill \cite{hill:2001}, Hill and Boratav \cite{hill-boratav:2001} and
Yakhot \cite{yakhot:2001} for the case of hydrodynamic turbulence.
To this end, we make use of the calculus of isotropic tensors in MHD turbulence
introduced by Chandrasekhar \cite{chandra:1951}.
We thus consider the MHD equations in the following form
\begin{eqnarray}
  \frac{\partial}{\partial t}u_i + u_n \frac{\partial}{\partial x_n} u_i -h_n
\frac{\partial}{\partial x_n}h_i &= -\frac{1}{\rho} \frac{\partial}{\partial x_i}\left(p +
\frac{1}{2} \rho |{\bf h}|^2 \right) + \nu \nabla^2_{\bf x} u_i,
\label{eq:uhill}\\
\frac{\partial }{\partial t}h_i + u_n \frac{\partial}{\partial x_n} h_i -h_n
\frac{\partial}{\partial x_n}u_i &= \lambda \nabla^2_{\bf x} h_i ,
\label{eq:hhill}
\end{eqnarray}
where summation over equal indices is implied. Here, $p({\bf x},t)$ denotes
the hydrodynamic pressure, $\rho$ the density, $\nu $ the kinematic
viscosity and $\lambda$ the magnetic diffusivity of the fluid.
In the following, the density of the fluid is set to one. Furthermore,
it should be noted that in this convenient form of the MHD equations, the magnetic field
${\bf h}({\bf x},t)$ has the dimensions of a velocity \cite{chandra:1951}.
From these equations (\ref{eq:uhill}-\ref{eq:hhill}), Chandrasekhar derived
the MHD analogon of the Friedmann-Keller correlation function
hierarchy \cite{keller1924} of hydrodynamic turbulence and made use of the
calculus of isotropic tensors \cite{Robertson1940,Chandrasekhar1950} in order
to derive, among other things, the corresponding von K\'{a}rm\'{a}n-Howarth equation
of MHD turbulence.

In our approach, which is more concerned with the \emph{local isotropy}
of MHD turbulence, we introduce the
velocity and magnetic field increments
${v}_i({\bf x},{\bf x}',t)={ u}_i({\bf x},t)-{ u}_i({\bf x}',t)= u_i-u_i'$ and
${b}_i({\bf x},{\bf x}',t)={ h}_i({\bf x},t)-{ h}_i({\bf x}',t)= h_i-h_i'$.
The evolution equations for
these increments can be derived in following a procedure
devised by Hill \cite{hill:2001} (see \ref{app:increment_evo} for further details)
and take the form
\begin{eqnarray}
\fl \qquad \frac{\partial }{\partial t}v_i + v_n \frac{\partial}{\partial r_n} v_i + U_n
\frac{\partial}{\partial X_n} v_i - b_n \frac{\partial }{\partial r_n} b_i - H_n \frac{\partial
}{\partial X_n} b_i &=& -P_i + \nu (\nabla_{\bf x}^2 +
\nabla_{{\bf x}'}^2) v_i,
\label{eq:v}\\
\fl \qquad \frac{\partial}{\partial t} b_i+ v_n \frac{\partial}{\partial r_n} b_i + U_n
\frac{\partial}{\partial X_n}b_i
- b_n \frac{\partial}{\partial r_n} v_i - H_n  \frac{\partial}{\partial X_n} v_i&=&\lambda
(\nabla_{\bf x}^2 + \nabla_{{\bf x}'}^2) b_i,
\label{eq:b}
\end{eqnarray}
where we have introduced the mean fields
${U}_i = \frac{{u}_i +{u}_i'}{2}$,
${H}_i = \frac{{h}_i +{h}_i'}{2}$,
and where we have switched to relative and center coordinates ${\bf r}= {\bf x} -{\bf x}'$ and
${\bf X}= \frac{{\bf x}+ {\bf x}'}{2}$. Furthermore, the total pressure gradient increment has
been introduced according to
\begin{equation}
 P_i= \frac{\partial}{\partial X_i} \left[ p-p' +
 \frac{1}{2} (|{\bf h}|^2 -|{\bf h}'|^2)\right].
 \label{eq:pressure_inc}
\end{equation}
Eqs. (\ref{eq:v}) and (\ref{eq:b}) are the point of departure for the
derivation of the structure function hierarchy, the main objective in this section.
The usual procedure consists in multiplying Eqs.
(\ref{eq:v}) and (\ref{eq:b}) by certain increment components and subsequently
taking the ensemble average
in order to make use of the assumption of homogeneity
and isotropy.

Before we address this issue, however, we want to address certain implications
of the additional magnetic field on the
structure function procedure along the lines of Hill \cite{hill:2001}. First-of-all,
it is important to
take notice of the influence of the mean magnetic field ${\bf H}$ in the increment evolution
equations (\ref{eq:v}) and (\ref{eq:b}): In contrast to the mean velocity field ${\bf U}$,
which can be removed in a system comoving with the mean velocity, the mean magnetic field
will be present in all moment equations derived from the increment equations.
The influence of these terms can only be removed
by certain combinations of the moments in addition to the assumption of homogeneity.
Furthermore, as far as the tensorial character
of the moments derived from Eqs. (\ref{eq:v}) and (\ref{eq:b}) is concerned,
we have to deal with certain
statistical quantities that are not invariant under the full rotation group
\cite{Robertson1940,Chandrasekhar1950}.
This difference arises due to ${\bf h}$ being an axial vector
which is unchanged under a reflexion, contrary to the true
polar vector ${\bf u}$ which changes signs. Therefore, tensorial quantities
that involve an odd number of magnetic field components exhibit a lack of
mirror symmetry and are thus skew tensors. By contrast, an even number of magnetic field components
leads to the usual tensorial forms encountered in hydrodynamic turbulence (see \ref{app:second-order}
for further discussion). Concerning the defining scalars of these tensors,
an important restriction emerges from the incompressibility condition of the velocity
and the magnetic field, i.e., $\frac{\partial}{\partial x_i} u_i({\bf x},t)=0$
and $\frac{\partial}{\partial x_i} h_i({\bf x},t)=0$. The incompressibility conditions
give rise to a first relation between the transverse
$S_{\,t\,t}^{{\bf v}{\bf v}}(r,t)$ and longitudinal $S_{r\,r}^{{\bf v}{\bf v}}(r,t)$
velocity (magnetic) field structure function of second order, namely
\begin{equation}
 \fl \qquad S_{\,t\,t}^{{\bf v}{\bf v}}(r,t)=\frac{1}{2 r} \frac{\partial}{\partial r}
 \left( r^2 S_{r\,r}^{{\bf v}{\bf v}}(r,t) \right)\qquad
\textrm{and} \qquad
S_{\,t\,t}^{{\bf b}{\bf b}}(r,t)=\frac{1}{2 r} \frac{\partial}{\partial r} \left( r^2 S_{r\,r}^{{\bf
b}{\bf b}}(r,t) \right).
\label{eq:KH}
\end{equation}
These relations are the well-known von K\'{a}rm\'{a}n-Howarth relations
and are a direct consequence of the incompressibility, homogeneity and isotropy of the MHD flow
discussed in \ref{app:kar3d}.
The existence of such direct relations between higher-order longitudinal and transverse
structure functions, however, is far less obvious. Therefore, we have to rely
on structure function relations that are directly derived from the evolution equations
of the increments, i.e., Eqs.
(\ref{eq:v}) and (\ref{eq:b}).

A first evolution equation for the symmetric
tensor of second order
$\langle v_i v_j +b_i b_j \rangle$ is derived in \ref{app:energy-bal} according to
\begin{eqnarray}\nonumber
\fl &~& \frac{\partial }{\partial t} \langle  v_i v_j+ b_i b_j \rangle + \frac{\partial }{\partial
r_n}\langle v_n (v_i v_j+b_i b_j)  \rangle  -  \frac{\partial }{\partial r_n} \langle b_n( v_i b_j +
v_j b_i)  \rangle \\
\nonumber
\fl  &~&+\frac{\partial }{\partial X_n}\langle U_n (v_i v_j+b_i b_j)  \rangle -
\frac{\partial }{\partial X_n} \langle H_n( v_i b_j + v_j b_i)\rangle
+\langle v_i P_j + v_j P_i \rangle \\
\fl &=&  2   \nu \left( \nabla^2_{\bf r} + \frac{1}{4}
\nabla^2_{\bf X} \right) \langle v_i v_j \rangle -2\langle \epsilon_{i\,j}^{{\bf u}{\bf u}}
\rangle + 2 \lambda \left( \nabla^2_{\bf r} + \frac{1}{4} \nabla^2_{\bf X} \right)
\langle b_i b_j \rangle -2 \langle \epsilon_{i\,j}^{{\bf h}{\bf h}} \rangle .
\label{eq:vivj}
\end{eqnarray}
Here, we have introduced the tensors of the local energy dissipation rates
\begin{eqnarray}
 \epsilon_{i\,j}^{{\bf u}{\bf u}}&=& \nu \sum_{n} \left[ \left(\frac{\partial u_i}{\partial x_n}
\right)\left(\frac{\partial u_j}{\partial x_n} \right)+\left(\frac{\partial u_i'}{\partial x_n'}
\right)\left(\frac{\partial u_j'}{\partial x_n'} \right)\right],
\label{eq:eps_u}\\
\epsilon_{i\,j}^{{\bf h}{\bf h}}&=& \lambda \sum_n \left[  \left(\frac{\partial h_i}{\partial x_n}
\right)\left(\frac{\partial h_j}{\partial x_n} \right)+\left(\frac{\partial h_i'}{\partial x_n'}
\right)\left(\frac{\partial h_j'}{\partial x_n'} \right)\right].
\label{eq:eps_h}
\end{eqnarray}
Eq. (\ref{eq:vivj}) is the first equation in a chain of transport equations and couples
to tensors of third order via the nonlinear terms. A further simplification
of Eq. (\ref{eq:vivj}) arises from the assumption of homogeneity which enables us to neglect
terms that involve a center derivative
acting on the ensemble average, i.e.,  $\frac{\partial }{\partial X_n}\langle ... \rangle=0$.
At this stage of the hierarchy,
pressure contributions also vanish on the basis of homogeneity
\cite{hill:1997}.

The averaged equation of energy balance
in MHD turbulence can be obtained from Eq. (\ref{eq:vivj}) in summing
over equal $i=j$, which is performed in \ref{app:energy-bal}.
The latter equation can be used to derive the
MHD analagon of Kolmogorov's 4/5-law of hydrodynamic turbulence.
In addition to the longitudinal velocity structure function of third order, the 4/5-law
in MHD turbulence involves
the mixed correlation function $C^{{\bf h} {\bf h} {\bf u}}_{ i\, j\, ,n}({\bf r},t)
= \langle h_i h_j u_n' \rangle$
that is symmetric in $i,j$ and the antisymmetric correlation function
\begin{equation}
  C^{{\bf u} {\bf h} {\bf h}}_{ \,i;j,n}({\bf r},t)=
  \langle (h_j u_i-u_j h_i) h_n' \rangle=
  C^{{\bf u} {\bf h} {\bf h}}_{ \,r;t\,t}(r,t)
  \left(\frac{r_j}{r}
\delta_{in} - \frac{r_i}{r} \delta_{jn}\right).
\label{eq:anti}
\end{equation}
In the following, tensorial forms of the antisymmetric
type (\ref{eq:anti}) will be indicated by a semicolon
between the antisymmetric indices. The notation
of the correlation functions follows a similar approach than
the notation used by Chandrasekhar in his seminal discussion
of correlation functions in MHD turbulence \cite{chandra:1951}.
Antisymmetric tensors such
as (\ref{eq:anti}) will be encountered throughout the entire
structure function hierarchy in MHD turbulence
and lead to modified scaling relations in comparison to the ordinary symmetric
tensors encountered in hydrodynamics.
This can be seen from the 4/5-law in MHD turbulence
(see \ref{app:45law}) in the inertial range
\begin{equation}
	S^{{\bf v} {\bf v} {\bf v}}_{ r\, r\, r} \left( r \right)
	- 12 C^{{\bf h} {\bf h} {\bf u}}_{ t\, \, t\, \,r} \left( r \right)-	\frac{24}{r^4}
\int_{0}^{r} \textrm{d} r' ~r'^3 C^{{\bf u} {\bf h} {\bf h}}_{ \,r;t\,t}(r')
	= - \frac{4}{5} \langle \varepsilon^{{\bf u}{\bf u}} +\varepsilon^{{\bf h}{\bf h}} \rangle r.
	\label{eq:four-fifths}
\end{equation}
The corresponding averaged local energy dissipation rates
$ \langle \varepsilon^{{\bf u}{\bf u}} \rangle$ and $\langle \varepsilon^{{\bf h}{\bf h}} \rangle$
can be recovered from
Eqs. (\ref{eq:eps_u},{\ref{eq:eps_h}), as it is further discussed in \ref{app:energy-bal}.
In the absence of the antisymmetric tensor, for instance in
the case of a vanishing electromotive force in Eq.
(\ref{eq:anti}), we recover the relation established by Politano
and Pouquet \cite{Politano1998}.

The next order equation relates structure functions of third and fourth order. It
is also the first order which provides a relation between the
longitudinal and transverse structure functions
based on the MHD equations. By contrast,
the von K\'{a}rm\'{a}n-Howarth relations (\ref{eq:KH})
are a pure statement
of the corresponding incompressibility conditions.
Furthermore, in the next order of the hierarchy we
have to deal for the first time with statistical quantities
that contain the pressure gradient increment $P_i$.
The evolution equation for the symmetric tensor of third order is derived in \ref{app:next-order}
and reads
\begin{eqnarray}\nonumber
\fl &~&\frac{\partial}{\partial t} \big \langle v_i v_j v_k + v_i b_j b_k + b_i v_j b_k + b_i b_j v_k
\big \rangle+ \frac{\partial}{\partial r_n} \big \langle v_i v_j v_k v_n - b_i b_j b_k b_n \big
\rangle \\ \nonumber
\fl &~&+   \frac{\partial}{\partial r_n} \langle v_n(v_i b_j b_k + b_i v_j b_k + b_i b_j v_k) \rangle -
\frac{\partial}{\partial r_n}  \langle b_n( v_i v_j b_k + b_i v_j v_k + v_i b_j v_k) \rangle \\
\fl &=& -  \left \langle(v_i v_j + b_i b_j)  P_k +   ( v_i v_k+ b_i b_k) P_j+ (v_j v_k + b_j b_k)  P_i
\right\rangle,
\label{eq:sym}
\end{eqnarray}
where dissipative terms have been neglected in the inertial range \cite{hill-boratav:2001}.
Obviously, this can only be a crude approximation since these terms contain
the joint statistics of the the velocity (magnetic) field increment
and the local energy dissipation rates (\ref{eq:eps_u},{\ref{eq:eps_h}),
which are known to contribute even in the vicinity of $\nu, \lambda \rightarrow 0 $.
Nonetheless, for now we proceed in introducing the
notations
\begin{eqnarray}
 S_{i\,j\,k\,n}^{{\bf v}{\bf v}{\bf v}{\bf v}}({\bf r},t) &=& \langle v_i v_j v_k v_n \rangle, \\
 S_{i\,j\,k\,n}^{{\bf b}{\bf b}{\bf b}{\bf b}}({\bf r},t) &=& \langle b_i b_j b_k b_n \rangle, \\
 S_{i\,j;k\,n}^{{\bf v}{\bf v}{\bf b}{\bf b}}({\bf r},t) &=& \langle v_i v_j b_k b_n - b_i b_j v_k
v_n \rangle,\\
 T_{ijk} ({\bf r},t) &=& \left \langle(v_i v_j + b_i b_j)  P_k +   ( v_i v_k+ b_i b_k) P_j+ (v_j v_k
+ b_j b_k)  P_i  \right\rangle,
\end{eqnarray}
which enable us to rewrite Eq. (\ref{eq:sym}) in shorter form as
\begin{eqnarray}\nonumber
 &~&\quad \frac{\partial }{\partial r_n} \left(  S_{i\,j\,k\,n}^{{\bf v}{\bf v}{\bf v}{\bf v}}({\bf r},t)-
S_{i\,j\,k\,n}^{{\bf b}{\bf b}{\bf b}{\bf b}}({\bf r},t) \right) \\
&~&-\frac{\partial}{\partial r_n} \left(  S_{i\,j;k\,n}^{{\bf v}{\bf v}{\bf b}{\bf
b}}({\bf r},t) + S_{j\,k;i\,n}^{{\bf v}{\bf v}{\bf b}{\bf b}}({\bf r},t)+ S_{i\,k;j\,n}^{{\bf v}{\bf
v}{\bf b}{\bf b}}({\bf r},t) \right)  = T_{ijk}({\bf r},t).
\label{eq:sym2}
\end{eqnarray}
Here, the tensors in the first line are ordinary symmetric tensors of fourth order,
described in \ref{app:fourth}, whereas the mixed tensor
possesses another tensorial form
\begin{equation}\label{anti2}
  S_{i\,j;k\,n}^{{\bf v}{\bf v}{\bf b}{\bf b}}({\bf r},t) =  S_{r\,r;t\,t}^{{\bf v}{\bf v}{\bf
b}{\bf b}}({ r},t) \left(\frac{r_i r_j}{r^2} \delta_{kn} -\frac{r_k r_n}{r^2} \delta_{ij} \right),
\end{equation}
since it is antisymmetric in exchanging $ij$ against $kn$.
It can therefore be considered as the next-order equivalent tensor of Eq. (\ref{eq:anti}).
Inserting the corresponding tensors (\ref{app:fourth}) yields
\begin{eqnarray} \nonumber
\fl\qquad &~& \frac{1}{r^2}\frac{\partial}{\partial r} \left[ r^2 \left( S_{r\,r\,r\,r}^{{\bf v}{\bf v}{\bf
v}{\bf v}}({ r}) - S_{r\,r\,r\,r}^{{\bf b}{\bf b}{\bf b}{\bf b}}({ r}) \right)  \right]
-\frac{6}{r}
\left(S_{r\,r\,t\,t}^{{\bf v}{\bf v}{\bf v}{\bf v}}({ r}) - S_{r\,r\,t\,t}^{{\bf b}{\bf b}{\bf
b}{\bf b}}({ r}) -S_{r\,r;t\,t}^{{\bf v}{\bf v}{\bf b}{\bf b}}({ r})\right) \\
\fl &~&= -T_{rrr}(r),
\label{eq:vier1}
\end{eqnarray}
for the longitudinal structure functions and
\begin{eqnarray}\ \nonumber
 \fl \qquad &~&\frac{1}{r^4} \frac{\partial}{\partial r} \left[r^4 \left(S_{r\,r\,t\,t}^{{\bf v}{\bf v}{\bf
v}{\bf v}}({ r}) - S_{r\,r\,t\,t}^{{\bf b}{\bf b}{\bf b}{\bf b}}({ r}) \right)  \right] - \frac{4
}{3 r} \left( S_{t\,\,t\,\,t\,\,t}^{{\bf v}{\bf v}{\bf v}{\bf v}}({ r}) - S_{t\,\,t\,\,t\,\,t}^{{\bf
b}{\bf b}{\bf b}{\bf b}}({ r})\right)+\frac{\partial}{\partial r} S_{r\,r;t\,t}^{{\bf v}{\bf v}{\bf
b}{\bf b}}({ r})\\
\fl &~&= -T_{rtt}(r),
\label{eq:vier2}
\end{eqnarray}
for the mixed structure functions.\\
These equations represent the generalization of the next-order structure function relations in hydrodynamic
turbulence derived
in \cite{hill-boratav:2001} and \cite{yakhot:2001} in the presence of a magnetic field. As in the hydrodynamic
case the longitudinal and transverse structure functions of fourth order
are
coupled via the mixed terms $S_{r\,r\,t\,t}(r)$ in Eq. (\ref{eq:vier1}) and (\ref{eq:vier2}). Furthermore,
the presence of the pressure contributions $T_{rrr}(r)$ and $T_{rtt}(r)$ in the inertial range
are considered to be responsible for the persistent different scaling behavior between
the longitudinal and transverse structure functions in hydrodynamic turbulence
\cite{grauer-homann-pinton:2012,hill-boratav:2001,gotoh-nakano:2003}.
Nevertheless, Eq. (\ref{eq:vier1}) and (\ref{eq:vier2}) distinguish themselves from their hydrodynamic counterparts
in several points: First-of-all, the presence of the antisymmetric tensor $S_{i\,j;k\,n}^{{\bf v}{\bf v}{\bf
b}{\bf b}}({\bf r})(r)$ leads to novel terms, especially in the mixed velocity-magnetic structure function equation
(\ref{eq:vier2}) that involves $S_{r\,r;t\,t}^{{\bf v}{\bf v}{\bf
b}{\bf b}}({ r})$ in a new differential relation compared to the other mixed structure functions. Furthermore,
only differences between pure velocity and magnetic structure functions enter the relation, which can lead
to cancellation effects, e.g., in the case of alignment solutions. In addition, the pressure contributions
on the right-hand side of (\ref{eq:vier1}) and (\ref{eq:vier2}) include contributions from the magnetic pressure
next to the hydrodynamic pressure which opens up the way to discuss the role of pressure in MHD turbulence
via a local Bernoulli law along the lines of Gotoh and Nakano \cite{gotoh-nakano:2003}
for the case of hydrodynamic turbulence.
\section{Rescaling relations between longitudinal and transverse structure functions}
\label{sec:rescale}
In this section, we briefly want to review rescaling relations between longitudinal
and transverse structure functions of all orders that were derived recently \cite{grauer-homann-pinton:2012}.
We shall then proceed to motivate such rescaling relations for MHD turbulence, namely by means of Eqs.
(\ref{eq:vier1}) and (\ref{eq:vier2}).
The point of departure for hydrodynamic turbulence
is the observation by Siefert and Peinke \cite{siefert-peinke:2004}
that $S^{{\bf v}{\bf v}}_{r\,r}(r)$ is
a smooth function of $r$ for which we can interpret the right-hand side of
the von K\'{a}rm\'{a}n-Howarth relation (\ref{eq:KH}) as the first two
terms of a Taylor expansion around $r$
\begin{equation}
 S^{{\bf v}{\bf v}}_{r\,r}\left(r+ \frac{r}{2} \right)\approx S^{{\bf v}{\bf v}}_{r\,r}(r)
+\frac{r}{2} \frac{\partial}{\partial r} S^{{\bf v}{\bf v}}_{r\,r}(r).
\label{eq:kh_rescale}
\end{equation}
Here, we neglect terms of higher order under the assumption that $r / 2$ is
much smaller than the integral length scale $L$.
In this approximation, the transverse structure function is simply the longitudinal structure
function rescaled by the factor $3 /2$ according to
\begin{equation}
 S^{{\bf v}{\bf v}}_{t\,t}(r)\approx S^{{\bf v}{\bf v}}_{r\,r}\left(\frac{3}{2}r\right).
\end{equation}
This procedure can be generalized
for structure functions of the order $n$ in
hydrodynamic turbulence \cite{grauer-homann-pinton:2012}, which is discussed
at the example of the fourth order equations
(\ref{eq:vier1}) and (\ref{eq:vier2}) in the hydrodynamic limit
\begin{eqnarray}
 \frac{1}{r^2}\frac{\partial}{\partial r} \left[ r^2 S_{r\,r\,r\,r}^{{\bf v}{\bf v}{\bf v}{\bf v}}({
r})  \right]- \frac{6}{r} S_{r\,r\,t\,t}^{{\bf v}{\bf v}{\bf v}{\bf v}}({ r}) &=& -T_{rrr}(r),
\label{eq:vier1hydro}\\
  \frac{1}{r^4} \frac{\partial}{\partial r} \left[r^4 S_{r\,r\,t\,t}^{{\bf v}{\bf v}{\bf v}{\bf
v}}({ r}) \right] - \frac{4 }{3 r}  S_{t\,\,t\,\,t\,\,t}^{{\bf v}{\bf v}{\bf v}{\bf v}}({ r}) &=&
-T_{rtt}(r).
\label{eq:vier2hydro}
\end{eqnarray}
These two equations are solely related by the mixed term $S_{r\,r\,t\,t}^{{\bf v}{\bf v}{\bf v}{\bf
v}}({ r}) $, in contrast to the MHD equations, where the antisymmetric mixed terms $S_{r\,r;t\,t}^{{\bf
v}{\bf v}{\bf b}{\bf b}}({ r})$ appear. The rescaling properties can be repeated under the neglect
of the pressure terms, so that we obtain
\begin{eqnarray}
 3S_{r\,r\,t\,t}^{{\bf v}{\bf v}{\bf v}{\bf v}}({ r}) &\approx& S_{r\,r\,r\,r}^{{\bf v}{\bf v}{\bf
v}{\bf v}}({ r}) + \frac{r}{2} \frac{\partial}{\partial r} S_{r\,r\,r\,r}^{{\bf v}{\bf v}{\bf v}{\bf
v}}({ r}) \approx S_{r\,r\,r\,r}^{{\bf v}{\bf v}{\bf v}{\bf v}}\left(\frac{3}{2 }r \right), \\
 \frac{1}{3}  S_{t\,\,t\,\,t\,\,t}^{{\bf v}{\bf v}{\bf v}{\bf v}}({ r}) &\approx&
S_{r\,r\,t\,t}^{{\bf v}{\bf v}{\bf v}{\bf v}}({ r}) + \frac{r}{4} \frac{\partial}{\partial r}
S_{r\,r\,t\,t}^{{\bf v}{\bf v}{\bf v}{\bf v}}({ r}) \approx S_{r\,r\,t\,t}^{{\bf v}{\bf v}{\bf
v}{\bf v}}\left(\frac{5}{4}r \right),
\end{eqnarray}
which yields the transverse structure function of fourth order in terms
of the rescaled longitudinal structure function of fourth order
according to
\begin{equation}
S_{t\,\,t\,\,t\,\,t}^{{\bf v}{\bf v}{\bf v}{\bf v}}({ r}) \approx S_{r\,r\,r\,r}^{{\bf v}{\bf v}{\bf
v}{\bf v}}\left({\frac{3}{2}\frac{5}{4} r}\right).
\label{eq:vier_rescaled}
\end{equation}
It is clear that neglecting the pressure terms in Eqs. (\ref{eq:vier1hydro})
and (\ref{eq:vier2hydro}) in combination with the corresponding
Taylor expansions has to be treated with caution.
Nevertheless, it has been shown \cite{grauer-homann-pinton:2012} that
the rescaling relations are an accurate method to map scales without altering
the corresponding scaling behavior. It is therefore an essential
tool for the investigation of possible
different intermittency behavior of the longitudinal
and transverse structure functions. In considering higher order structure function
equations \cite{hill:2001}, a general relation between even order
transverse structure function of order $n$ can be derived according to
\begin{equation}
 S_{nt}^{n{\bf v}}({ r}) \approx S_{nr}^{n{\bf v}}\left(\frac{3}{2} \frac{5}{4}... \frac{n+1}{n} r
\right)= S_{nr}^{n{\bf v}}\left(\frac{(n+1)!!}{n!!} r\right).
\label{eq:rescale_gen}
\end{equation}
Turning to the case of MHD turbulence, the
von K\'{a}rm\'{a}n-Howarth relation  (\ref{eq:KH})
suggests that Eq. (\ref{eq:kh_rescale})
is also valid for the magnetic field structure functions
of second order. The next order equations (\ref{eq:vier1}) and (\ref{eq:vier2}),
however, are more complicated due
to the additional appearance of the anti-symmetric tensor
$S_{r\,r;t\,t}^{{\bf
v}{\bf v}{\bf b}{\bf b}}({ r})$. Nevertheless, assuming that the contribution
of the latter are rather small, we can - in a first approximation -
establish the same relation (\ref{eq:vier_rescaled}) between
$ S_{t\,\,t\,\,t\,\,t}^{{\bf v}{\bf v}{\bf v}{\bf v}}({ r}) - S_{t\,\,t\,\,t\,\,t}^{{\bf
b}{\bf b}{\bf b}{\bf b}}({ r})$ and
$ S_{r\,r\,r\,r}^{{\bf v}{\bf v}{\bf v}{\bf v}}({ r}) - S_{r\,r\,r\,r}^{{\bf
b}{\bf b}{\bf b}{\bf b}}({ r})$. Moreover, if velocity
and magnetic structure functions possess a unique power law
in the inertial range, it is appropriate to assume that the general
rescaling relation
(\ref{eq:rescale_gen}) holds independently for
velocity and magnetic structure functions of all orders, given that
higher order relations similar to Eqs.
(\ref{eq:vier1}) and (\ref{eq:vier2}) exist.

\section{Direct Numerical Simulations of 3D MHD turbulence and comparison
of longitudinal and transverse structure functions}
\label{sec:num}
In the previous section, we argued that rescaling relations of
the form (\ref{eq:rescale_gen}) apply for all even order
velocity and magnetic structure functions. We have examined
these relations in direct numerical simulations of 3D MHD turbulence.
Tab. \ref{tab:1} summarizes the corresponding characteristic parameters
of the simulations.
\begin{table}[h]
\centering
 \resizebox{\textwidth}{!}{
\begin{tabular}[h]{l l l l l l l l l l  l }
 run & Re$_{\lambda}$ & $u_{rms}$ & $h_{rms}$
 & $\langle \varepsilon^{{\bf u}{\bf u}} \rangle $
 & $\nu= \lambda$  & $\eta$ & $\tau_{\eta}$ & $L$ & $T_L$ & $N^3$ \\
\hline 3D NS&  460  & 0.19 & 0 & $3.6 \cdot 10^{-3}$
& $2.5 \cdot 10^{-5}$ & $1.45 \cdot 10^{-3}$
& 0.083 & 1.85 & 9.9 & $2048^3$\\
3D MHD & 430 & 0.23 & 0.36 &  $1.2 \cdot 10^{-2}$ & $7 \cdot 10^{-3}$  & $2.3 \cdot 10^{-3}$
& 0.078 & 2.75 & 6.5 & $2048^3$
\end{tabular}}
\caption{Characteristic parameters of the direct
numerical simulations of 3D MHD and hydrodynamic turbulence:
Taylor-Reynolds number $\textrm{Re}_{\lambda}=
\sqrt{\frac{15u_{rms}L}{\nu}}$, root mean square velocity $u_{rms}= \sqrt{ \langle{\bf u
}^2\rangle}$, root mean square magnetic field $h_{rms}= \sqrt{\langle{\bf h }^2\rangle}$,
averaged kinetic energy dissipation rate $\langle \varepsilon^{{\bf u}{\bf u}} \rangle $,
kinematic viscosity $\nu$ and magnetic diffusivity $\lambda$,
dissipation length
$\eta=\left(\frac{\nu^3}{\langle \varepsilon^{{\bf u}{\bf u}} \rangle}\right)^{1/4}$,
dissipation time $\tau_{\eta}= \left( \frac{\nu}{\eta} \right)^{1/2}$,
integral length scale $L= \frac{\left(\frac{1}{2} (u_{rms}^2 +
h_{rms}^2) \right)^{\frac{3}{2}}}
{\langle \varepsilon^{{\bf u}{\bf u}} \rangle + \langle \varepsilon^{{\bf h}{\bf h}} \rangle}$,
large-eddy turn-over time
$T_L=\frac{L}{u_{rms}}$ and resolution $N$. For the sake of completeness, the
averaged magnetic energy dissipation rate
$\langle \varepsilon^{{\bf h}{\bf h}} \rangle $
in the MHD simulations is $1.6 \cdot 10^{-2}$.}
\label{tab:1}
\end{table}
\begin{figure}[p]
 \includegraphics[width=0.48 \textwidth]{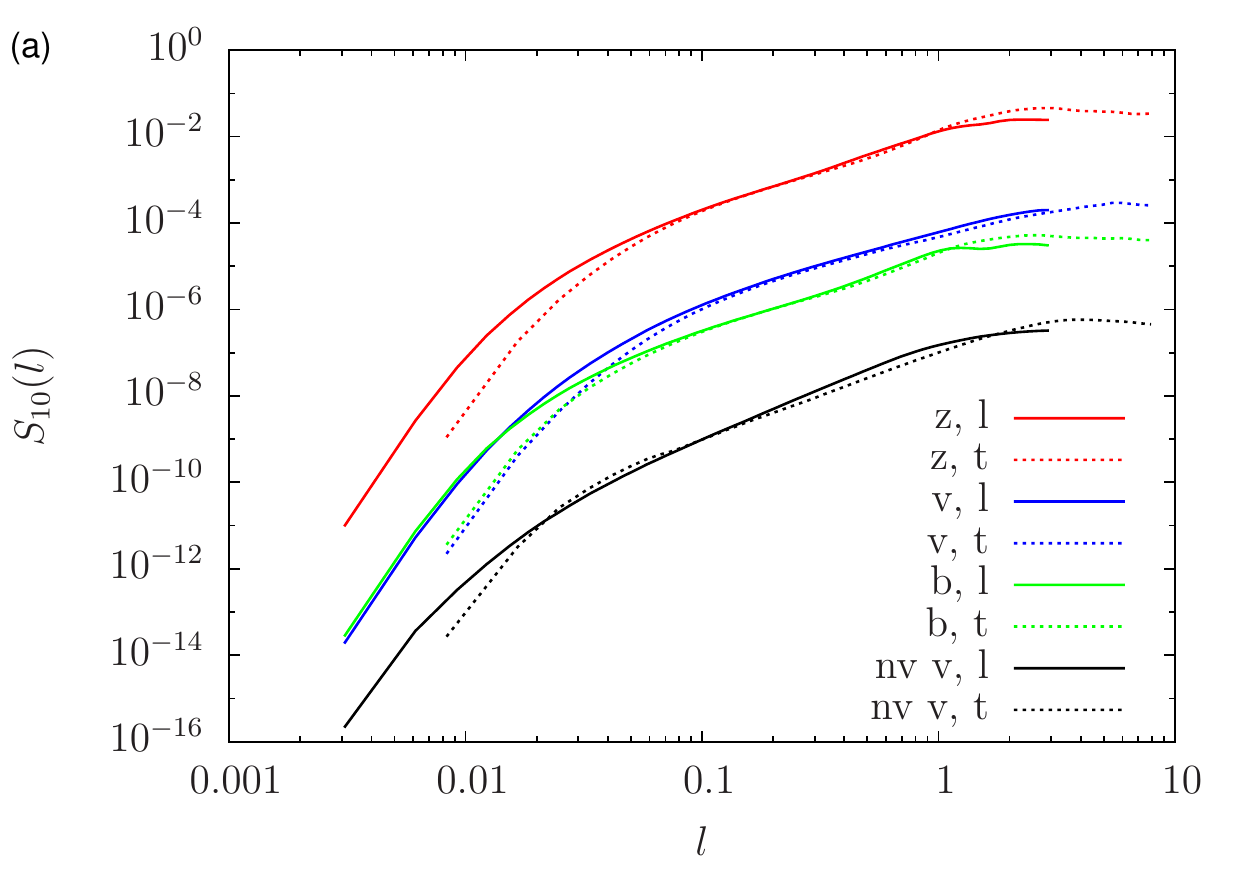}
 \includegraphics[width=0.48 \textwidth]{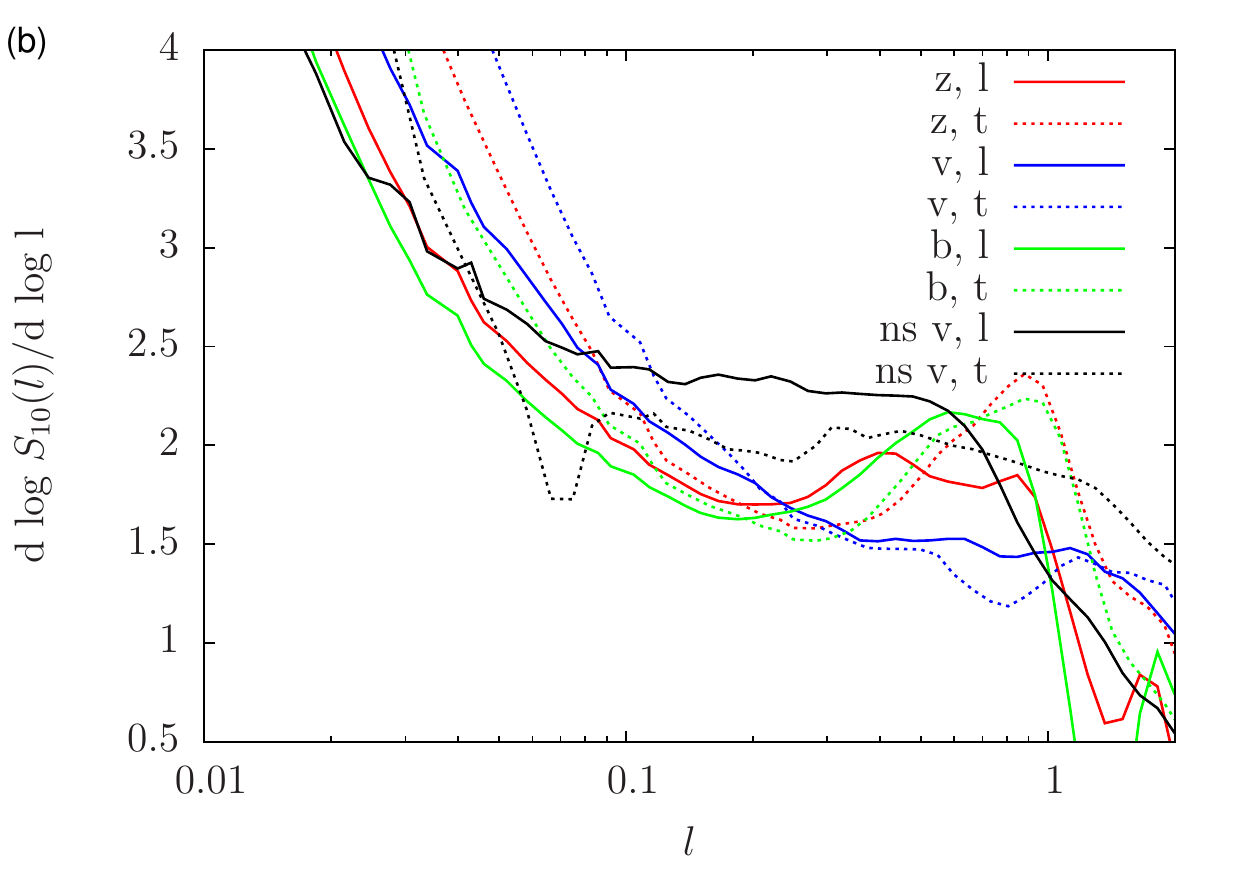}
 \caption{(a) Rescaled longitudinal and transverse structure functions in 3D MHD
 (green: magnetic field, blue: velocity field, Els\"{a}sser field ${\bf z}^+$: red)
 and hydrodynamic turbulence (black).
 The scales are mapped accurately via the rescaling relation (\ref{eq:rescale_gen}).
 The hydrodynamic transverse structure function (dashed black line) possesses a slightly more intermittent
 character than the longitudinal structure function (straight black line)
 whereas the rescaled MHD structure functions do not show considerable differences.\\
 (b)  Logarithmic derivatives of the structure functions from (a). Power law scaling
 should manifest itself as a flat curve.}
 \label{fig:rescale}
\end{figure}
\begin{figure}[p]
 \includegraphics[width=0.48 \textwidth]{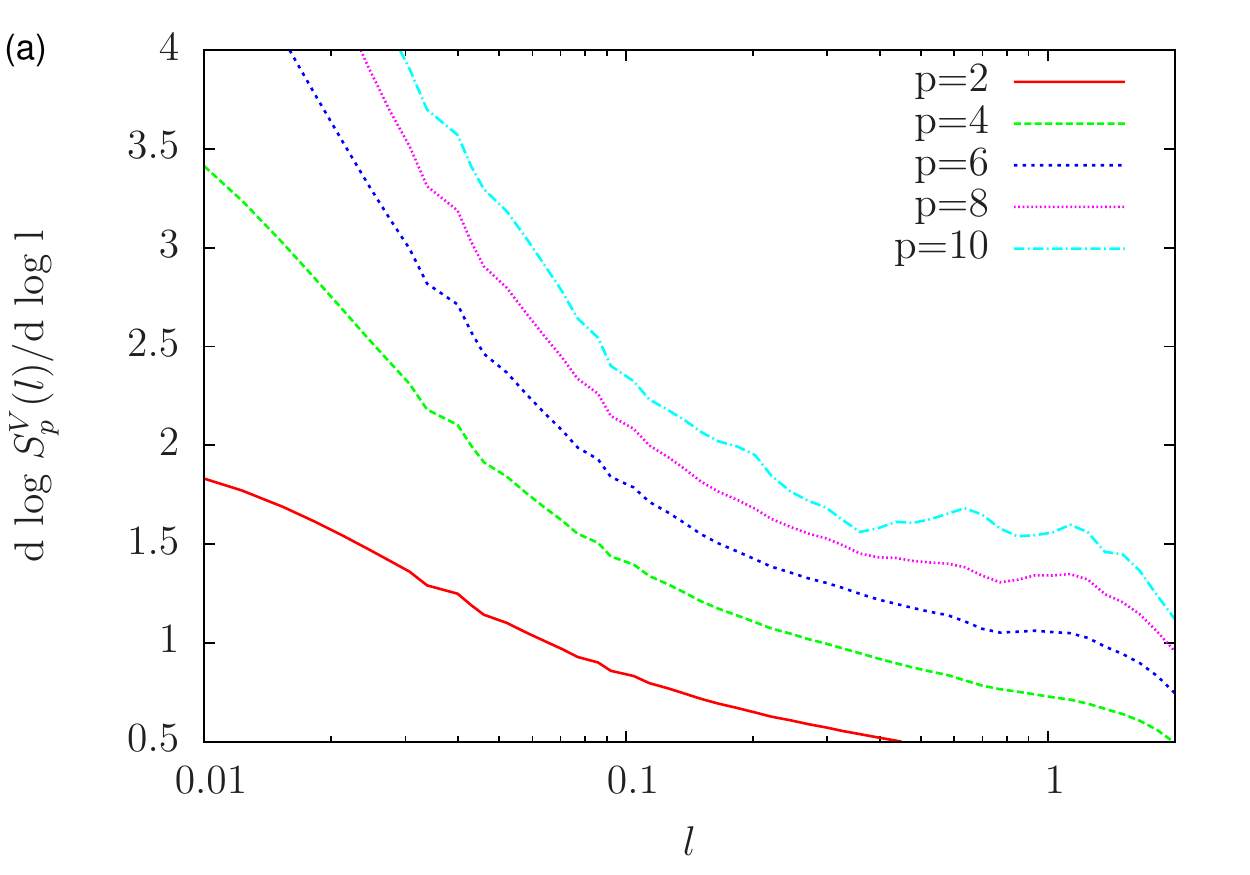}
  \includegraphics[width=0.48 \textwidth]{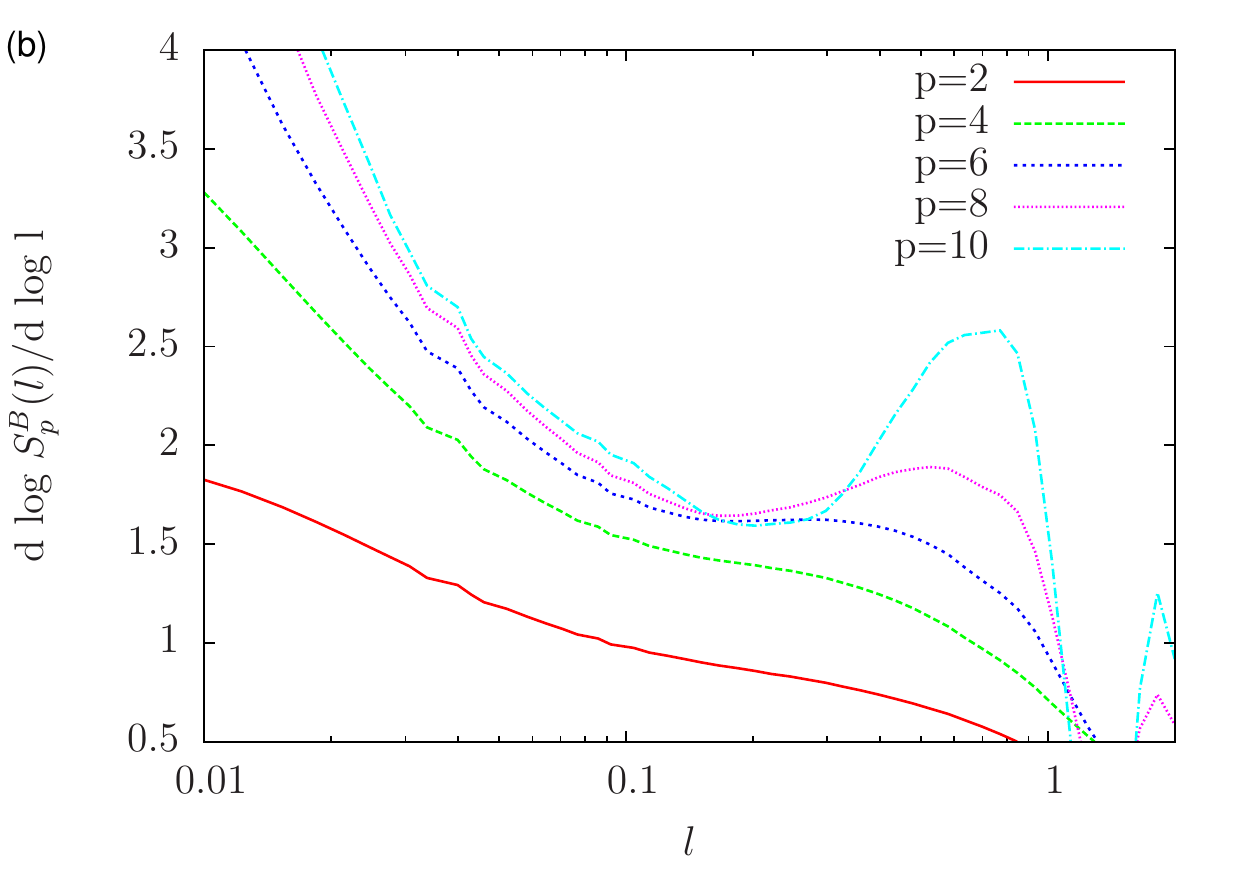}
   \caption{(a) Logarithmic derivative of the longitudinal velocity field structure functions
   of order $2-10$.\\
   (b) Logarithmic derivative of the longitudinal magnetic field structure functions
   of order $2-10$.}
   \label{fig:power}
\end{figure}
As an example, Fig. \ref{fig:rescale} (a) shows
the rescaling between longitudinal and transverse structure functions of
order 10. The scales of
the velocity field structure functions (blue) and the magnetic field structure functions
(green) are mapped accurately onto each other. This is also true for the structure functions
of the Els\"{a}sser field ${\bf z}^+ ({\bf x},t) = {\bf u}({\bf x},t)+{\bf h}({\bf x},t)$.
Furthermore, Fig. \ref{fig:rescale} shows no substantial difference between rescaled structure
functions of order 10 in MHD turbulence. In opposition to the finding
from hydrodynamic turbulence (see the black line in Fig. \ref{fig:rescale})
which suggests that the rescaled transverse structure function
possesses a slightly more intermittent character than the longitudinal one,
we thus arrive at a somewhat contradictory conclusion: Although that structure
functions in MHD turbulence are known to show a pronounced intermittency
in comparison to their hydrodynamic counterparts \cite{biskamp}, the differences
in scaling behavior between longitudinal and transverse structure functions
in MHD turbulence are less pronounced than in hydrodynamic turbulence.
This becomes even more apparent from the logarithmic derivative plot
of the longitudinal and transverse structure functions in Fig. \ref{fig:rescale} (b)
which shows that the scaling exponents of the structure functions of order 10 in
hydrodynamic turbulence lie above the ones in MHD turbulence. The latter fact
is commonly attributed to the dissipative structures in MHD turbulence
that consist mainly of vortex and current sheets and are believed to possess
a more singular character than the vortex tubes in hydrodynamic turbulence \cite{biskamp}. Nevertheless, Fig. \ref{fig:rescale} shows that these geometric considerations
are not an adequate explanation for the differences of scaling behavior
between longitudinal and transverse structure functions in general.\\
The negative finding from MHD turbulence thus suggests that
a more subtle mechanism is
at the heart of the problem of the longitudinal and transverse structure function
differences. Before we discuss such a mechanism on the basis of the pressure contributions
that enter in Eqs. (\ref{eq:vier1}) and (\ref{eq:vier2}), it is in order to briefly
discuss the notion of scaling in MHD turbulence in general. The logarithmic
derivative plot in Fig. \ref{fig:rescale} (b) shows
that a clear power law scaling of the structure functions
is somehow hard to anticipate. In opposition, the structure functions
from hydrodynamic turbulence (black)
manifest themselves by rather flat curves (power law behavior) in the logarithmic derivative plot.
Especially the magnetic longitudinal
structure functions  show a pronounced bump on larger scales $r$ and are thus missing
a clear power law behavior as it can be seen from Fig. \ref{fig:power} (b).
Moreover, the bump is increasing with increasing order of the structure function
and seems to hint at the existence of two different power law behaviors in the inertial range.
Since at this point, we are only interested in relative differences between longitudinal
and transverse structure functions, we only take note of this problem of power laws
in MHD turbulence and leave its evaluation to further work.

\section{Alignment of dissipative structures and depletion of pressure}
\label{sec:alingment}
Apparently, the identical scaling behavior of the longitudinal and
transverse structure functions in MHD turbulence discussed in
the previous section is
not directly related to the singular structures of the MHD flow.
Therefore, we want to address a different mechanism that considers
the influence of the pressure gradient on the longitudinal and
transverse structure functions. In hydrodynamic
turbulence, the relations (\ref{eq:vier1hydro})
and (\ref{eq:vier2hydro}) suggest that
the scaling behavior of the longitudinal and transverse structure
functions is altered solely by pressure contributions
represented by $T_{rrr}(r)$ and $T_{rtt}(r)$.
\begin{table}[t]
\centering
 \resizebox{\textwidth}{!}{
\begin{tabular}[h]{l l l l l l l l l l  l }
run & Re$_{\lambda}$ & $u_{rms}$ & $h_{rms}$ & $\nu^{(2)}=\lambda^{(2)}$ & dx & $\eta$ & $\tau_{\eta}$ &
$L$ & $T_L$ & $N^2$\\
\hline 2D MHD & 110-170  & 0.831 & 0.759 & $2 \cdot 10^{-10}$ & $2,05\cdot 10^{-3}$ & $1,64 \cdot 10^{-3}$
& 0.037 & 4.741 & 5.705 & $(3072)^2$
\end{tabular}}
\caption{Characteristic parameters of the
numerical simulations: Taylor-Reynolds number $\textrm{Re}_{\lambda}=
u_{rms} \left(\frac{ L^2}{\nu^{(2)}
(\langle \varepsilon^{{\bf u}{\bf u}} \rangle
+ \langle \varepsilon^{{\bf h}{\bf h}} \rangle)}\right)^{1/4}$, root mean square velocity $u_{rms}= \sqrt{ \langle{\bf u
}^2\rangle}$, root mean square magnetic field $h_{rms}= \sqrt{\langle{\bf h }^2\rangle}$,
dissipation length $\eta=\left(\frac{(\nu^{(2)})^3}{\langle \varepsilon^{{\bf u}{\bf u}} \rangle}
\right)^{\frac{1}{10}}$,  dissipation time $\tau_{\eta} = \left( \frac{\nu^{(2)}}
{\langle \varepsilon^{{\bf u}{\bf u}} \rangle^2} \right)^{\frac{1}{5}}$,
integral length scale  $L= \frac{\left(\frac{1}{2} (u_{rms}^2 +
h_{rms}^2) \right)^{\frac{3}{2}}}
{\langle \varepsilon^{{\bf u}{\bf u}} \rangle + \langle \varepsilon^{{\bf h}{\bf h}} \rangle}$
and large-eddy turn-over time
$T_L=\frac{L}{u_{rms}}$.}
\label{tab:2}
\end{table}
\begin{figure}[p]
 \includegraphics[width=0.49 \textwidth]{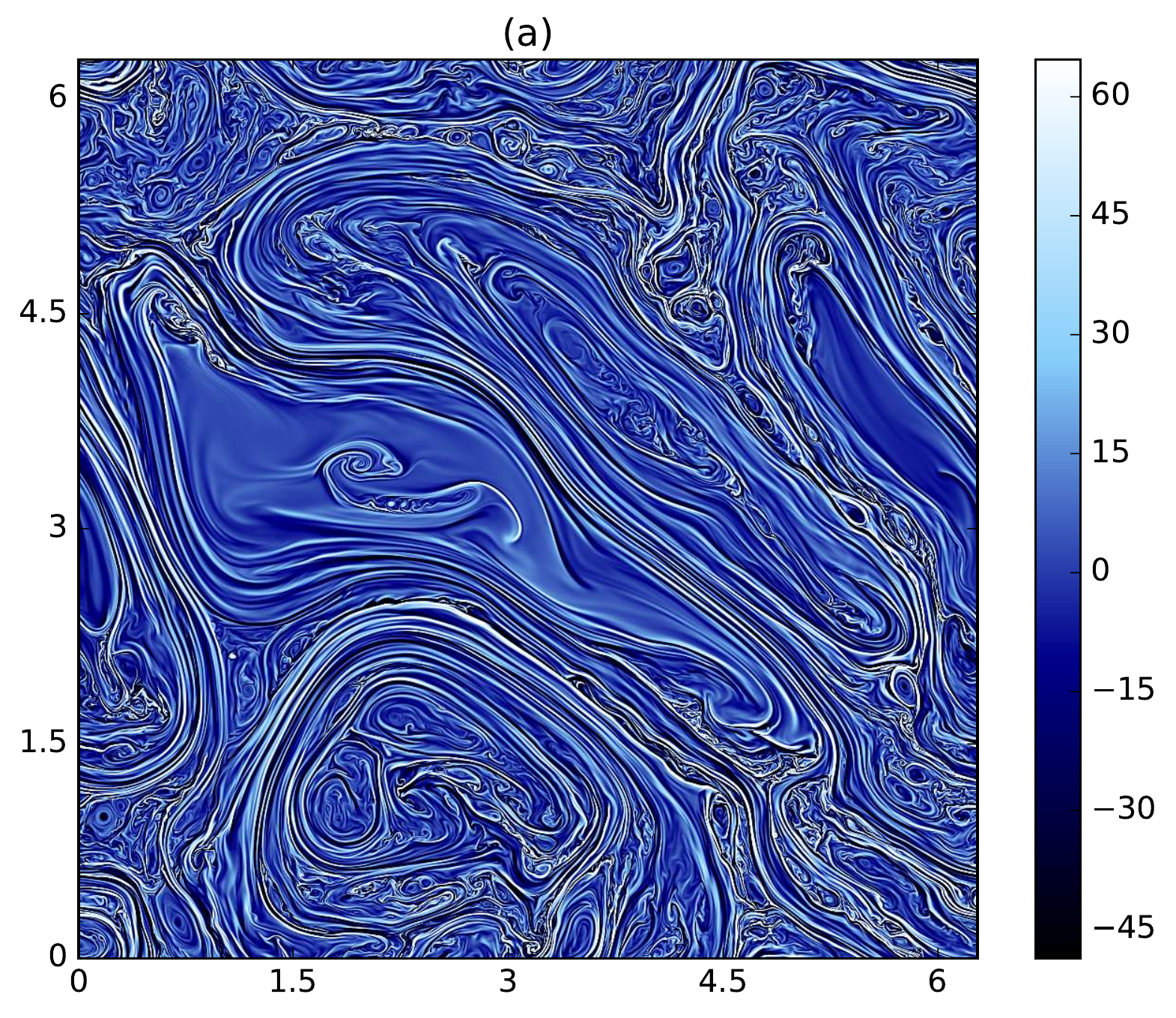}
 \includegraphics[width=0.49 \textwidth]{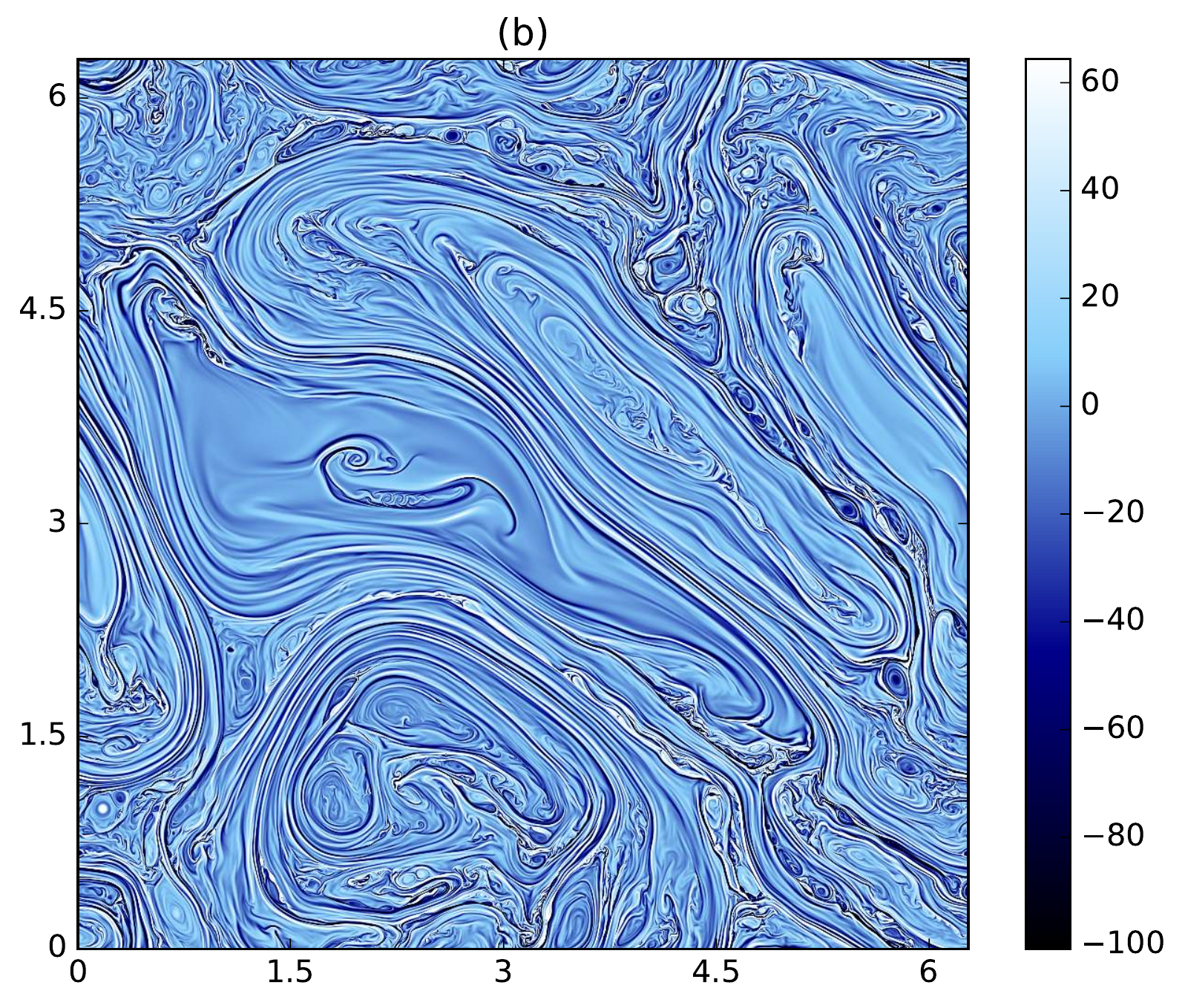}
 \caption{(a) Snapshot no. 1 of the vorticity field $\omega({\bf x},t)$
 from direct numerical simulations of 2D MHD turbulence. (b) Same snapshot
 no. 1 of the current density $j({\bf x},t)$.}
 \label{fig:om_j_1}
\end{figure}
\begin{figure}[p]
 \includegraphics[width=0.49 \textwidth]{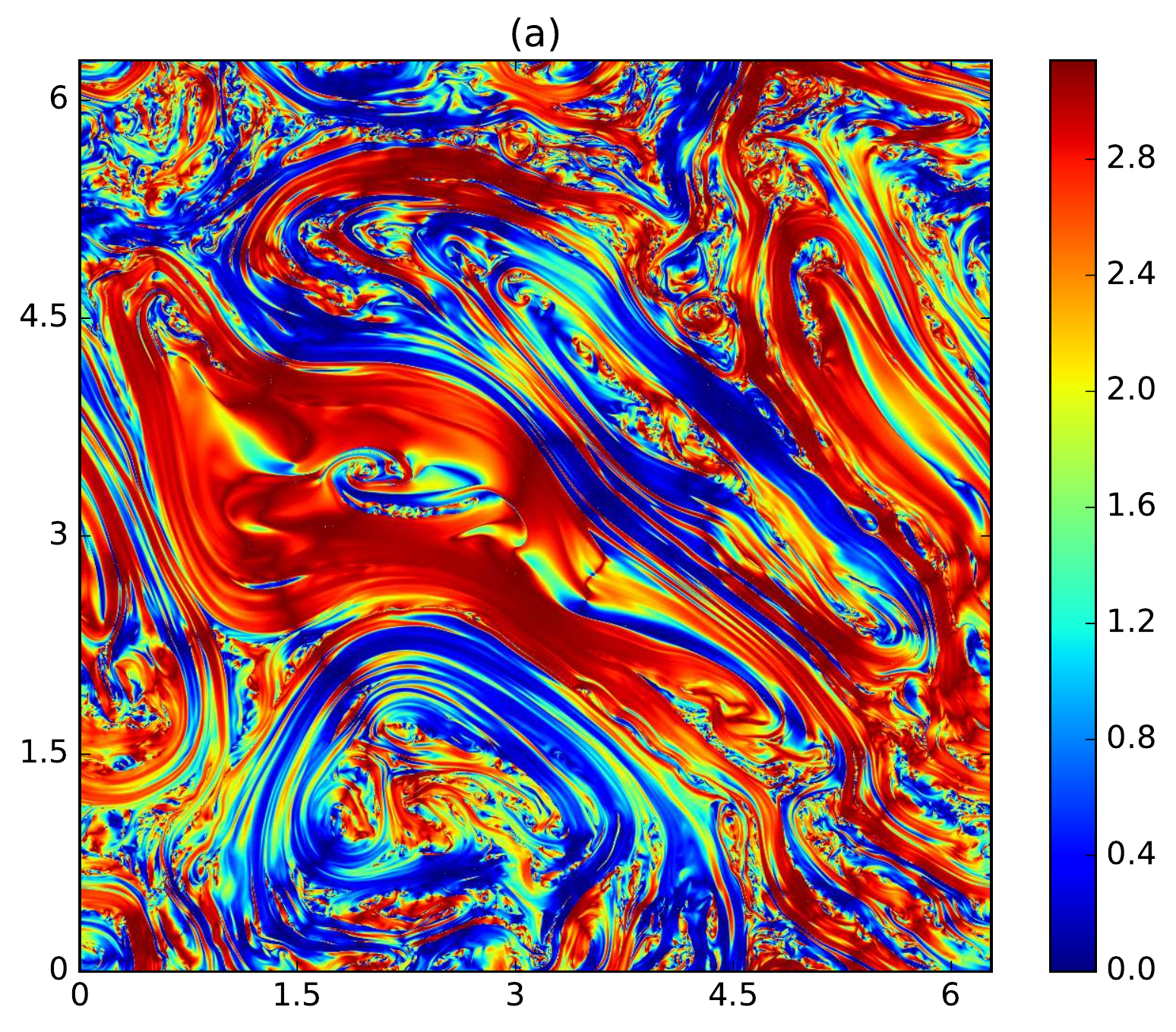}
  \includegraphics[width=0.49 \textwidth]{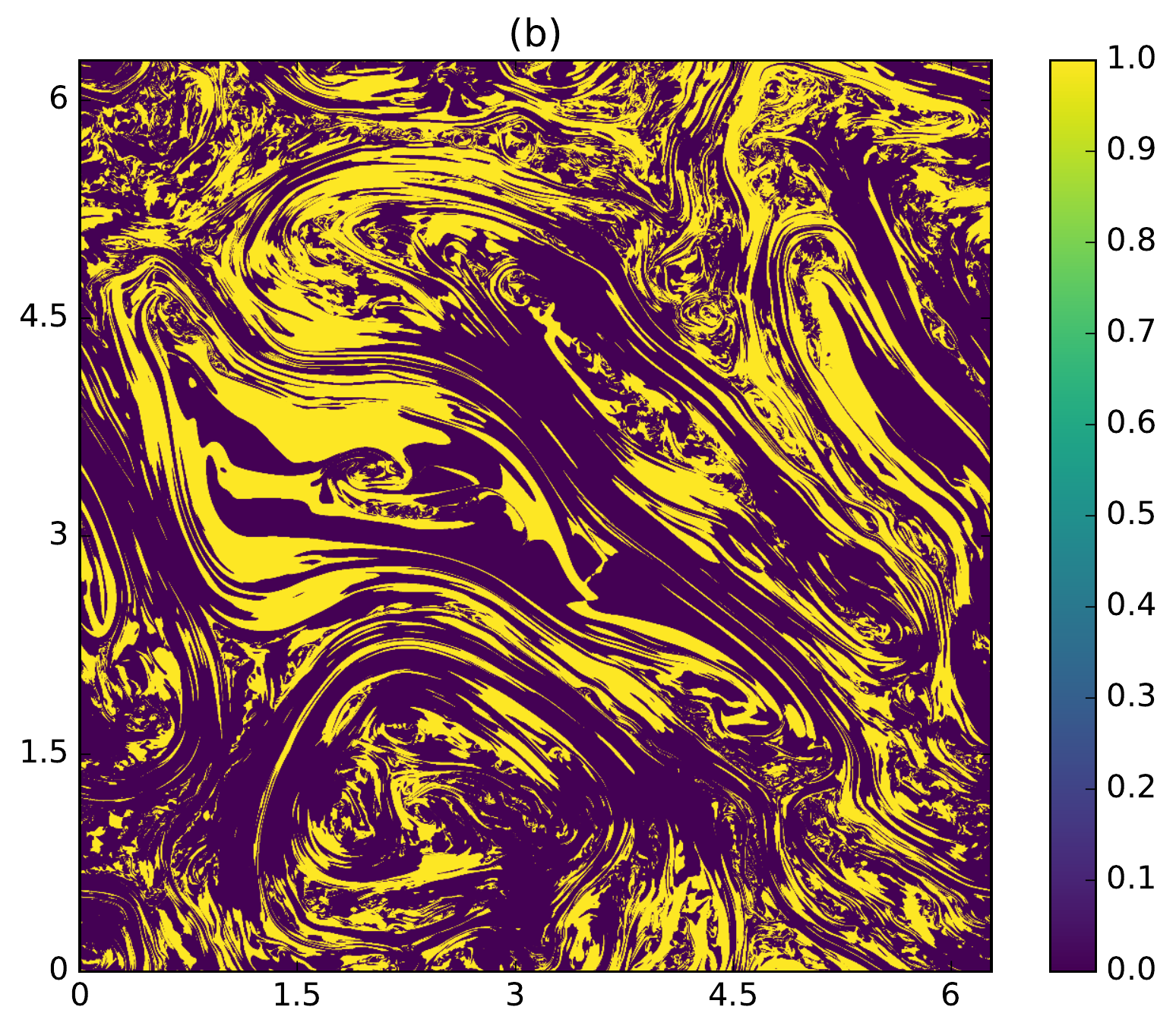}
  \caption{(a) Snapshot no. 1 of the alignment angle $\varphi({\bf x},t)$
  between velocity field and magnetic field.\\
  (b) Snapshot no. 1 of the
  filter function $\chi({\bf x},t)$ introduced in Eq. (\ref{eq:filter}).
  The covered area is 35.33 per cent of the total area.}
   \label{fig:phi_fil_1}
\end{figure}
\begin{figure}[p]
 \includegraphics[width=0.49 \textwidth]{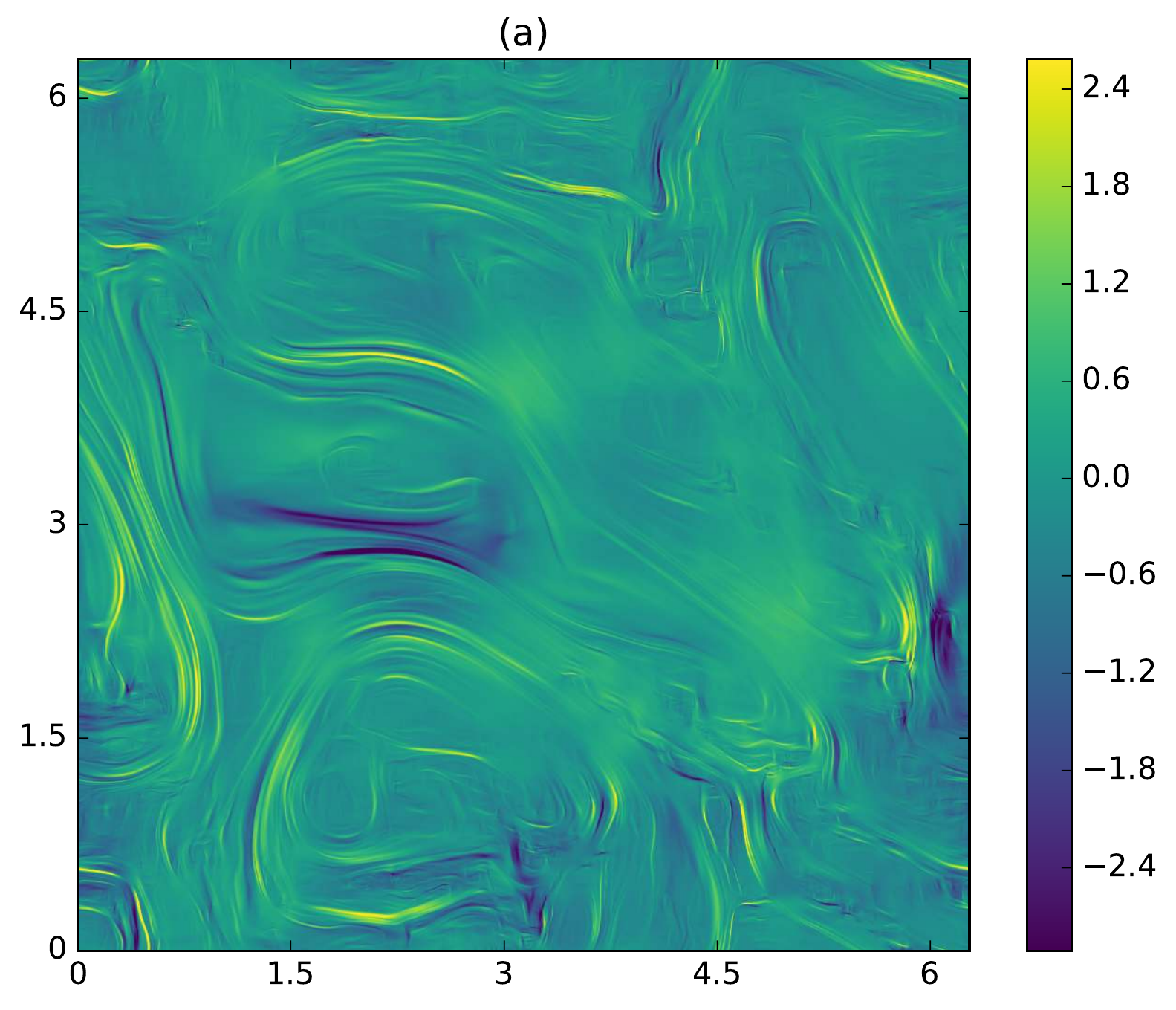}
  \includegraphics[width=0.49 \textwidth]{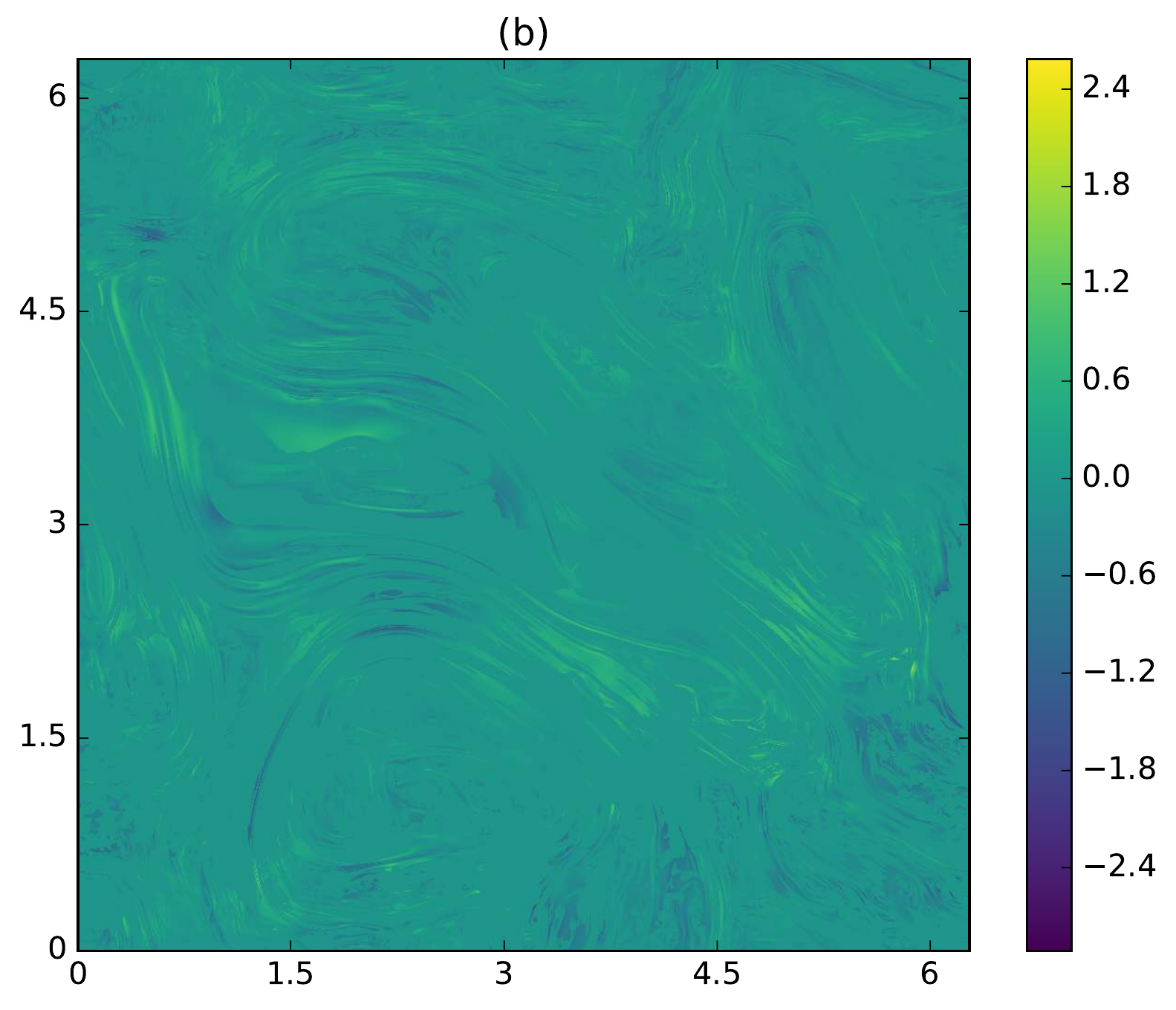}
  \caption{(a) Snapshot no. 1 of the total pressure field $p_{tot}({\bf x},t)$. \\
  (b) Snapshot no. 1 of the filtered pressure field via the filter function
  $\chi({\bf x},t)$ from Eq. (\ref{eq:filter}).
  The pressure field is effectively reduced.}
   \label{fig:pressure_1}
\end{figure}
\begin{figure}[p]
 \includegraphics[width=0.49 \textwidth]{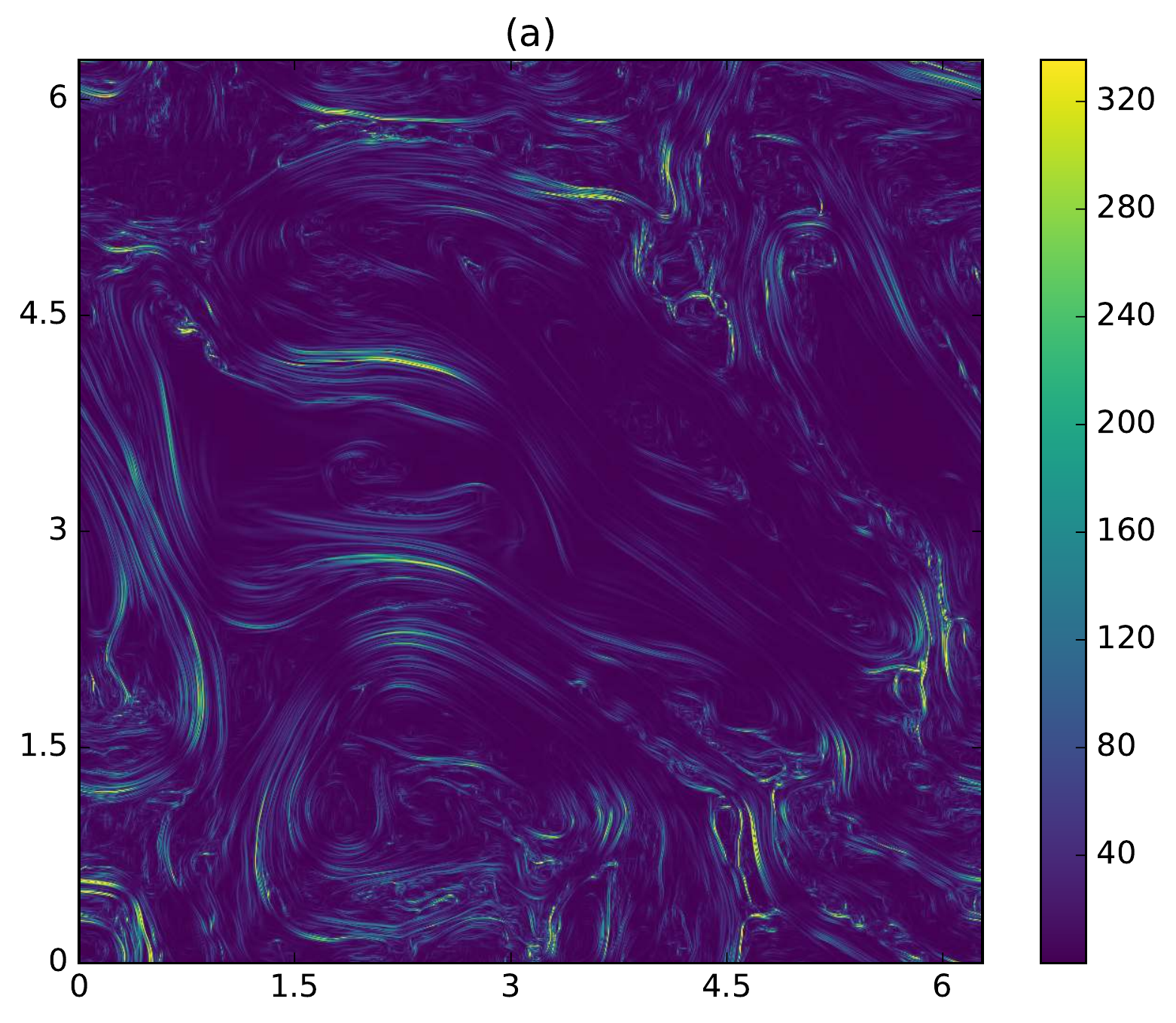}
  \includegraphics[width=0.49 \textwidth]{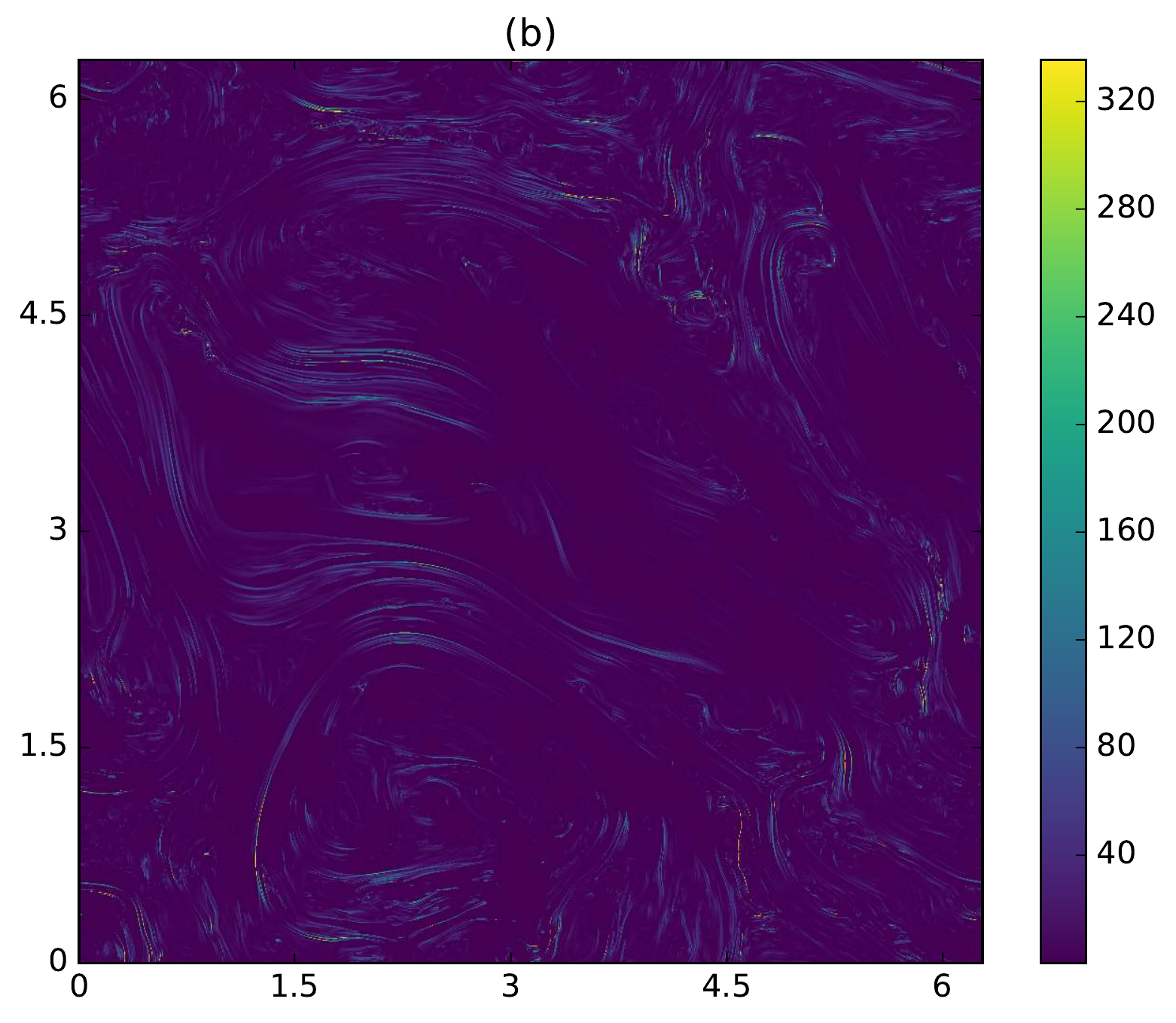}
 \caption {(a) Snapshot no. 1 of the norm of the total
 pressure gradient field $|\nabla p_{tot}({\bf x},t)|$. \\
  (b) Snapshot no. 1 of the filtered norm of the pressure gradient field.}
   \label{fig:absgradpressure_1}
\end{figure}
\begin{figure}[p]
 \includegraphics[width=0.49 \textwidth]{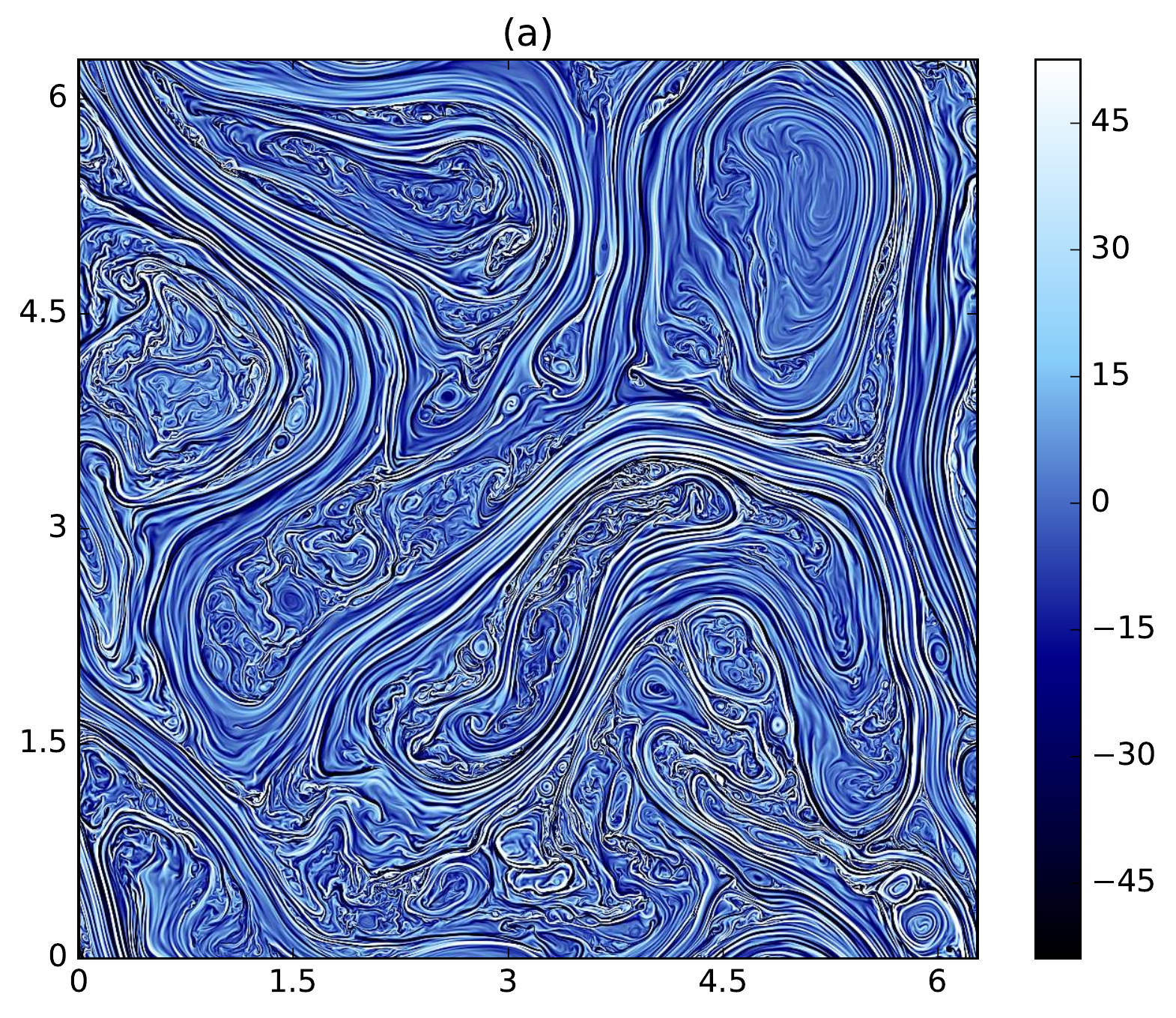}
 \includegraphics[width=0.49 \textwidth]{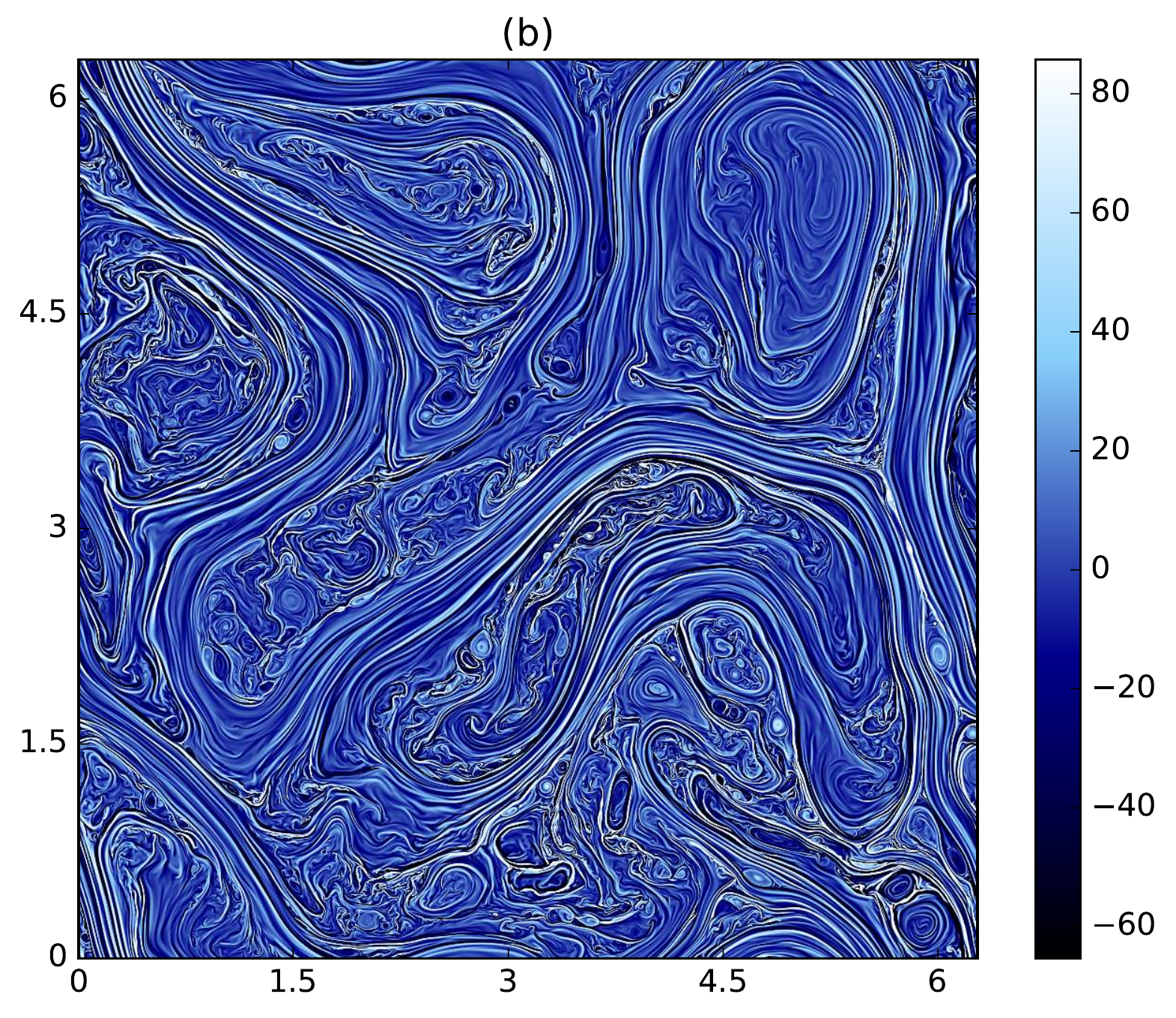}
 \caption{(a) Snapshot no. 2 of the vorticity field $\omega({\bf x},t)$
 from direct numerical simulations of 2D MHD turbulence.\\
 (b) Same snapshot
 no. 2 of the current density $j({\bf x},t)$.}
 \label{fig:om_j_690}
\end{figure}
\begin{figure}[p]
 \includegraphics[width=0.49 \textwidth]{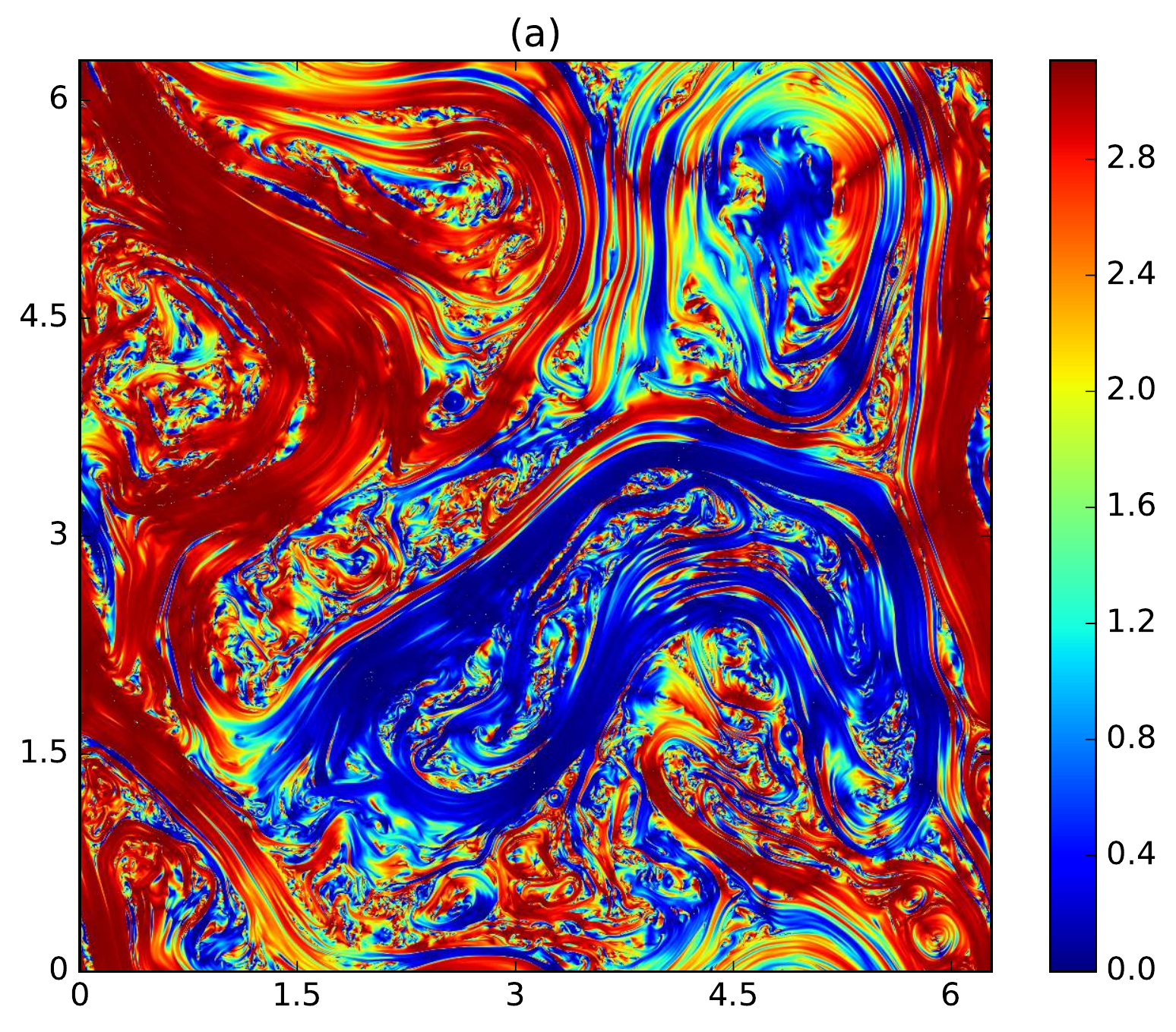}
  \includegraphics[width=0.49 \textwidth]{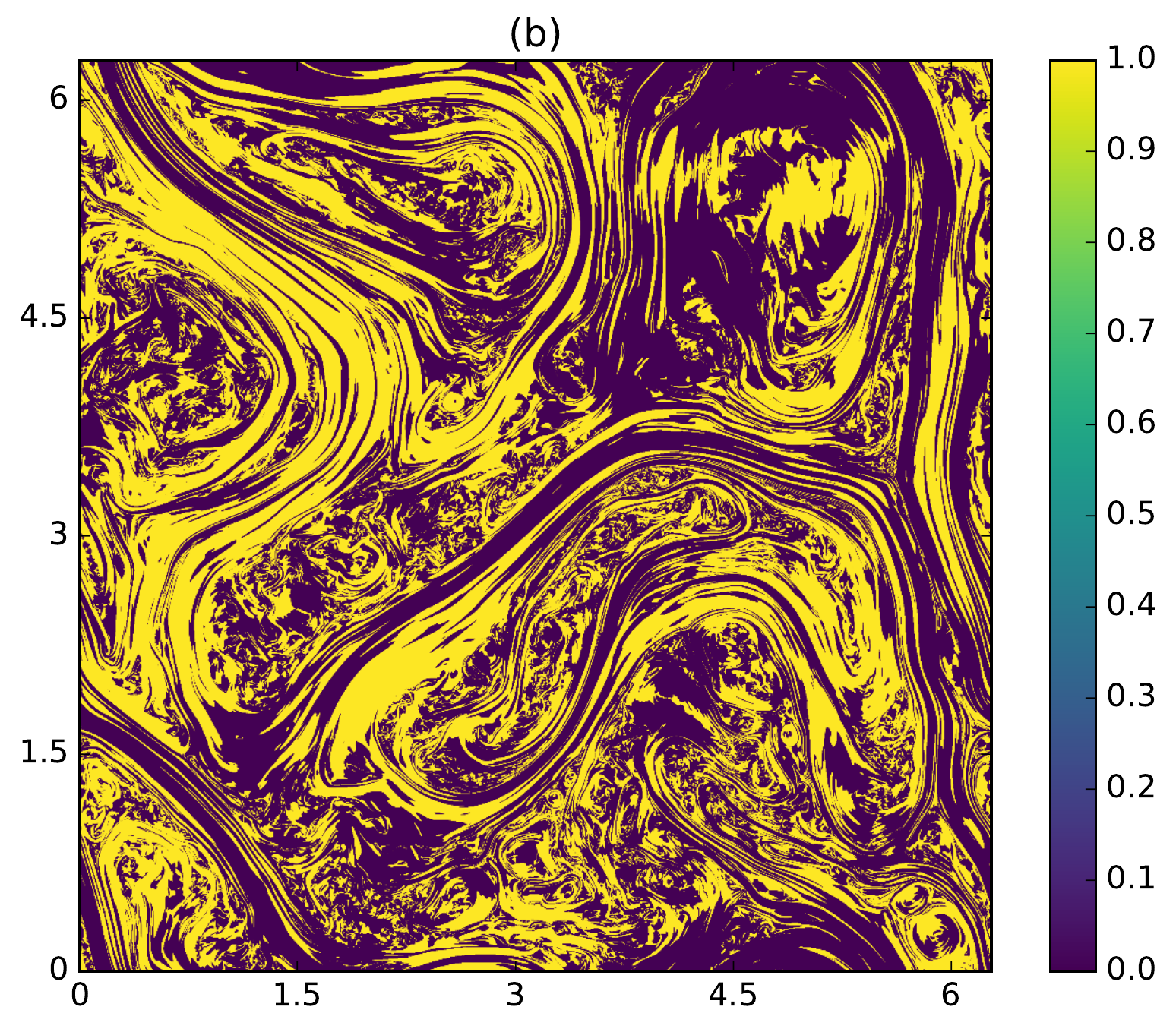}
  \caption{(a) Snapshot no. 2 of the alignment angle $\varphi({\bf x},t)$
  between velocity field and magnetic field. (b) Snapshot no. 2 of
  filter function $\chi({\bf x},t)$ in Eq. (\ref{eq:filter}).
  The covered area is 44.57 per cent of the total area.}
   \label{fig:phi_fil_690}
\end{figure}
\begin{figure}[p]
 \includegraphics[width=0.49 \textwidth]{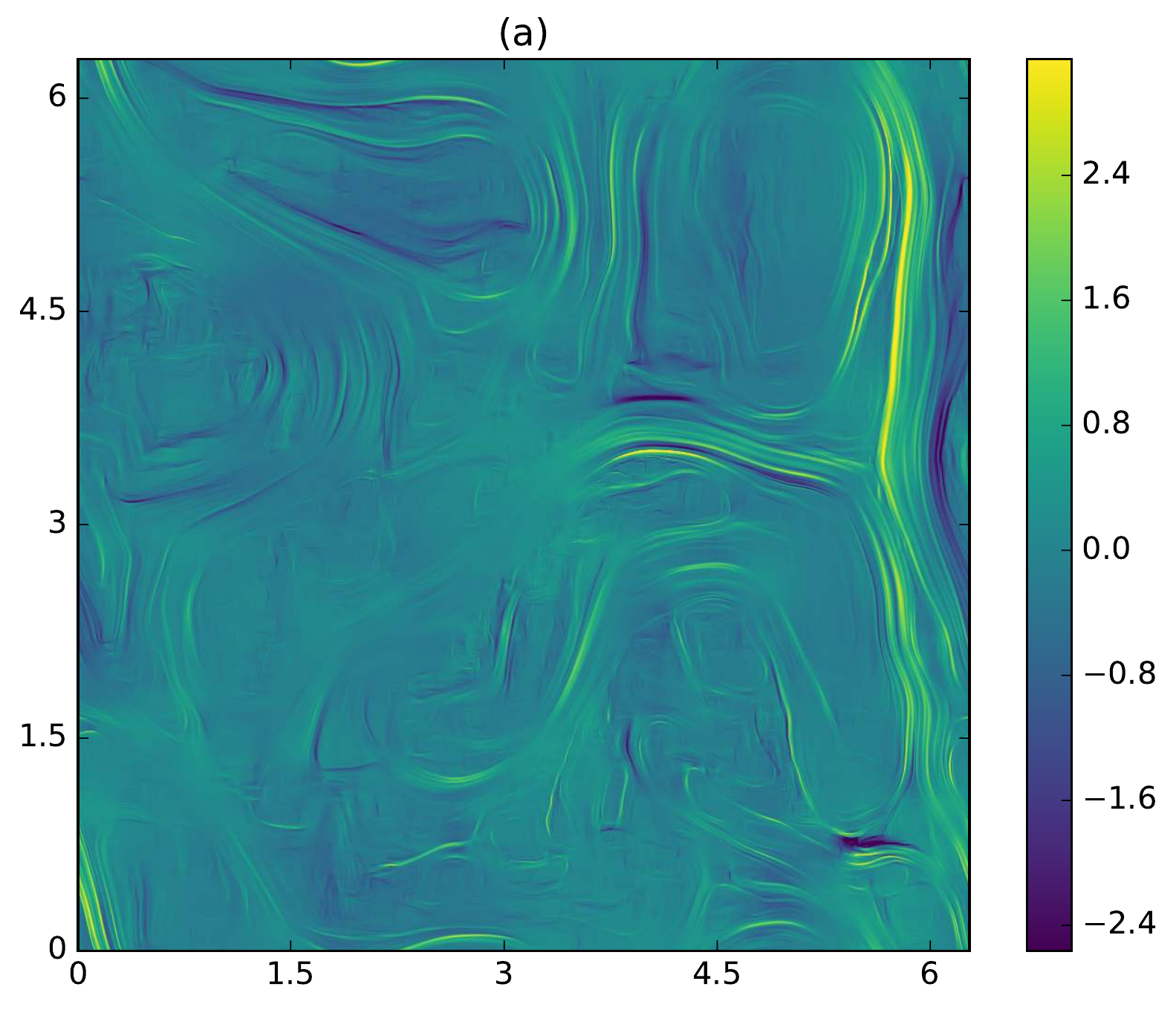}
  \includegraphics[width=0.49 \textwidth]{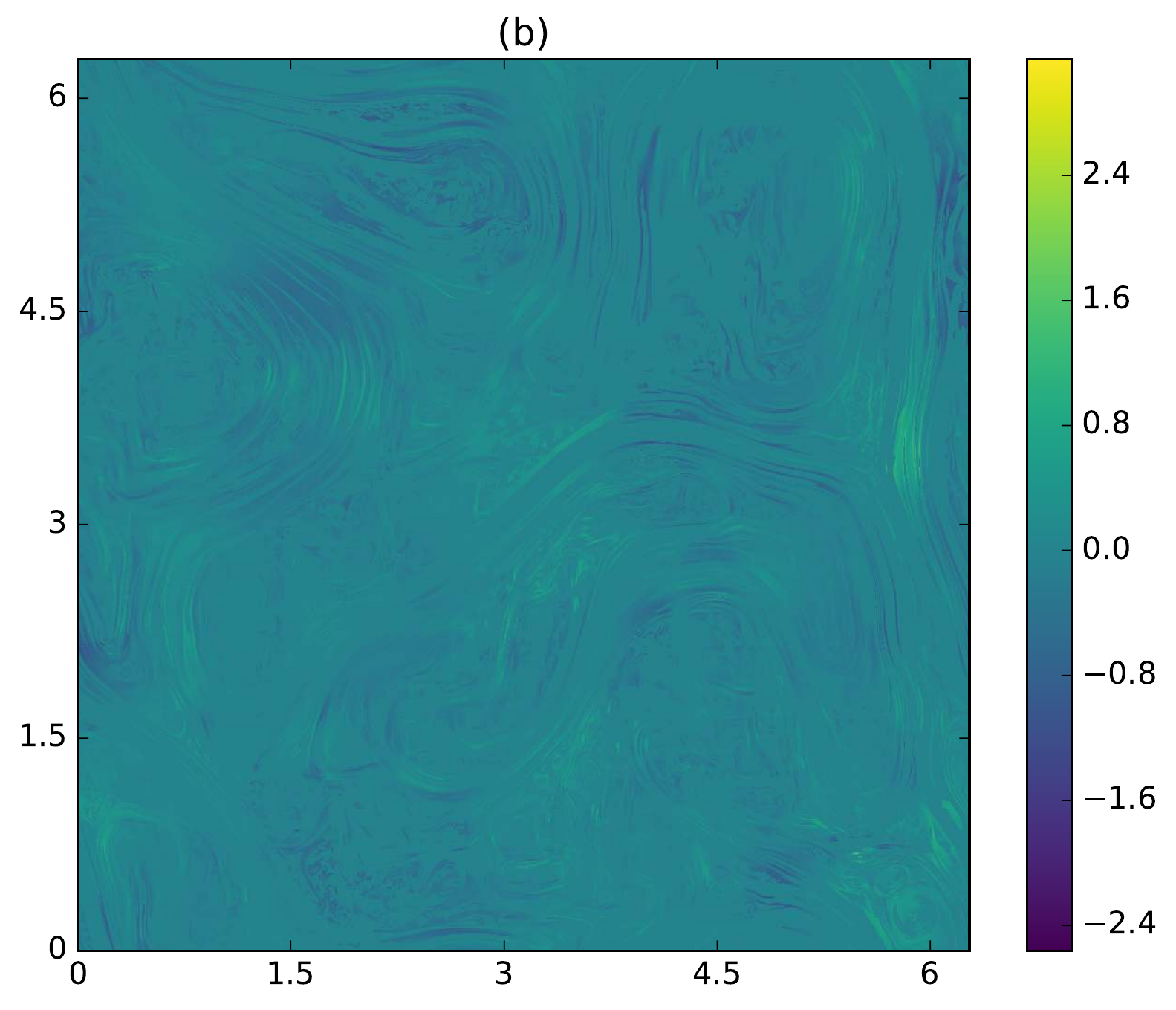}
  \caption{(a) Snapshot no. 2 of the total pressure field $p_{tot}({\bf x},t)$. \\
  (b) Snapshot no. 2 of the filtered pressure field via the filter function
  $\chi({\bf x},t)$ from Eq. (\ref{eq:filter}.)
  Again, the pressure field is effectively reduced.}
   \label{fig:pressure_690}
\end{figure}
\begin{figure}[p]
 \includegraphics[width=0.49 \textwidth]{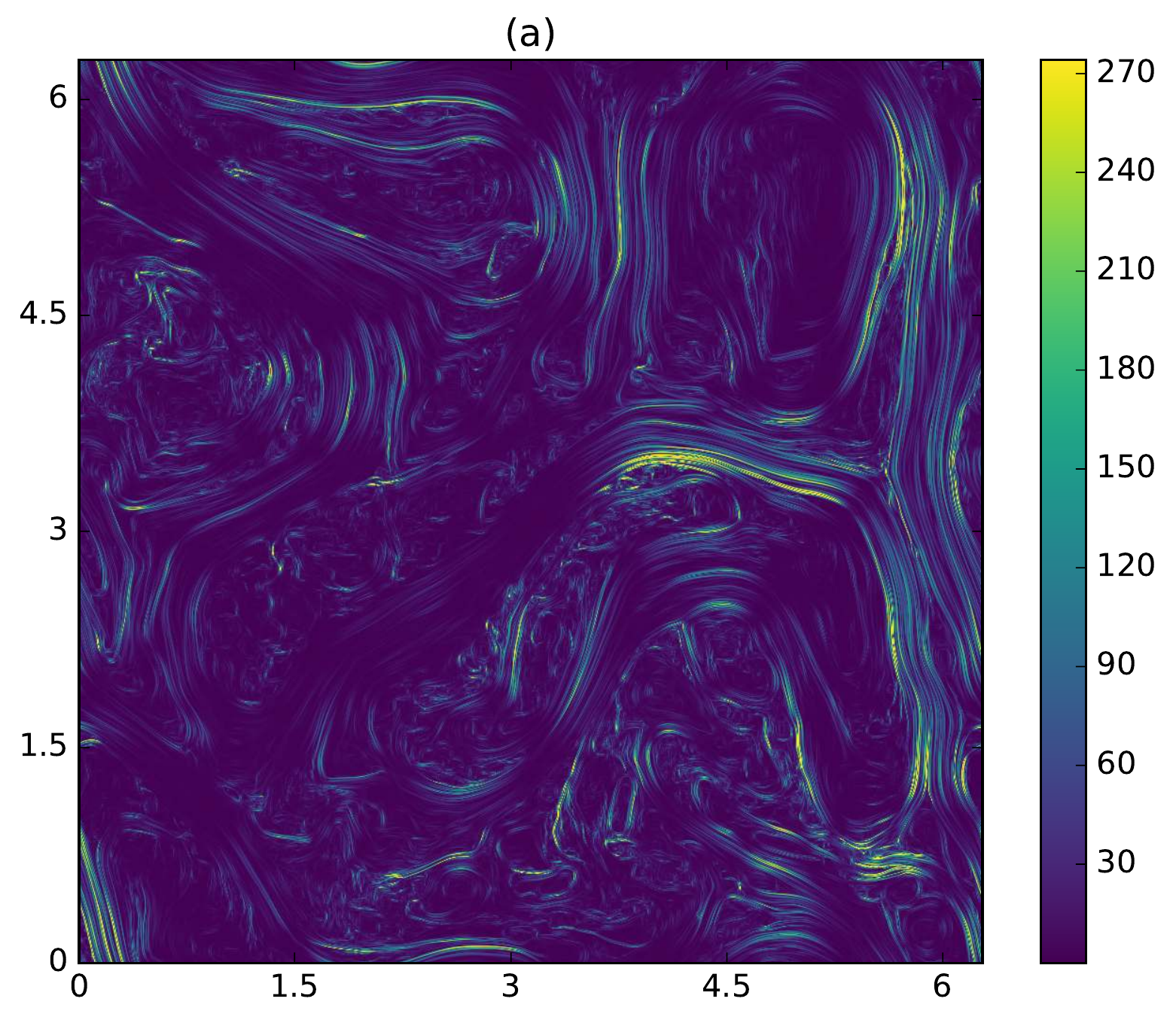}
  \includegraphics[width=0.49 \textwidth]{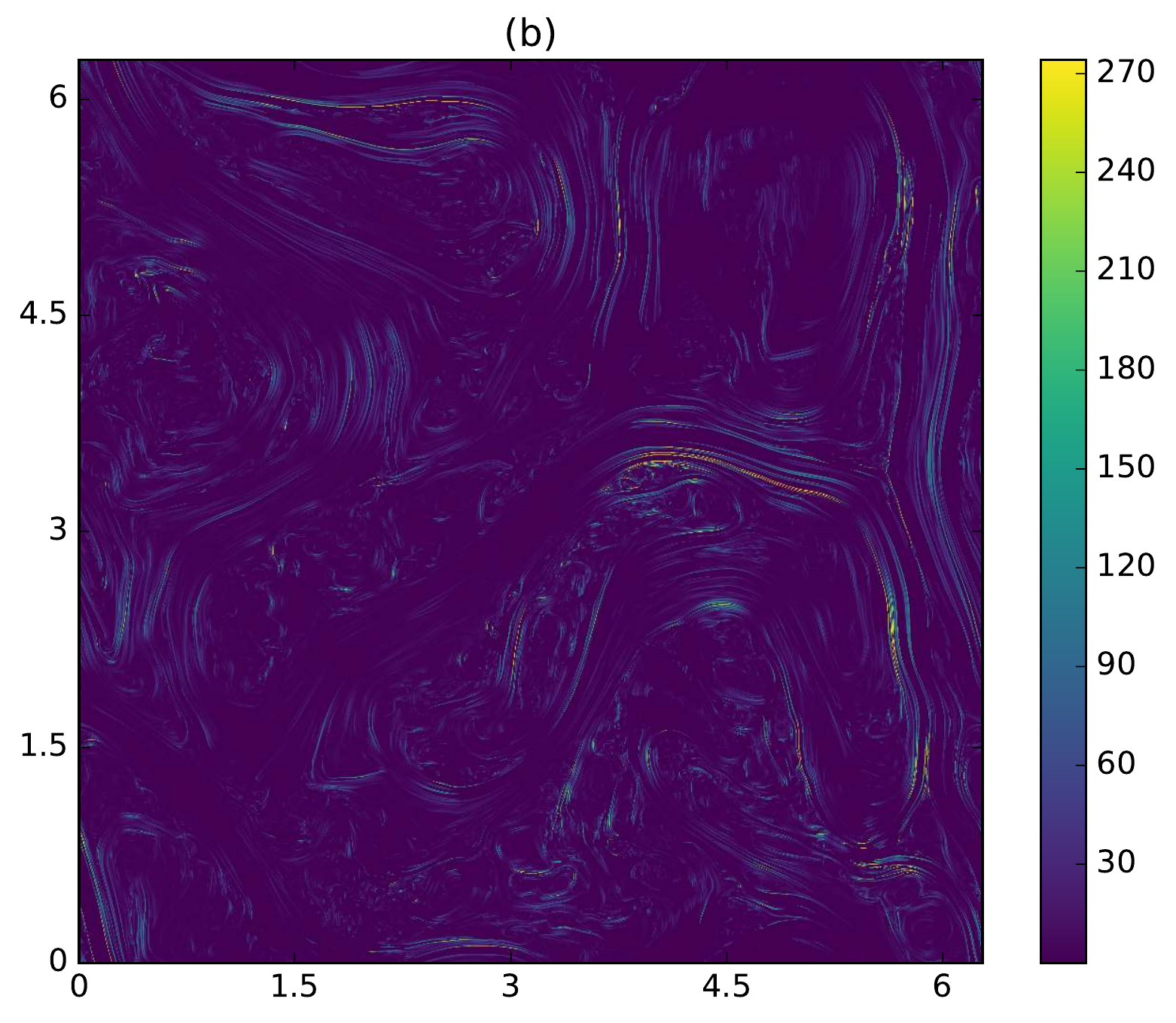}
  \caption{(a) Snapshot no. 2 of the norm of the total
 pressure gradient field $|\nabla p_{tot}({\bf x},t)|$. \\
  (b) Snapshot no. 2 of the filtered norm of the pressure gradient field.}
   \label{fig:absgradpressure_690}
\end{figure}
Consequently, the pressure contributions in MHD turbulence
that enter on the r.h.s
of Eqs. (\ref{eq:vier1})
and (\ref{eq:vier2})
have to be slightly milder than the ones in hydrodynamic turbulence.
In order to quantitatively discuss this behavior
it is convenient to consider the total pressure
$p_{tot}({\bf x},t) =p({\bf x},t) +
\frac{1}{2} |{\bf h}({\bf x},t)|^2$. Taking
the divergence of Eq. (\ref{eq:uhill}) results in a Poisson equation
for $p_{tot}({\bf x},t)$ that can be solved with the usual method
of Green's functions according to
\begin{eqnarray}
 p_{tot}({\bf x},t)
=\frac{1}{4\pi} \int \frac{\textrm{d} {\bf x}'}
 {|{\bf x}-{\bf x}'|} \frac{\partial^2 (u_i({\bf x}',t)u_j({\bf x}',t)
 -h_i({\bf x}',t)h_j({\bf x}',t))}{\partial x_i' \partial x_j'}.
 \label{eq:p_tot}
\end{eqnarray}
As in hydrodynamic turbulence, the total pressure
is thus determined \emph{nonlocally}
from the nonlinear terms in Eq. (\ref{eq:uhill}). Therefore,
if the MHD flow possesses \emph{local} regions
of alignment and equal magnitude, i.e.,
\begin{equation}
 {\bf u}({\bf x},t) = \pm {\bf h}({\bf x},t),
 \label{eq:equi}
\end{equation}
this results in an effective depletion of pressure contributions
that are mainly located in these regions due to the
$1/ |{\bf x}-{\bf x}'|$ dependence in Eq. (\ref{eq:p_tot})
(we refer the reader to the monograph \cite{chandra:1961}
for a further discussion of these so-called
equipartition solutions in MHD turbulence).
As a consequence, the pressure gradient increment (\ref{eq:pressure_inc})
that enters in the pressure contributions
of Eqs. (\ref{eq:vier1})
and (\ref{eq:vier2}) is also reduced.
\begin{figure}[p]
  \centering
 \includegraphics[width=0.49 \textwidth]{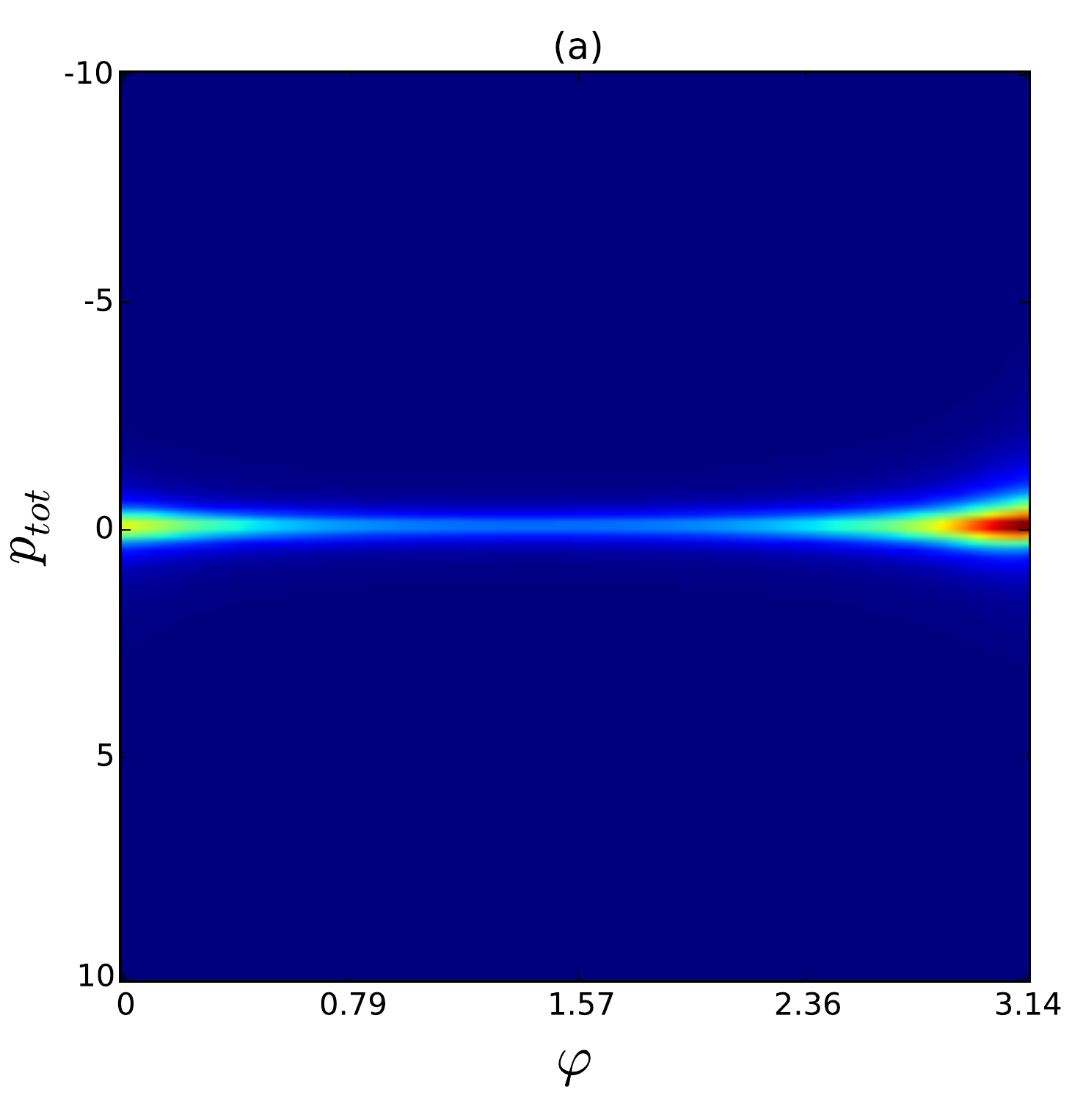}
  \includegraphics[width=0.49 \textwidth]{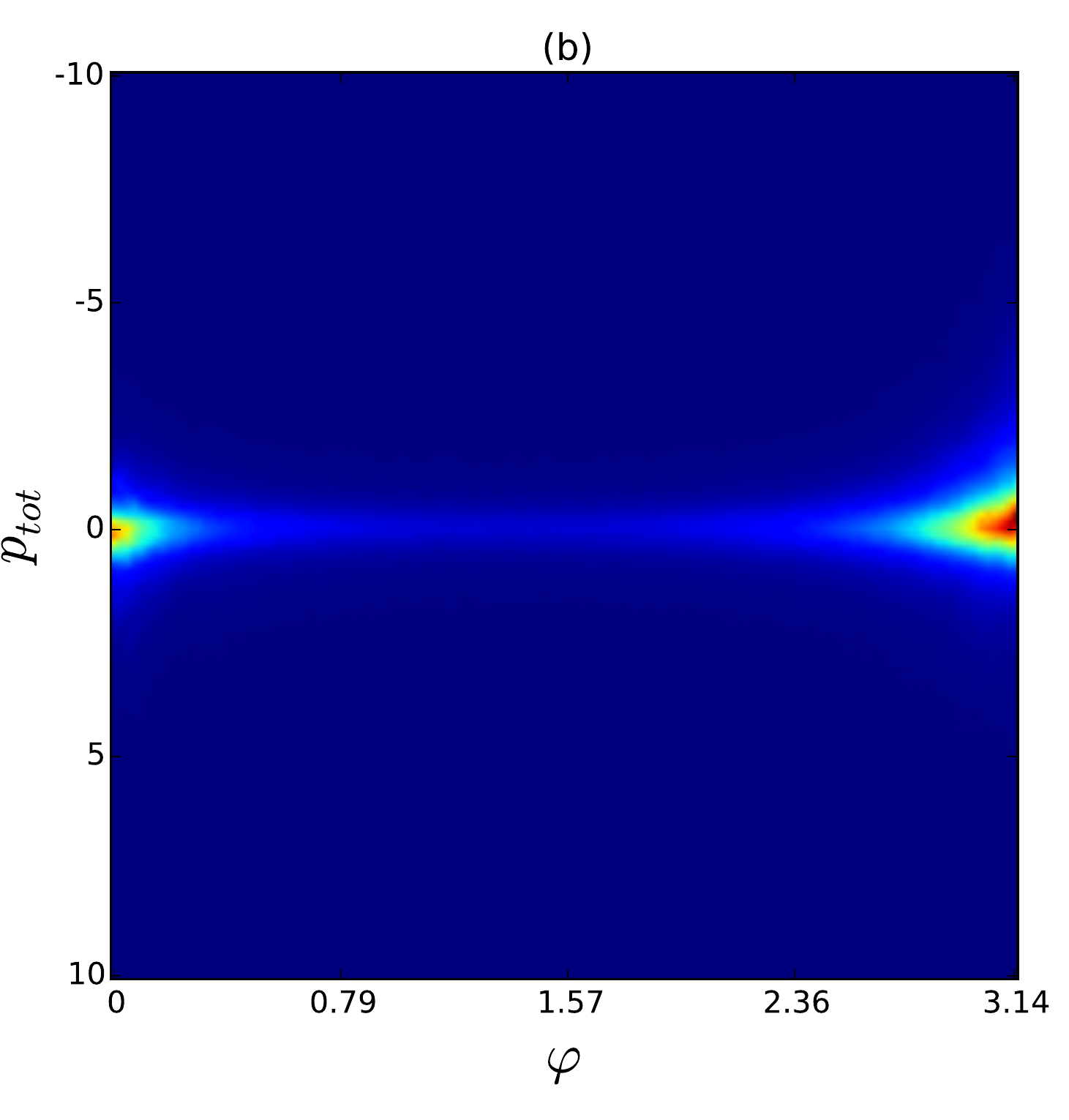}
   \caption{(a) Joint probability density function $PDF(p_{tot},\varphi)$ of pressure
                and alignment angle. Averages are performed over 700 snapshots. Vanishing pressure is found with high probability if the velocity and the magnetic field are aligned (anti-aligned). Note that in this particular time span of the simulations, anti-alignment occurs with a higher probability than alignment.  \\
            (b) Joint probability density function $PDF(p_{tot},\varphi)$ of pressure
                and alignment angle conditioned on the current density (20\%
                of the maximum). }
   \label{fig:2Dhistogram}
\end{figure}
\begin{figure}[p]
\centering
 \includegraphics[width=0.95 \textwidth]{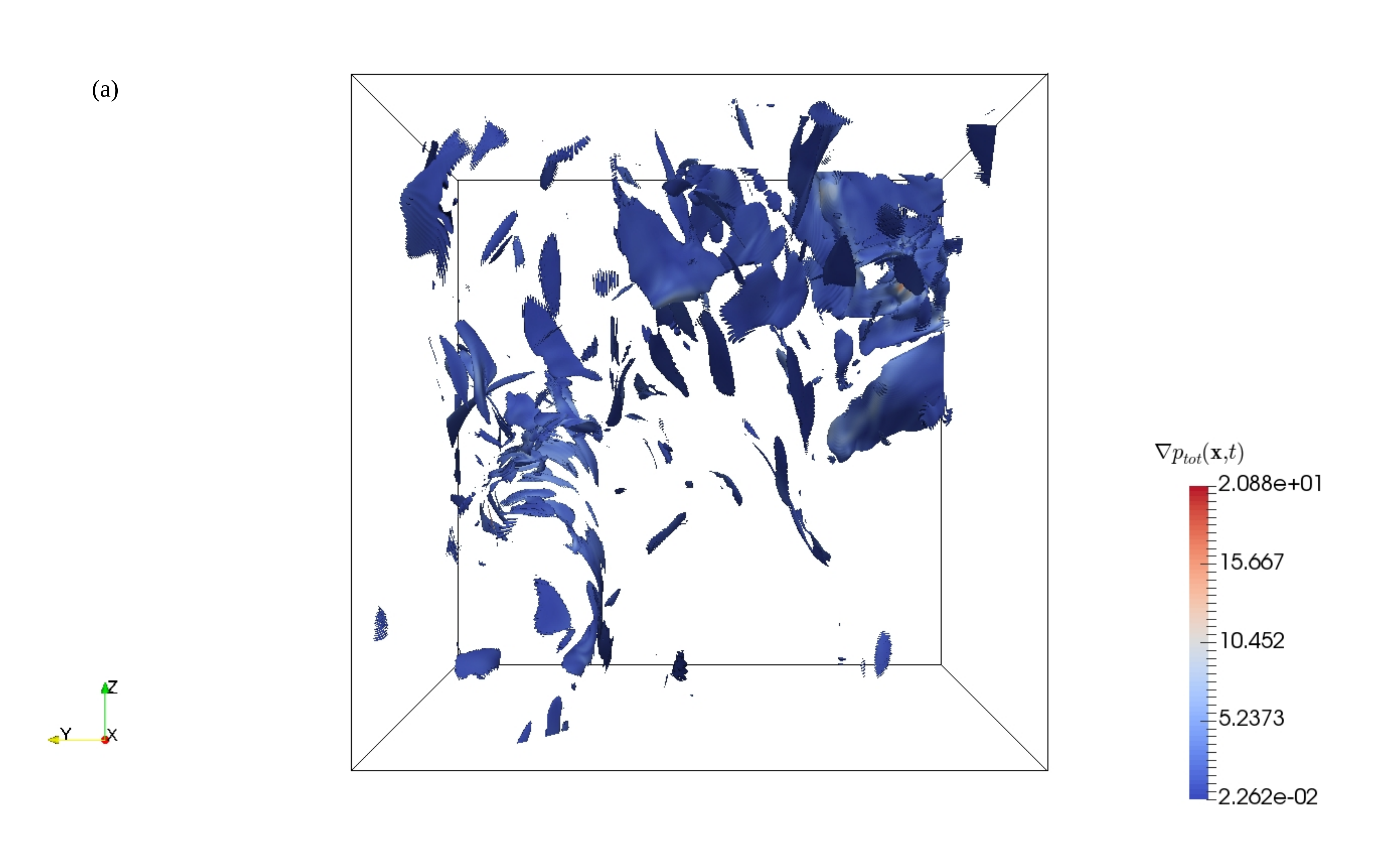}
 \includegraphics[width=0.95 \textwidth]{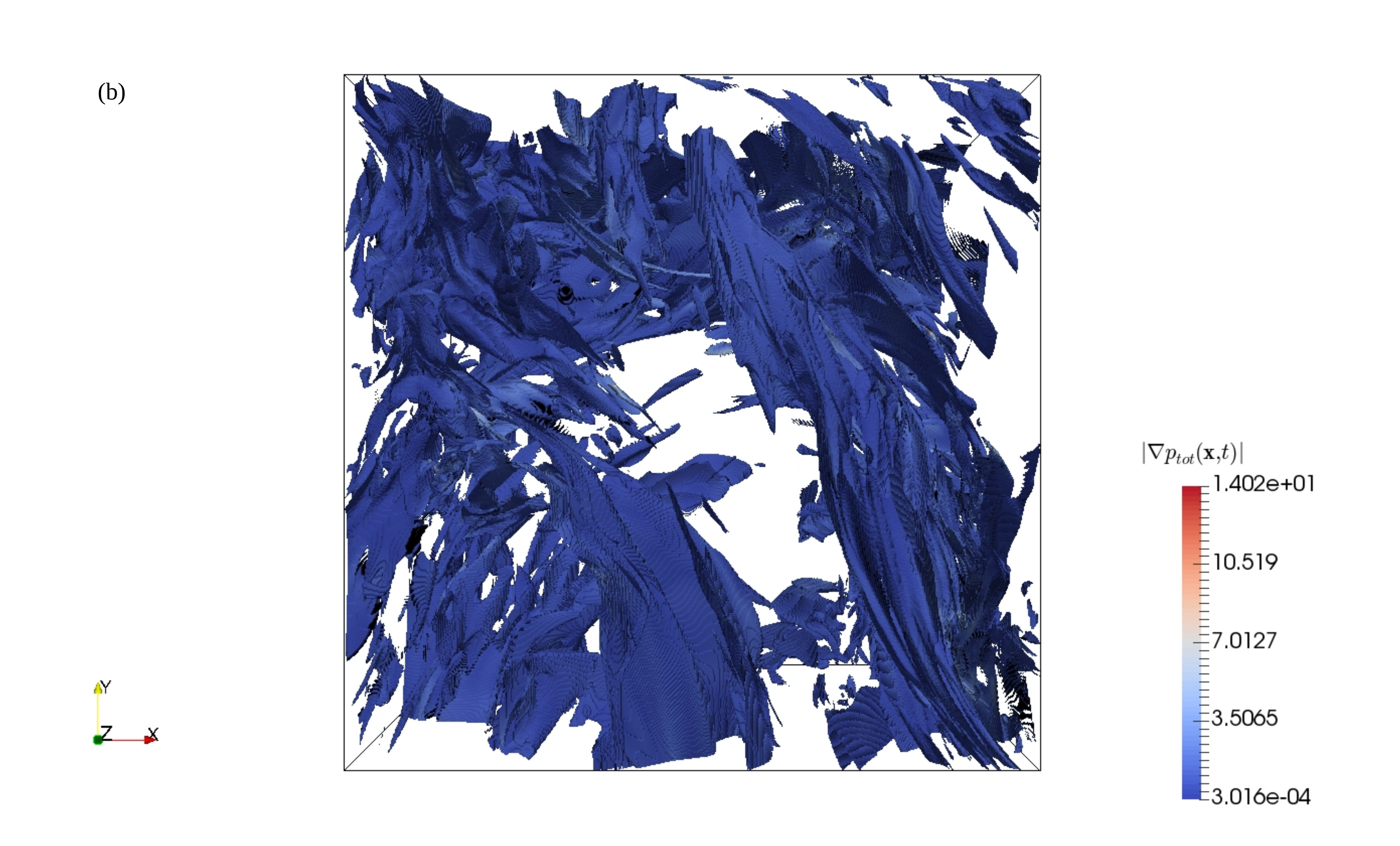}
 \caption{(a) Isosurface plot of the vorticity field $|\boldsymbol \omega({\bf x},t)|$
 of a $512^3$-segment
 from direct numerical simulations of 3D MHD turbulence ($2048^3$). The isosurface lies at
 $0.35 \,\omega_{max}$.  The volumes are color coded with the
 magnitude of the  total pressure gradient  $|\nabla p_{tot}({\bf x},t)|$.
 (b) Isosurface plot of the filter function  $\chi({\bf x},t)$ (\ref{eq:filter})
  of a $512^3$-segment
 from direct numerical simulations of 3D MHD turbulence ($2048^3$). The surface lies at 1 and
  is color coded with the
 magnitude of the total pressure gradient  $|\nabla p_{tot}({\bf x},t)|$. The total pressure gradient
 vanishes in regions of alignment and same magnitude of velocity and magnetic field.}
 \label{fig:om_3D}
\end{figure}
\begin{figure}[p]
 \includegraphics[width=0.49 \textwidth]{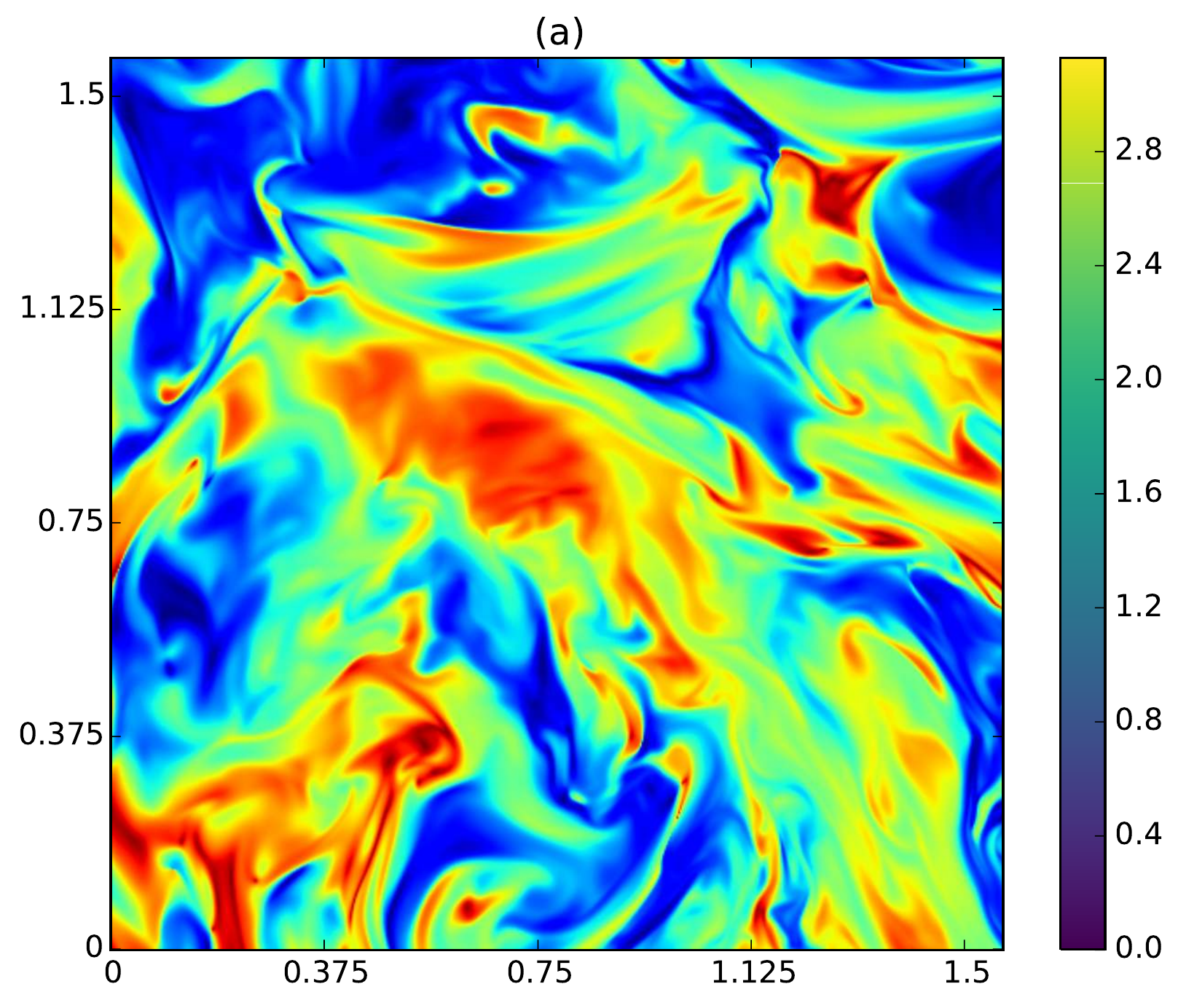}
  \includegraphics[width=0.49 \textwidth]{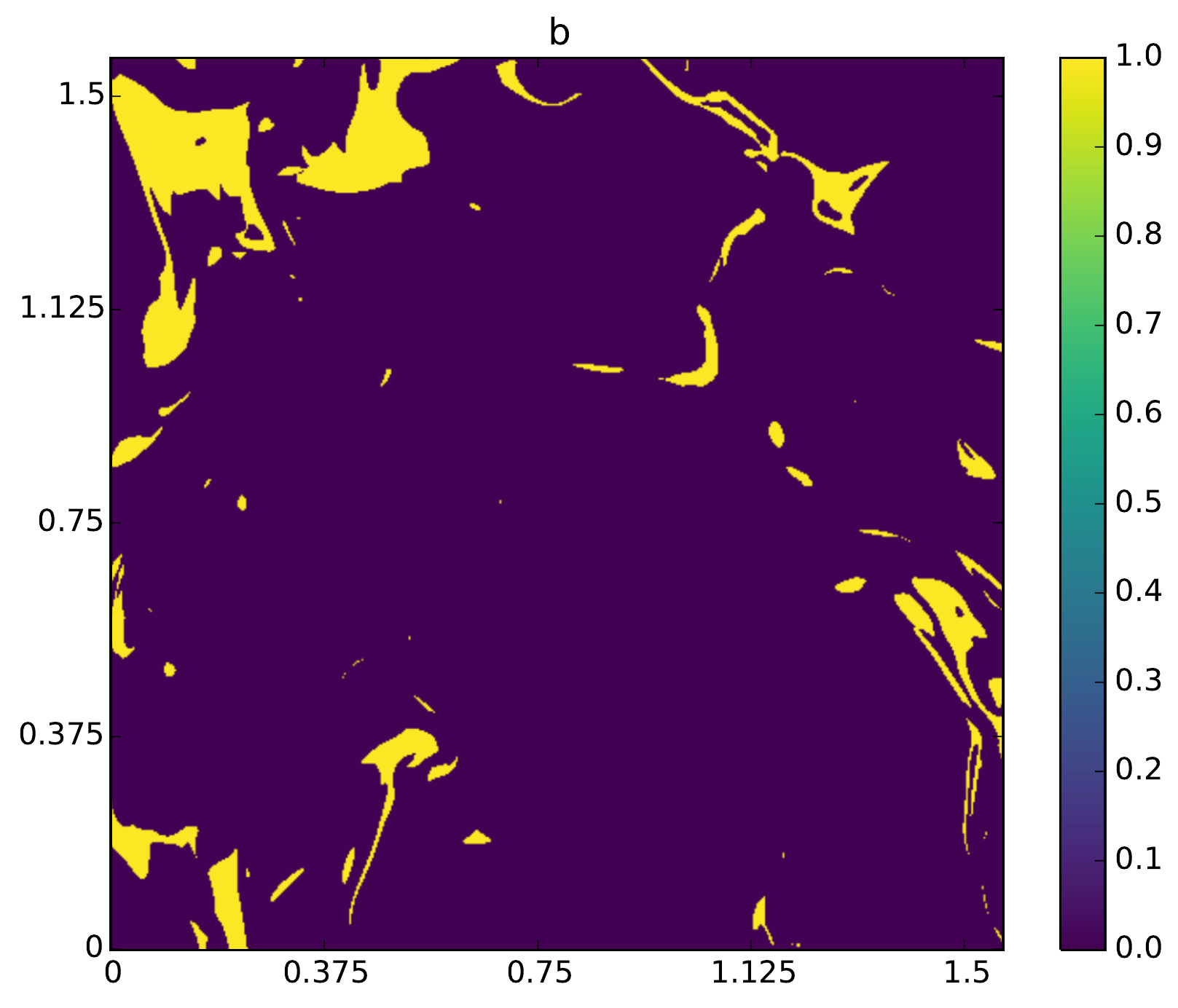}
  \caption{(a) Slice of the the alignment angle $\varphi({\bf x},t)$
  between velocity field and magnetic field at constant $z=\pi/4$ for 3D MHD turbulence.
  (b) Slice of
  filter function $\chi({\bf x},t)$ in Eq. (\ref{eq:filter}). The covered volume of the $512^3$-segment
  is  9.14 per cent.}
   \label{fig:phi_fil_3D}
\end{figure}
\begin{figure}[p]
 \includegraphics[width=0.49 \textwidth]{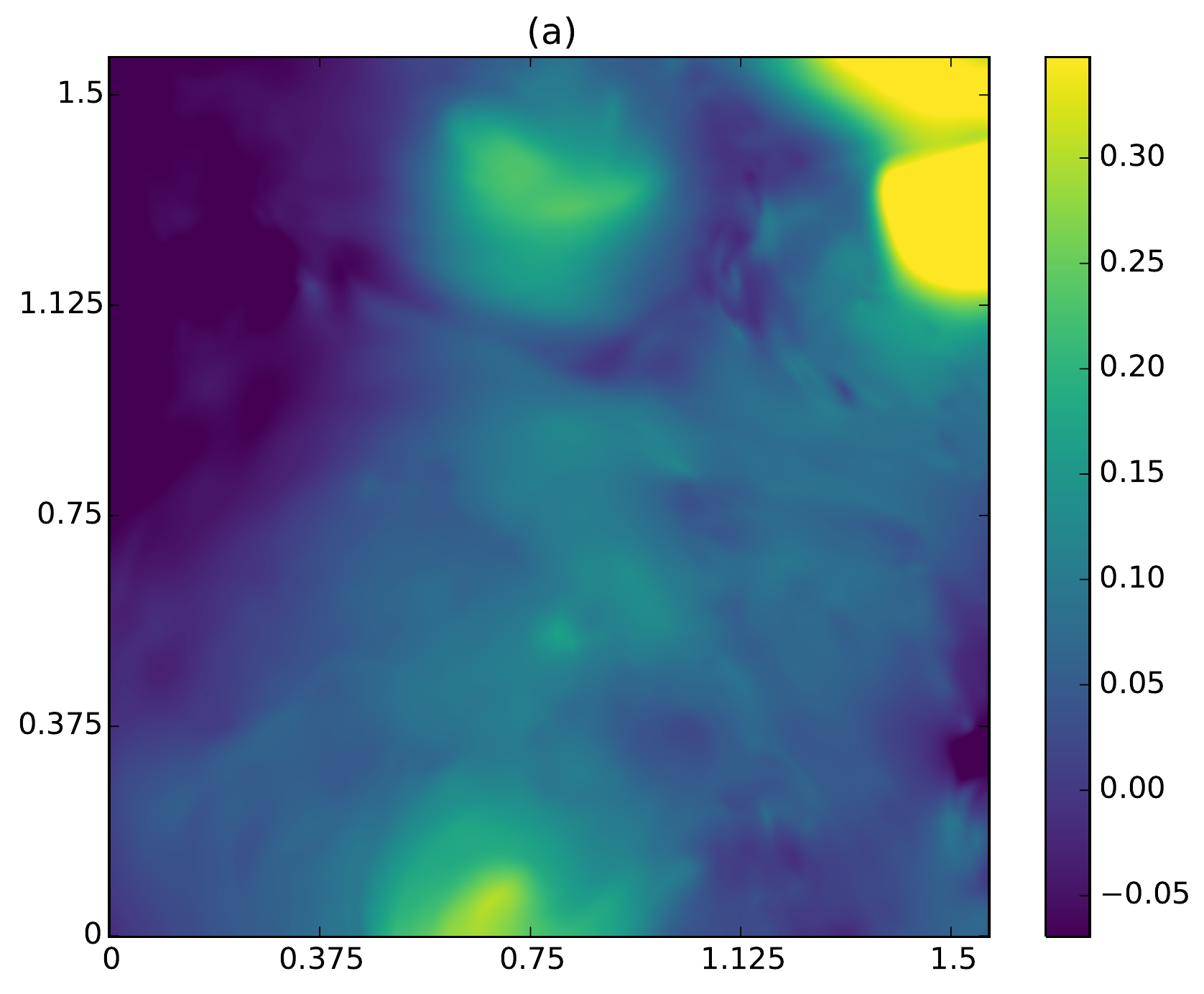}
  \includegraphics[width=0.49 \textwidth]{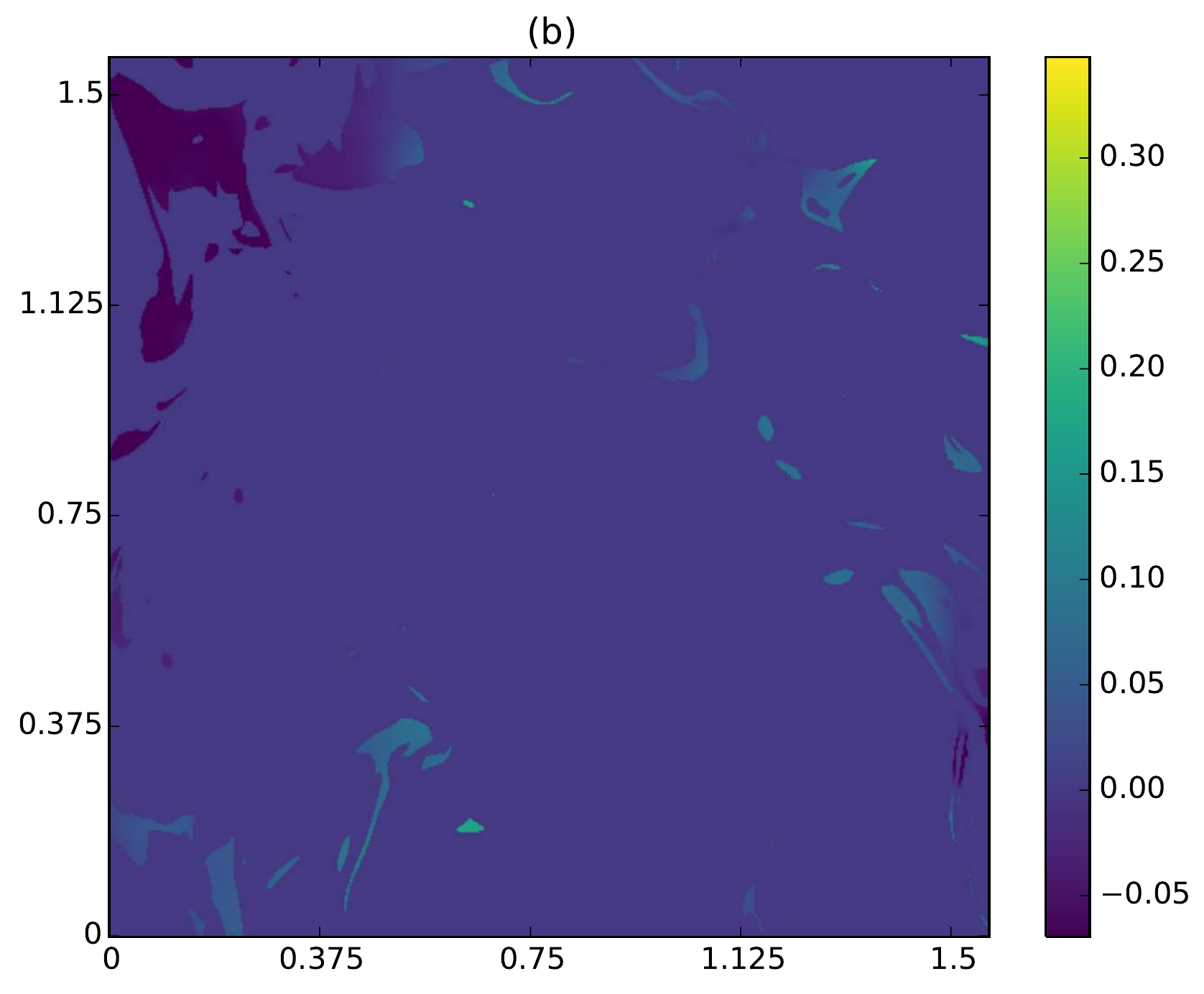}
  \caption{(a) Slice of the pressure field at constant $z=\pi/4$ from Fig. \ref{fig:om_3D} (b).  \\
  (b) Slice of the filtered pressure field via the filter function
  $\chi({\bf x},t)$ from Eq. (\ref{eq:filter}).}
   \label{fig:pressure_slice}
\end{figure}
\begin{figure}[p]
 \includegraphics[width=0.49 \textwidth]{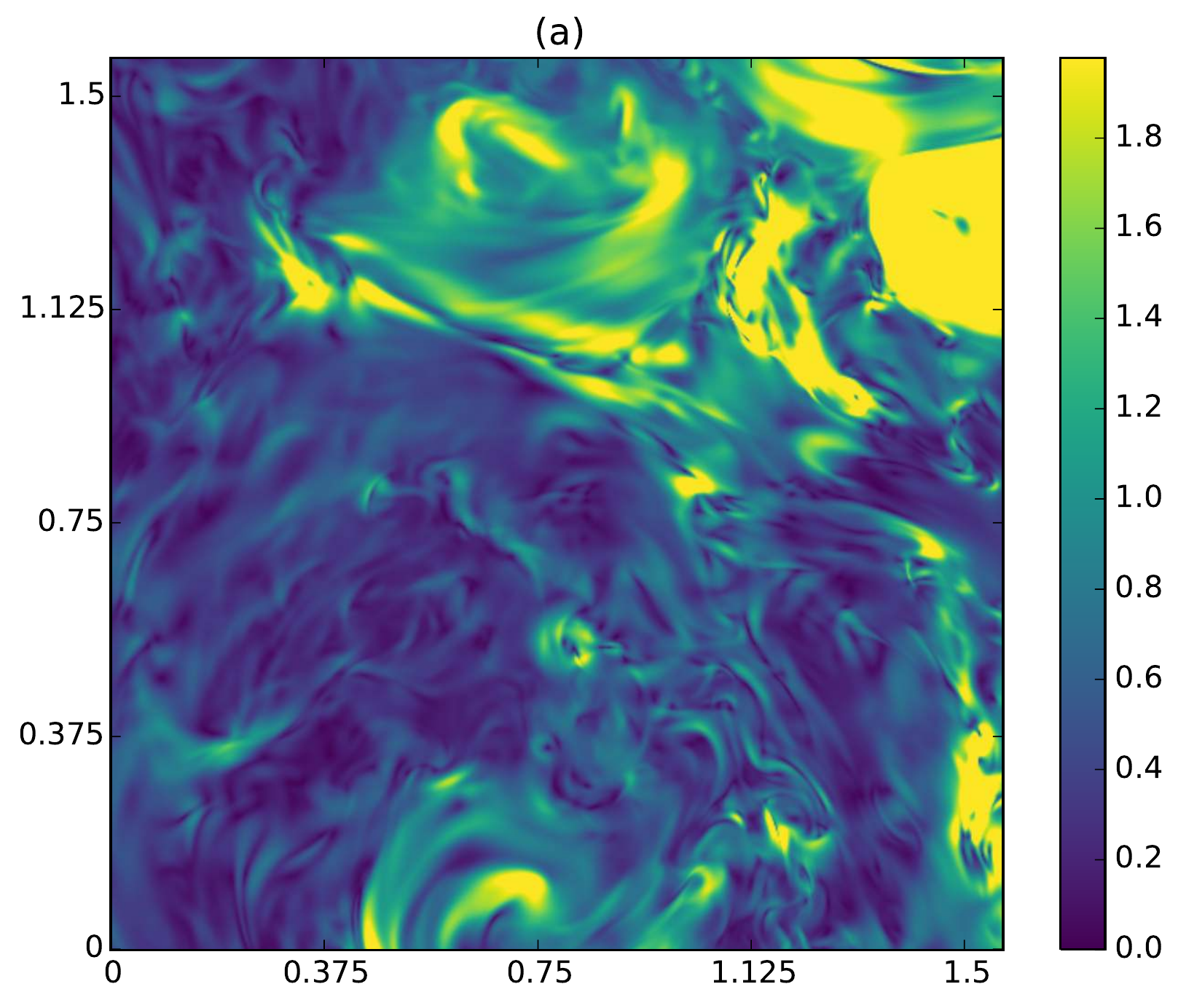}
  \includegraphics[width=0.49 \textwidth]{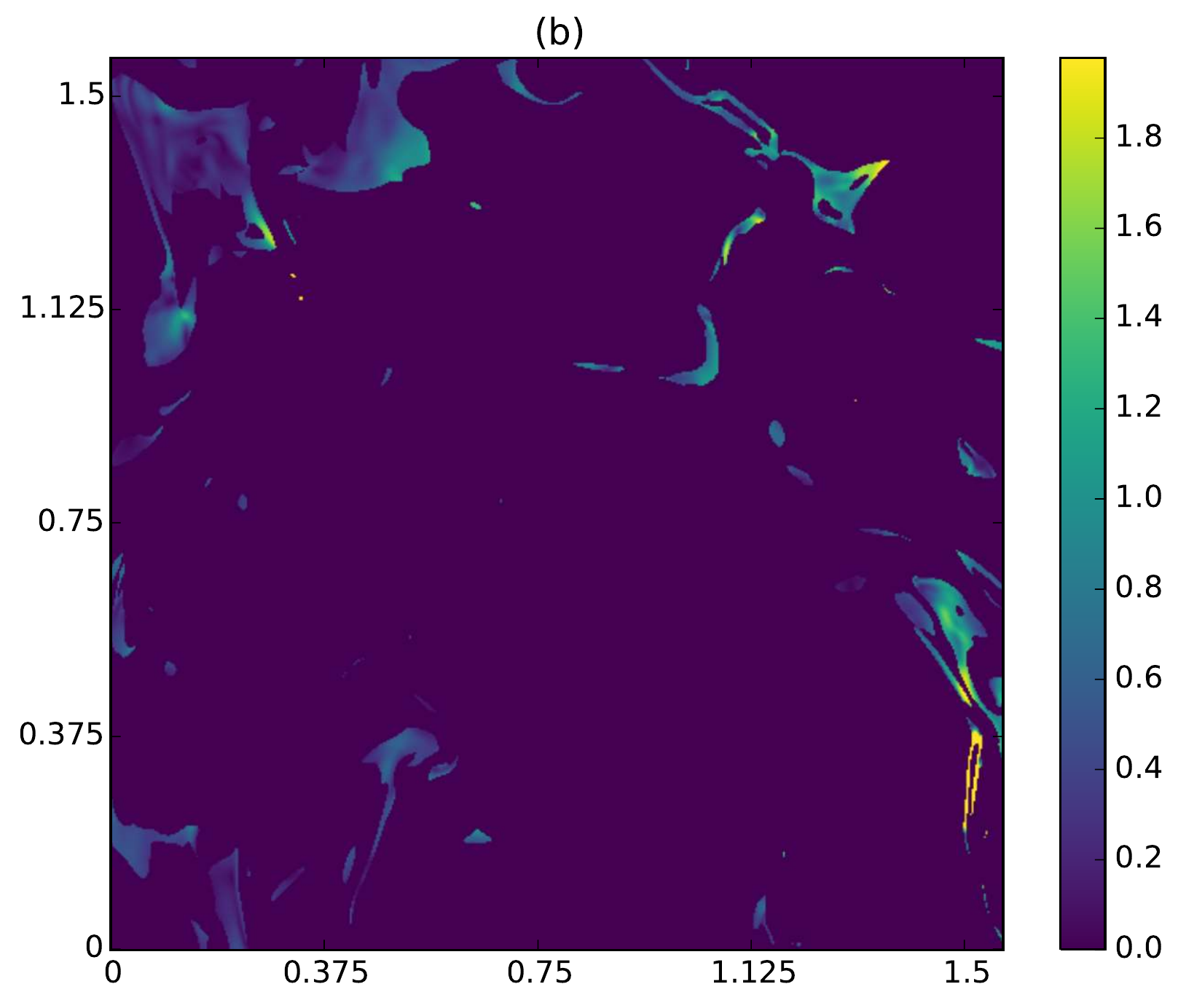}
  \caption{(a) Slice of the norm of the pressure gradient field $|\nabla p_{tot}({\bf x},t)|$
  at constant $z=\pi/4$ from Fig. \ref{fig:om_3D} (b).  \\
  (b) Slice of the filtered norm of the pressure gradient field via the filter function
  $\chi({\bf x},t)$ from Eq. (\ref{eq:filter}).}
   \label{fig:absgradpressure_slice}
\end{figure}
The described mechanism can thus
be considered as a
first explanation of the observations made in section \ref{sec:num}.
We emphasize that our alignment mechanism bears similarities
with the phenomenological theory of scale-dependent alignment proposed by Boldyrev
\cite{boldyrev:2005,Boldyrev2006} which is in accordance
with experimentally obtained energy spectra of MHD turbulence. Boldyrev
argued that small-scale turbulent eddies spontaneously develop
alignment of magnetic and velocity field polarizations with respect to
a large-scale magnetic field. The latter does not necessarily have to be
an external magnetic field, but can also be apprehended
as an effective magnetic field formed by the large scale vortical motions.
Similar to the case of decaying MHD turbulence which ultimately reaches the
equilibrium state in Eq. (\ref{eq:equi}), the alignment causes a depletion
of the nonlinearities in the MHD equations. In externally forced MHD turbulence, however,
the tendency for dynamic alignment is reduced due to the energy flux across scales
that preserves the nonlinear interactions. Therefore, at each scale $r$ of the
energy cascade the alignment should reach a maximum that is consistent with the constant
energy flux through this scale. In imposing the latter condition, Boldyrev
derived the field-perpendicular energy spectrum according to
$E(k_{\perp}) \sim k_{\perp}^{-3/2}$.

In order to get a first idea and for reasons
of graphical representation, we examine
the impact of local regions of
alignment on the pressure
field by direct numerical
simulations of forced 2D MHD turbulence.
The characteristic parameters
can be found in Table \ref{tab:2}. Here, we made use of
the concept of hyperviscosity by replacing $\nabla_{{\bf x}}^2
\rightarrow (\nabla_{{\bf x}}^2)^2$ in Eqs. (\ref{eq:uhill}) and (\ref{eq:hhill})
in order to attain higher Reynolds numbers and to extend the inertial range.
As a first impression, we have depicted a snapshot of
the vorticity $\omega({\bf x},t)= [\nabla
\times {\bf u}({\bf x},t)]_z$
and the current density $j({\bf x},t)= [\nabla \times {\bf h}({\bf x},t)]_z$
in Fig. \ref{fig:om_j_1}. The singular structures
in the direct cascading range mainly consist of current
and vortex sheets.
The alignment angle $\varphi({\bf x},t)$ between ${\bf u}({\bf x},t)$ and
${\bf h}({\bf x},t)$
is depicted in Fig. \ref{fig:phi_fil_1} (a). It can already be seen that
regions of preferential alignment and anti-alignment do in fact exist.
In this particular snapshot no. 1
they occur in a nearly dipolar manner, for instance
in the dipole on the left bottom.
The clustering
of like-signed alignment angles seems therefore to be in agreement with the arguments
established by Boldyrev \cite{boldyrev:2005,Boldyrev2006}.
The total pressure $p_{tot}({\bf x},t)$ of snapshot no. 1 is depicted in Fig. \ref{fig:pressure_1}.
and is mainly concentrated in dipolar boundary regions
of the alignment angle, where
flips out of the alignment (\ref{eq:equi}) occur. This seems to support
our hypothesis that pressure contributions are depleted in regions of alignment
and comparable magnitude of the velocity and magnetic field.
In order to further underline our hypothesis, we introduce a filter function
\begin{equation}
 \chi({\bf x},t)= \left \{\begin{array}{ll}
		1 & \textrm{where} ~~|\cos (\varphi({\bf x},t))| > 0.88 ~~\textrm{and}\\
		 & |u({\bf x},t)^2 - h({\bf x},t)^2| < 0.1 \times
		 \textrm{max} |u({\bf x},t)^2 - h({\bf x},t)^2|,
		\\
		0 & \textrm{ else}.
	   \end{array} \right.
	   \label{eq:filter}
\end{equation}
The filter function can roughly be interpreted as a measure of
a deviation of ten per cent of the alignment relation
(\ref{eq:equi}). Applying this simple filter function to
the total pressure in Fig. (\ref{fig:pressure_1})
shows that it is indeed depleted in
regions of approximate alignment and equal magnitude. Since regions of
maximal pressure contributions in Fig. \ref{fig:pressure_1} (b)
are filtered, they can be assigned to regions
where Eq. ({\ref{eq:equi}) is not fulfilled, i.e., ${\bf u}$ and
${\bf h}$ are not aligned.
A second snapshot no. 2 is presented in Figs.
\ref{fig:om_j_690}-{\ref{fig:absgradpressure_690}.
The snapshot no. 2 reveals an increased preference of aligned regions in comparison
to snapshot no. 1. The area covered by the filter function is 44.57 per cent (snapshot no. 2)
in comparison to 35.33 (snapshot no. 1). The alignment angle of snapshot no. 2 in
Fig. \ref{fig:phi_fil_690} shows a pronounced dipole structure of alignment (anti-alignment).
Pressure contributions in Fig. \ref{fig:pressure_690}
-{\ref{fig:absgradpressure_690} are also filtered to a great extend.
In order to quantify and support our statement that the alignment of velocity
and magnetic field is the cause for the absence of differences of longitudinal
and transverse structure function scaling we constructed a two-dimensional
histogram, where the joint probability density function of pressure and
alignment angle is shown. Fig. \ref{fig:2Dhistogram} (a) demonstrates that
with high probability the pressure is nearly zero if the alignment angle is
either $0$ or $\pi$. Moreover, Fig. \ref{fig:2Dhistogram} (b) shows the same
histogram but now conditioned  on the current density above a threshold (20\%
of the maximum). Here, the vanishing of the pressure at angles $0$ and $\pi$ is
even more pronounced and supports the above statement statement even stronger.

In the final part of this section, we extend our investigations to 3D
MHD turbulence which was the basis of our findings in section \ref{sec:num}.
To this end, we have depicted an isosurface plot of the vorticity in Fig. \ref{fig:om_3D} (a).
The singular structures can be perceived as vortex sheets, in contrast to the hydrodynamic
case that is dominated by tube-like vortical structures. Here, the plot consists of a $512^3$-segment
of the full $2048^3$-box from a snapshot of the simulation in Tab. \ref{tab:1}. Moreover, the volumes
have been colored with the magnitude of the total pressure gradient. The total pressure gradient
is seen to be small on strong vortex sheets. Fig. \ref{fig:om_3D} (b)
shows an isosurface plot of the filter function from Eq. (\ref{eq:filter}).
The isosurface has additionally been colored with the magnitude of the total pressure gradient.
Again, the regions of preferential alignment and same magnitude of velocity and magnetic field
directly correspond to regions of depleted total pressure gradient.
A slice of the filter function at constant $z=\pi/4$ is depicted in Fig. \ref{fig:phi_fil_3D} (b).
The covered volume of the filter function is 9.14 per cent of the entire $512^3$-box, which
is far less than in the 2D case.
Fig. \ref{fig:pressure_slice}} (a)
shows a slice for constant
$z=\pi/4$ of the total pressure field from Fig. \ref{fig:om_3D} (b).
The filtered total pressure in Fig. \ref{fig:pressure_slice}
(b) and the filtered total pressure gradient Fig. \ref{fig:absgradpressure_slice} (b)
supports our hypothesis that alignment regions effectively reduce pressure contributions.
\section{Conclusion and Outlook}
We derived a hierarchy of structure functions from the basic MHD equations.
The deployed procedure yielded exact equations between longitudinal and
transverse structure functions similar to the next order equations
by Hill and Boratav \cite{hill-boratav:2001} for the purely hydrodynamic
limit case. Furthermore, neglecting the pressure contributions
in these equations
allowed us to establish rescaling relations between longitudinal and
transverse structure functions in MHD turbulence. The latter were used
to directly compare the scaling behavior of the longitudinal
and transverse structure functions in the inertial range of direct numerical
simulations of
3D MHD turbulence. In contrast to hydrodynamic
turbulence, no clear scaling differences could be observed. This
rather unexpected
finding was explained by an effective reduction of the total pressure gradient
due to regions of preferential alignment and same magnitude of velocity
field and magnetic field fluctuations. In a first attempt, this
potential mechanism was tested with the help of direct numerical simulations
of 2D and 3D MHD turbulence. It could be shown that pressure contributions
are indeed depleted in such regions of preferential equipartition.
Further work will be dedicated to the identification of a similar mechanism
in hydrodynamic turbulence.

\section*{Acknowledgment}

Parts of this research were supported by the DFG-Research Unit FOR 1048,
project B2, and the French Agence Nationale de la Recherche
under grant ANR-11-BLAN-045, projet SiCoMHD.
Access to the IBM BlueGene/P computer JUGENE at the FZ
J\"ulich was made available through the project HBO36.
J.F. is grateful for discussions with H. Politano.

\appendix
\section{Derivation of the evolution equations for velocity and magnetic increments}
\label{app:increment_evo}
The derivation of the evolution equations for velocity and magnetic field increments
from the MHD equations (\ref{eq:uhill}) and (\ref{eq:hhill}) follows Hill's
procedure \cite{hill:2001} for the
case of hydrodynamic turbulence.
We introduce the velocity and magnetic field increments
\begin{eqnarray}
 { v}_i({\bf x},{\bf x}',t)&=&{ u}_i({\bf x},t)-{ u}_i({\bf x}',t),\\
{ b}_i({\bf x},{\bf x}',t)&=&{ h}_i({\bf x},t)-{ h}_i({\bf x}',t).
\end{eqnarray}
For brevity, let us denote $u_i({\bf x}',t)=u_i'$ and $h_i({\bf x}',t)=h_i'$. Furthermore, it is required that
${\bf x}$ and ${\bf x}'$ have no relative motion so that terms like $\frac{\partial}{\partial x_n}
u_i'$ and $\frac{\partial}{\partial x_n'} u_i$ vanish.\\
Subtracting the evolution equation of the velocity field (\ref{eq:uhill}) at point ${\bf x}'$
from the same equation at point ${\bf x}$ and performing
the same procedure for the induction equation (\ref{eq:hhill}) yields
\begin{eqnarray}
\fl \qquad \frac{\partial}{\partial t}v_i + u_n \frac{\partial}{\partial x_n} v_i + u_n'
\frac{\partial}{\partial x_n'} v_i -h_n \frac{\partial}{\partial x_n}b_i- h_n'
\frac{\partial}{\partial x_n'}b_i = -P_i + \nu (\nabla^2_{\bf x}+\nabla^2_{{\bf x}'}) v_i,\\
\label{eq:uhillneu}
\fl \qquad \frac{\partial }{\partial t}b_i + u_n \frac{\partial}{\partial x_n} b_i + u_n'
\frac{\partial}{\partial x_n'} b_i -h_n \frac{\partial}{\partial x_n}v_i- h_n'
\frac{\partial}{\partial x_n'}v_i  = \lambda (\nabla^2_{\bf x}+\nabla^2_{{\bf x}'}) b_i,
\label{eq:hhillneu}
\end{eqnarray}
where
\begin{equation}
 \fl P_i({\bf x},{\bf x}',t)=
 \left ( \frac{\partial}{\partial x_i}+ \frac{\partial}{\partial x_i'} \right)
 \left[ p({\bf x},t)-p({\bf
x}',t) + \frac{1}{2} (|{\bf h}({\bf x},t)|^2 -|{\bf h}({\bf x}',t)|^2)\right],
\end{equation}
is the total pressure increment.\\
In the following, we introduce the mean velocity and magnetic fields
\begin{equation}
 \qquad U_i({\bf x},{\bf x}',t) = \frac{u_i({\bf x},t) +u_i({\bf x}',t)}{2},
\end{equation}
\begin{equation}
 \qquad H_i({\bf x},{\bf x}',t) = \frac{h_i({\bf x},t) +h_i({\bf x}',t)}{2},
\end{equation}
and perform a coordinate transform to relative and center coordinates according to
\begin{equation}
 \qquad \quad~~ {\bf r}= {\bf x} -{\bf x}' \qquad
 \textrm{and} \qquad {\bf X}= \frac{{\bf x}+ {\bf x}'}{2} .
\end{equation}
The coordinate transform to relative and center coordinate implies the following
relations for the derivatives
\begin{equation}
 \frac{\partial}{\partial x_i}= \frac{1}{2} \frac{\partial}{\partial X_i}
  - \frac{\partial}{\partial r_i} \qquad \textrm{and} \qquad
  \frac{\partial}{\partial x_i'}= \frac{1}{2} \frac{\partial}{\partial X_i}
  + \frac{\partial}{\partial r_i}.
  \label{eq:transform}
\end{equation}
In making use of these relations, we obtain the equations of motion for the velocity and
the magnetic field increment
\begin{equation}\nonumber
\fl  \frac{\partial }{\partial t}v_i + v_n \frac{\partial}{\partial r_n} v_i + U_n
\frac{\partial}{\partial X_n} v_i - b_n \frac{\partial }{\partial r_n} b_i - H_n \frac{\partial
}{\partial X_n} b_i = - \frac{\partial}{\partial X_i} P + \nu (\nabla_{\bf x}^2 +
\nabla_{{\bf x}'}^2) v_i,
\label{eq:v_i}
\end{equation}
\begin{equation}
\fl \frac{\partial}{\partial t} b_i+ v_n \frac{\partial}{\partial r_n} b_i + U_n
\frac{\partial}{\partial X_n}b_i
- b_n \frac{\partial}{\partial r_n} v_i - H_n  \frac{\partial}{\partial X_n} v_i= \lambda
(\nabla_{\bf x}^2 + \nabla_{{\bf x}'}^2) b_i.
\label{eq:b_i}
\end{equation}
By means of these conversions, the implication of homogeneity for the statistical interpretation of
the increment equations (\ref{eq:v_i}) and (\ref{eq:b_i}) can be seen directly: If $\frac{\partial}{\partial
X_n}$ acts on a statistical quantity, then $ \frac{\partial}{\partial
X_n} \langle...\rangle=0 $, since this quantity should be independent of
the rate of change with respect to the place where the measurement is performed \cite{hill:2001}.
\section{Structure functions of second order}
\label{app:second-order}
As it has been pointed out by Chandrasekhar \cite{chandra:1951}, the MHD equations provide the evolution
equations for three different correlation functions of second order, namely $
\langle u_i({\bf x},t) u_j({\bf x}',t) \rangle$, $\langle h_i({\bf x},t) h_j({\bf x}',t) \rangle$
and the cross helicity correlation function $\langle u_i({\bf x},t) h_j({\bf x}',t) \rangle$.
Under the assumption
of homogeneity, the first two correlation functions possess the ordinary form of a tensor of order two,
i.e., the tensorial form that is going to be introduced in Eq. (\ref{eq:S_bb}).
The third tensor, however, is a quantity that is not invariant under the full rotation group.
This difference arises due to ${\bf h}$ being an axial vector
which is unchanged
under a reflexion whereas
the true polar vector ${\bf u}$ changes signs. Due to this lack of mirror symmetry, the cross
helicity correlation function possesses the following tensorial form
\begin{equation}
 C_{i\,j}^{{\bf u}{\bf h}}({\bf r},t)= \langle u_i({\bf x},t) h_j({\bf x}',t) \rangle =  C^{{\bf u}{\bf h}}(r,t)
 \epsilon_{ijn} \frac{r_n}{r}.
\end{equation}
In the following, our aim is to discuss a similar treatment for the structure functions of second order
in MHD turbulence. To this end, let us consider the magnetic structure function of order two
(the same treatment applies
to the velocity structure function of order two):
\begin{eqnarray}\nonumber
\fl \qquad &~&S_{i\,j}^{{\bf b}{\bf b}}({\bf x},{\bf x}',t)
 = \langle (h_i({\bf x},t) - h_i({\bf x}',t))
(h_j({\bf x},t) - h_j({\bf x}',t)) \rangle \\ \nonumber
 \fl &~&= \langle h_i({\bf x},t) h_j({\bf x},t) \rangle +  \langle h_i({\bf x}',t) h_j({\bf x}',t)
\rangle-  \langle h_i({\bf x},t) h_j({\bf x}',t) \rangle-  \langle h_i({\bf x}',t) h_j({\bf x},t)
\rangle\\
\fl &~&= 2  \langle h_i({\bf x},t) h_j({\bf x},t) \rangle  - 2  \langle h_i({\bf x},t) h_j({\bf x}',t)
\rangle ,
\end{eqnarray}
where homogeneity and isotropy were used in the last step, so that
\begin{eqnarray}\nonumber
  \langle h_i({\bf x}',t) h_j({\bf x}',t) \rangle&=&  \langle h_i({\bf x},t) h_j({\bf x},t) \rangle
= C_{i\,j}^{{\bf h}{\bf h}}(0,t)\\
  \langle h_i({\bf x},t) h_j({\bf x}',t) \rangle&=&  \langle h_i({\bf x}',t) h_j({\bf x},t) \rangle=
C_{i\,j}^{{\bf h}{\bf h}}({\bf r},t).
\end{eqnarray}
Therefore, the structure function  $S_{i\,j}^{{\bf b}{\bf b}}({\bf r},t)$ can be written in terms of
the correlation function $C_{i\,j}^{{\bf h}{\bf h}}({\bf r},t)$ according to
\begin{equation}
 S_{i\,j}^{{\bf b}{\bf b}}({\bf r},t)=2 (C_{i\,j}^{{\bf h}{\bf h}}(0,t)- C_{i\,j}^{{\bf h}{\bf
h}}({\bf r},t)),
\label{eq:Dij}
\end{equation}
whereas the cross helicity structure function behaves in another way due to the lack of mirror
symmetry
\begin{eqnarray}\nonumber
 \fl \qquad &~& S_{i\,j}^{{\bf v}{\bf b}}({\bf x},{\bf x}',t)
 = \langle (u_i({\bf x},t) - u_i({\bf x}',t))
(h_j({\bf x},t) - h_j({\bf x}',t)) \rangle \\ \nonumber
\fl &~& =\langle u_i({\bf x},t) h_j({\bf x},t) \rangle +  \langle u_i({\bf x}',t) h_j({\bf x}',t)
\rangle-  \langle u_i({\bf x},t) h_j({\bf x}',t) \rangle-  \langle u_i({\bf x}',t) h_j({\bf x},t)
\rangle\\
\fl &~&=2  \langle u_i({\bf x},t) h_j({\bf x},t) \rangle.
\end{eqnarray}
In the last step we made use of the assumption of homogeneity
\begin{equation}
 \langle u_i({\bf x},t) h_j({\bf x},t) \rangle = \langle u_i({\bf x}',t) h_j({\bf x}',t) \rangle
 =C_{i\,j}^{{\bf u}{\bf h}}(0,t),
\end{equation}
and
\begin{equation}
 \langle u_i({\bf x},t) h_j({\bf x}',t) \rangle = -\langle u_i({\bf x}',t) h_j({\bf x},t) \rangle
 =C_{i\,j}^{{\bf u}{\bf h}}({\bf r},t).
\end{equation}
This last relation is responsible for the different decomposition behavior of the second order tensor of
the cross helicity structure function, which can be expressed in terms of the cross helicity
correlation function according to
 \begin{equation}
 S_{i\,j}^{{\bf v}{\bf b}}({\bf r},t)=2 C_{i\,j}^{{\bf u}{\bf h}}(0,t).
\end{equation}
The  cross helicity structure function thus becomes a purely local quantity.
\subsection{The von K\'{a}rm\'{a}n-Howarth relation}\label{app:kar3d}
In general we are interested in relations between longitudinal and transverse structure functions.
Under the assumption of isotropy and homogeneity, the tensorial form of the
second order magnetic structure function reads
\begin{equation}
 S_{i\,j}^{{\bf b}{\bf b}}({\bf r},t)=(S_{r\,r}^{{\bf b}{\bf b}}(r,t) - S_{t\,\,t}^{{\bf b}{\bf b}}
(r,t)) \frac{r_i r_j}{r^2} + S_{t\,\,t}^{{\bf b}{\bf b}}(r,t) \delta_{ij},
\label{eq:S_bb}
\end{equation}
where the subscript $rr$ denotes the longitudinal and $tt$ the transverse structure function.
Eq. (\ref{eq:Dij}) implies that
\begin{equation}
 \frac{\partial}{\partial r_i }  S_{i\,j}^{{\bf b}{\bf b}}({\bf r},t)=0,
\end{equation}
due to the incompressibility condition for the magnetic field.
Inserting the tensorial form (\ref{eq:S_bb}) yields
\begin{eqnarray}
 \fl\qquad &~&\frac{\partial}{\partial r_i} S_{i\,j}^{{\bf b}{\bf b}}({\bf r},t)\\ \nonumber
\fl &=&\frac{\partial}{\partial r}
(S_{r\,r}^{{\bf b}{\bf b}}(r,t)-S_{t\,\,t}^{{\bf b}{\bf b}}(r,t)) \frac{r_j}{r} + \frac{2}{r}
(S_{r\,r}^{{\bf b}{\bf b}}(r,t)-S_{t\,\,t}^{{\bf b}{\bf b}}(r,t)) \frac{r_j}{r} +
\frac{\partial}{\partial r} S_{t\,\,t}^{{\bf b}{\bf b}}(r,t) \frac{r_j}{r}
=0,
\end{eqnarray}
 where we made use of $\frac{\partial}{\partial r_i}=\frac{r_i}{r} \frac{\partial}{\partial
r}$.
We finally obtain a first relation between the
longitudinal and the transverse structure function of second order
\begin{equation}
 S_{t\,\,t}^{{\bf b}{\bf b}}(r,t)=\frac{1}{2 r} \frac{\partial}{\partial r} \left( r^2 S_{r\,r}^{{\bf
b}{\bf b}}(r,t) \right),\label{eq:kuu}
\end{equation}
 which is known as the von K\'{a}rm\'{a}n-Howarth relation. The same relation holds also
 for the velocity structure function of second order whereas
  the cross helicity structure function bears no such relation.\\
Summing over equal $i$ and $j$ in (\ref{eq:S_bb}) and making use of relation (\ref{eq:kuu}) yields
\begin{equation}
 \langle v^2({r},t)\rangle = S_{r\,r}^{{\bf v}{\bf v}}(r,t)+ 2
S_{t\,\,t}^{{\bf v}{\bf v}}(r,t) =  \frac{1}{r^2}\frac{\partial}{\partial r} \left( r^3
S_{r\,r}^{{\bf v} {\bf v} } (r,t)  \right) ,
\end{equation}
and similarly
\begin{equation}
 \langle b^2({r},t)\rangle=  S_{r\,r}^{{\bf b}{\bf b}}(r,t)+ 2
S_{t\,\,t}^{{\bf b}{\bf b}}(r,t) =  \frac{1}{r^2}\frac{\partial}{\partial r} \left( r^3
S_{r\,r}^{{\bf b} {\bf b} } (r,t)  \right) ,
\end{equation}
which is needed for the averaged equation of energy balance in MHD turbulence (\ref{eq:energy_spherical}).
\section{Structure functions of third order}\label{app:Kol}
The velocity structure function of third order reads
\begin{equation}
 \fl \quad S_{i\,j\,n}^{{\bf v}{\bf v}{\bf v}}({\bf x},{\bf x}',t)= \langle (u_i({\bf x},t) - u_i({\bf
x}',t)) (u_j({\bf x},t) - u_j({\bf x}',t))(u_n({\bf x},t) - u_n({\bf x}',t))\rangle.
\end{equation}
 Since $\langle u_i({\bf x},t)u_j({\bf x},t)u_n({\bf x},t) \rangle=0$, the structure function
decomposes under the assumption of homogeneity and isotropy according to
\begin{equation}
  S_{i\,j\,n}^{{\bf v}{\bf v}{\bf v}}({\bf r},t)= -2( C_{\,i\,j,n}^{{\bf u} {\bf u} {\bf u}} ({\bf
r},t)+ C_{\,j\,n,i}^{{\bf u} {\bf u} {\bf u}} ({\bf r},t)+ 2 C_{\,n\,i,j}^{{\bf u} {\bf u} {\bf u}}
({\bf r},t)),
\label{eq:Dijn}
\end{equation}
where $C_{\,i\,j,n}^{{\bf u} {\bf u} {\bf u}} ({\bf r},t)$ is a third order tensor
that is symmetric in $ij$ and solenoidal in $n$, as it is described in \cite{Chandrasekhar1950}.
It can thus solely be expressed by
its longitudinal part
\begin{eqnarray}
  C_{\,i\,j,n}^{{\bf u} {\bf u} {\bf u}} ({\bf r},t)= &-&\frac{r^2}{2} \frac{\partial}{\partial r}
\left( \frac{  C_{r\,r\,r}^{{\bf u} {\bf u} {\bf u}} (r,t)}{r} \right) \frac{r_i r_j
r_n}{r^3}\\ \nonumber
 &+& \frac{1}{4 r}\frac{\partial}{\partial r} \left ( r^2 C_{r\,r\,r}^{{\bf u} {\bf u} {\bf u}}
(r,t)  \right) \left( \frac{r_i}{r} \delta_{jn} + \frac{r_j}{r} \delta_{in} \right) -
                                                               \frac{C_{r\,r\,r}^{{\bf u} {\bf u}
{\bf u}} (r,t)}{2} \frac{r_n}{r} \delta_{ij}.
\label{eq:Cijk}
\end{eqnarray}
Inserting this tensorial form in Eq. (\ref{eq:Dijn}) yields
\begin{equation*}
\fl\quad S_{i\,j\,n}^{{\bf v}{\bf v}{\bf v}}({\bf r},t)= 3 r^2 \frac{\partial}{\partial r}\left(
\frac{C_{r\,r\,r}^{{\bf u} {\bf u} {\bf u}} (r,t)}{r} \right) \frac{r_i r_j r_n}{r^3} -
\frac{\partial}{\partial r} ( r C_{r\,r\,r}^{{\bf u} {\bf u} {\bf u}} (r,t)) \left(\frac{r_n}{r}
\delta_{ij} +\frac{r_i}{r} \delta_{jn}  +\frac{r_j}{r} \delta_{in} \right) .
\end{equation*}
Now, we contract this tensor to a tensor of second order and obtain
 \begin{eqnarray}\nonumber
\fl \quad S_{r\,j\,n}^{{\bf v}{\bf v}{\bf v}}({\bf r},t)&=&  \frac{r_i}{r}  S_{i\,j\,n}^{{\bf v}{\bf v}{\bf
v}}({\bf r},t) \\
&=& \left( r\frac{\partial }{\partial r} C_{r\,r\,r}^{{\bf u} {\bf u} {\bf u}} (r,t)-
5C_{r\,r\,r}^{{\bf u} {\bf u} {\bf u}} (r,t) \right) \frac{r_j r_n}{r^2} -\frac{\partial}{\partial
r}( r C_{r\,r\,r}^{{\bf u} {\bf u} {\bf u}} (r,t)) \delta_{jn}.
\label{eq:rjn}
 \end{eqnarray}
Comparing this relation to the general form of a tensor of second order, for instance (\ref{eq:S_bb})
gives
\begin{eqnarray}
 \fl \quad S_{r\,t\,t}^{{\bf v}{\bf v}{\bf v}}({ r},t)&=&  -\frac{\partial}{\partial r}( r C_{r\,r\,r}^{{\bf
u} {\bf u} {\bf u}} (r,t)), \\
 \fl \quad S_{r\,r\,r}^{{\bf v}{\bf v}{\bf v}}({ r},t) &=& \left( r\frac{\partial }{\partial r}
C_{r\,r\,r}^{{\bf u} {\bf u} {\bf u}} (r,t)- 5C_{r\,r\,r}^{{\bf u} {\bf u} {\bf u}} (r,t) \right)+
 S_{r\,t\,t}^{{\bf v}{\bf v}{\bf v}}({ r},t) = - 6C_{r\,r\,r}^{{\bf u} {\bf u} {\bf u}} (r,t).
\end{eqnarray}
From these equations, a relation between the mixed and the
longitudinal velocity structure function of third order
can be derived as
\begin{equation}
  S_{r\,t\,t}^{{\bf v}{\bf v}{\bf v}}({ r},t)= \frac{1}{6} \frac{\partial}{\partial r} ( r
S_{r\,r\,r}^{{\bf v}{\bf v}{\bf v}}({ r},t)).
\label{eq:rtt}
\end{equation}
Summing (\ref{eq:rjn}) over equal indices $j=n$ gives
\begin{equation}
  S^{{\bf v}{\bf v}{\bf v}}({ r},t)=  \langle v_r(r,t) {\bf v}(r,t)^2 \rangle = S_{r\,r\,r}^{{\bf
v}{\bf v}{\bf v}}({ r},t)+2 S_{r\,t\,t}^{{\bf v}{\bf v}{\bf v}}({ r},t),
\end{equation}
which can be rewritten with the relation (\ref{eq:rtt}) as
\begin{equation}
 S^{{\bf v}{\bf v}{\bf v}}({ r},t)= \frac{1}{3r^3}\frac{\partial}{\partial r}\left( r^4 S^{{\bf v}
{\bf v} {\bf b}}_{r\, r\, r\,}(r,t) \right).
\end{equation}
This is the average which is needed for the averaged equation of energy balance in MHD turbulence
(\ref{eq:energy_spherical}).\\
The other average stems from the mixed third order tensor
\begin{equation}
  S^{{\bf b} {\bf b} {\bf v}} _{i\,j\,n} \left( {\bf r} , t \right) -
S^{{\bf v} {\bf b} {\bf b}} _{i\,j\,n} \left( {\bf r} , t \right) = U_{ijn}({\bf r},t) +H_{ijn}({\bf r},t),
\end{equation}
from equation (\ref{eq:decomp}) and
can be divided into
\begin{equation}
\fl \qquad U_{ijn}({\bf r},t) =-2 ( - \langle h_j h_n u_i' \rangle -\langle h_i h_n u_j' \rangle + \langle h_ih_j
u_n' \rangle),
\label{eq:U}
\end{equation}
and
\begin{equation}
\fl \qquad H_{ijn}({\bf r},t)=-2 (\langle ( u_n h_j-u_j h_n) h_i' \rangle +\langle (u_n h_i -u_i h_n) h_j' \rangle
- \langle (u_jh_i+ u_i h_j ) h_n' \rangle).
\end{equation}
Turning first to $U_{ijn}({\bf r},t)$, we have to evaluate correlation functions like $C^{{\bf h} {\bf h}
{\bf u}}_{\,i\,j,n} ({\bf r},t)= \langle h_i h_ j u_n' \rangle$. Since this tensor is again symmetric
in $ij$ and solenoidal in $n$, it has the same tensorial form as $C^{{\bf u} {\bf u}
{\bf u}}_{\,i\,j,n} ({\bf r},t)= \langle u_i u_ j u_n' \rangle$ in Eq. (\ref{eq:Cijk}) and
can solely be expressed in terms of $C_{r\,r\,r}^{{\bf h} {\bf h} {\bf u}} (r,t)$
\begin{eqnarray}
\fl \qquad U_{ijn}({\bf r},t)&=& \left(C_{r\,r\,r}^{{\bf h} {\bf h} {\bf u}} (r,t)- r
\frac{\partial}{\partial r} C_{r\,r\,r}^{{\bf h} {\bf h} {\bf u}} (r,t) \right)\frac{r_i r_j
r_n}{r^3}\\ \nonumber
\fl &~&- C_{r\,r\,r}^{{\bf h} {\bf h} {\bf u}} (r,t) \left( \frac{r_i}{r} \delta_{jn} + \frac{r_j}{r}
\delta_{in} \right)+ \left(3 C_{r\,r\,r}^{{\bf h} {\bf h} {\bf u}} (r,t) + r
\frac{\partial}{\partial r}C_{r\,r\,r}^{{\bf h} {\bf h} {\bf u}} (r,t)\right) \frac{r_n}{r}
\delta_{ij}.
\end{eqnarray}
This yields the following relations between the structure function $U_{ijn}(r,t)$ and the
correlation function $C_{r\,r\,r}^{{\bf h} {\bf h} {\bf u}} (r,t)$
\begin{eqnarray}
 U_{rrr}(r,t)&=&2 C_{r\,r\,r}^{{\bf h} {\bf h} {\bf u}} (r,t),\\
 U_{rtt}(r,t)&=&- C_{r\,r\,r}^{{\bf h} {\bf h} {\bf u}} (r,t),\\
 U_{ttr}(r,t)&=& 3 C_{r\,r\,r}^{{\bf h} {\bf h} {\bf u}} (r,t) +r \frac{\partial}{\partial r}
C_{r\,r\,r}^{{\bf h} {\bf h} {\bf u}} (r,t).
\end{eqnarray}
For $H_{ijn}(r,t)$ we need the antisymmetric tensor
\begin{equation}
  C^{{\bf u} {\bf h} {\bf h}}_{ \,i;j,n}({\bf r},t)=
  \langle (h_j u_i-u_j h_i) h_n' \rangle=
  C^{{\bf u} {\bf h} {\bf h}}_{ \,r;t\,t}(r,t)
  \left(\frac{r_j}{r}
\delta_{in} - \frac{r_i}{r} \delta_{jn}\right).
\end{equation}
and its symmetric counterpart
\begin{equation}
  C^{{\bf u} {\bf h} {\bf h}}_{ \,jn,i}({\bf r},t)=\langle (h_j u_n+u_j h_n) h_i' \rangle,
\end{equation}
which again fulfills the same equation as $C_{\,i\,j,n}^{{\bf u} {\bf u} {\bf u}} ({\bf r},t)$, namely
(\ref{eq:Cijk}).\\
We obtain
\begin{eqnarray}\nonumber
\fl \qquad \qquad  H_{ijn}({\bf r},t)=&~& \left(C_{r\,r\,r}^{{\bf u} {\bf h} {\bf h}} (r,t)- r
\frac{\partial}{\partial r} C_{r\,r\,r}^{{\bf u} {\bf h} {\bf h}} (r,t) \right)\frac{r_i r_j
r_n}{r^3}\\ \nonumber
\fl &+& \left(\frac{1}{2r} \frac{\partial}{\partial r} (r^2 C_{r\,r\,r}^{{\bf u} {\bf h} {\bf h}}
(r,t))+2  C^{{\bf u} {\bf h} {\bf h}}_{ \,r;t\,t}(r,t) \right) \left( \frac{r_i}{r} \delta_{jn} + \frac{r_j}{r}
\delta_{in} \right)\\
\fl &+& (-C_{r\,r\,r}^{{\bf u} {\bf h} {\bf h}} (r,t)
- 4 C^{{\bf u} {\bf h} {\bf h}}_{ \,r;t\,t}(r,t)) \frac{r_n}{r}
\delta_{ij} .
\end{eqnarray}
This yields the following relations
\begin{eqnarray}
 H_{rrr}(r,t)&=&2 C_{r\,r\,r}^{{\bf u} {\bf h} {\bf h}} (r,t),\\
 H_{rtt}(r,t)&=&\frac{1}{2 r} \frac{\partial}{\partial r} (r^2 C_{r\,r\,r}^{{\bf u} {\bf h} {\bf h}} (r,t)) + 2
C^{{\bf u} {\bf h} {\bf h}}_{ \,r;t\,t}(r,t),\\
 H_{ttr}(r,t)&=&-C_{r\,r\,r}^{{\bf u} {\bf h} {\bf h}}  -4 C^{{\bf u} {\bf h} {\bf h}}_{ \,r;t\,t}(r,t).
\end{eqnarray}
We need the function $S^{{\bf b} {\bf b} {\bf v}} \left( r , t \right) - S^{{\bf v} {\bf b} {\bf b}}
\left( r , t \right)$ for the averaged equation of energy balance in spherical coordinates.\\
 By making use of
\begin{equation}
 C_{r\,r\, r}^{{\bf h} {\bf h} {\bf u}} (r,t)=-2 C_{t\, t\, r}^{{\bf h} {\bf h} {\bf u}} (r,t),
\end{equation}
 we get
\begin{eqnarray}\nonumber
  S^{{\bf b} {\bf b} {\bf v}} \left( r , t \right) - S^{{\bf v} {\bf b} {\bf b}} \left( r , t
\right)&=& \langle v_r(r,t) b^2(r,t) \rangle - 2 \langle b_r(r,t) {\bf v}(r,t) \cdot {\bf b}(r,t)
\rangle \\ \nonumber
 &=& U_{rrr}(r,t) + 2U_{ttr}(r,t) + H_{rrr}(r,t) + 2H_{ttr}(r,t) \\
&=& -\frac{4}{r^3} \frac{\partial}{\partial r} (r^4 C_{t\,\, t\,\,r}^{{\bf h}{\bf h} {\bf u}}(r,t)
)- 8  C^{{\bf u} {\bf h} {\bf h}}_{ \,r;t\,t}(r,t).
\end{eqnarray}
Interestingly, the contributions from the symmetric correlation tensor $\langle (u_jh_i+ u_i h_j )
h_n' \rangle)$ vanish from this expression.
\section{Structure functions of fourth order}\label{app:fourth}
The tensor of fourth order, symmetric in all four indices is given by Monin and Yaglom \cite{monin}
in Vol. II by
formula (13.82). It has the form
\begin{eqnarray}\nonumber \label{vier}
\fl &~&S_{ijkn}({\bf r},t)= (S_{rrrr}(r,t) - 6 S_{rrtt}(r,t) +S_{tttt}(r,t) )
\frac{r_i r_j r_k r_n}{r^4}\\ \nonumber
\fl& +& (S_{rrtt}(r,t)-\frac{1}{3}S_{tttt}(r,t)) \left[ \frac{r_i r_j}{r^2} \delta_{kn}
+\frac{r_i r_k}{r^2} \delta_{jn}+\frac{r_i r_n}{r^2} \delta_{jk}
+\frac{r_j r_k}{r^2} \delta_{in}+\frac{r_j r_n}{r^2} \delta_{ik}+ \frac{r_k r_n}{r^2}
\delta_{ij} \right]\\
\fl &+& \frac{1}{3} S_{rrtt}(r,t) [\delta_{ij} \delta_{kn} +\delta_{ik}\delta_{jn}
+ \delta_{in}\delta_{jk} ].
\end{eqnarray}
If we calculate its divergence we get a tensor of third order, whose tensorial
form is given by Monin and Yaglom \cite{monin}
in Vol. II by formula (13.80), therefore we obtain
\begin{eqnarray}\nonumber
 \frac{\partial}{\partial r_n} S_{ijkn}({\bf r},t)&=& \left(\frac{\partial}{\partial r_n} S_{rrrn}({\bf r},t)
 - 3 \frac{\partial}{\partial r_n} S_{rttn}({\bf r},t) \right) \frac{r_i r_j r_n}{r^3}\\
 &+&\frac{\partial}{\partial r_n} S_{rttn}({\bf r},t)\left[ \frac{r_i}{r} \delta_{jk}+\frac{r_j}{r}
 \delta_{ik}+\frac{r_k}{r} \delta_{ij}\right].
\end{eqnarray}
This tensorial form can be compared with the original calculations. We get
\begin{eqnarray}\label{hill}\nonumber
 \fl \frac{\partial}{\partial r_n} S_{rrrn}({\bf r},t)
 - 3 \frac{\partial}{\partial r_n} S_{rttn}({\bf r},t)&=& \left(  \frac{\partial}{\partial r}
 + \frac{2}{r}\right) (S_{rrrr}(r,t) - 6 S_{rrtt}(r,t) +S_{tttt}(r,t) )\\
\fl &+&\left( 3 \frac{\partial}{\partial r} - \frac{6}{r}
\right)\left(S_{rrtt}(r,t)-\frac{1}{3}S_{tttt}(r,t)\right) ,
\end{eqnarray}
and
\begin{equation}\label{hill1}
 \frac{\partial}{\partial r_n} S_{rttn}({\bf r},t)= \left(\frac{\partial}{\partial r}+\frac{4}{r}
\right) S_{rrtt}(r,t) - \frac{4}{3r} S_{tttt}(r,t).
\end{equation}
The last equation has to be inserted into (\ref{hill}) in order to get the equation for
the longitudinal structure function $\frac{\partial}{\partial r_n} S_{rrrn}({\bf r},t)$.\\
For the antisymmetric tensor
\begin{equation}
  S_{i\,j;k\,n}^{{\bf v}{\bf v}{\bf b}{\bf b}}({\bf r},t) = \langle v_i v_j b_k b_n - b_i b_j v_k
v_n \rangle =S_{r\,r;t\,t}^{{\bf v}{\bf v}{\bf b}{\bf b}}({ r},t) \left(\frac{r_i r_j}{r^2}
\delta_{kn} -\frac{r_k r_n}{r^2} \delta_{ij} \right).
\end{equation}
we calculate the divergence as
\begin{eqnarray}\nonumber
  \frac{\partial}{\partial r_n} S_{i\,j;k\,n}^{{\bf v}{\bf v}{\bf b}{\bf b}}({\bf r},t)& =
&\frac{\partial }{\partial r}   S_{r\,r;t\,t}^{{\bf v}{\bf v}{\bf b}{\bf b}}({ r},t)\left[\frac{r_i
r_j r_k}{r^3} -  \frac{r_k}{r} \delta_{ij} \right]\\
 &+& \frac{ S_{r\,r;t\,t}^{{\bf v}{\bf v}{\bf b}{\bf b}}({ r},t)}{r} \left[\frac{r_i }{r} \delta_{jk}
 + \frac{r_j}{r} \delta_{ik} -2 \frac{r_i r_j r_k}{r^3} -2 \frac{r_k}{r} \delta_{ij} \right] .
\end{eqnarray}
Let us define $A_{ijk,n}({\bf r},t)$ as
\begin{equation}
A_{ijk,n}({\bf r},t) =  S_{i\,j;k\,n}^{{\bf v}{\bf v}{\bf b}{\bf b}}({\bf r},t)
+ S_{j\,k;i\,n}^{{\bf v}{\bf v}{\bf b}{\bf b}}({\bf r},t)
+ S_{i\,k;j\,n}^{{\bf v}{\bf v}{\bf b}{\bf b}}({\bf r},t).
\end{equation}
We obtain
\begin{eqnarray}\nonumber
\frac{\partial}{\partial r_n} A_{ijk,n}({\bf r},t)
&=& \left( 3 \frac{\partial}{\partial r}  S_{r\,r;t\,t}^{{\bf v}{\bf v}{\bf b}{\bf b}}({ r},t)
-\frac{6}{r} S_{r\,r;t\,t}^{{\bf v}{\bf v}{\bf b}{\bf b}}({ r},t) \right) \frac{r_i r_j r_k}{r^3}\\
&-& \frac{\partial}{\partial r} S_{r\,r;t\,t}^{{\bf v}{\bf v}{\bf b}{\bf b}}({ r},t) \left[
\frac{r_i}{r} \delta_{jk}+\frac{r_j}{r} \delta_{ik}+\frac{r_k}{r} \delta_{ij}\right],
\end{eqnarray}
Therefore, the coefficients can be read of analogous to (\ref{hill}) and (\ref{hill1}) as
\begin{equation}
 \frac{\partial}{\partial r_n} A_{rrrn}({\bf r},t)- 3 \frac{\partial}{\partial r_n} A_{rttn}({\bf
r},t)= \left(3 \frac{\partial}{\partial r}- \frac{6}{r}\right) S_{r\,r;t\,t}^{{\bf v}{\bf v}{\bf
b}{\bf b}}({ r},t),
\end{equation}
and
\begin{equation}
 \frac{\partial}{\partial r_n} A_{rttn}({\bf r},t)= - \frac{\partial}{\partial r}
 S_{r\,r;t\,t}^{{\bf v}{\bf v}{\bf b}{\bf b}}({ r},t).
\end{equation}
\section{The averaged equation of energy balance for velocity and magnetic field increments
in MHD turbulence}\label{app:energy-bal}
In this section, we derive an evolution equation for the symmetric tensor of second order
$\langle v_i v_j + b_i b_j \rangle$.
To this end, let us multiply (\ref{eq:v_i}) by $v_j$ and then do the same
procedure for interchanged indices. Adding the corresponding equations together yields
\begin{eqnarray}\nonumber
\fl &~&\frac{\partial}{\partial t}  v_i v_j   + \frac{\partial}{\partial r_n} v_n v_i v_j  +
\frac{\partial}{\partial X_n}  U_n v_i v_j   - b_n v_j \frac{\partial}{\partial r_n} b_i - b_n v_i
\frac{\partial}{\partial r_n} b_j - H_n v_j \frac{\partial}{\partial X_n} b_i - H_n v_i
\frac{\partial}{\partial X_n} b_j \\
\fl &~&= - v_i  P_j -  v_j  P_i + \nu v_j (\nabla^2_{\bf x} + \nabla^2_{{\bf x}'}) v_i
 +\nu v_i (\nabla^2_{\bf x} + \nabla^2_{{\bf x}'}) v_j,
 \label{eq:v_inc}
\end{eqnarray}
where we made use of the incompressibility condition for $v_n$ and $U_n$ in order to pull the
divergences in front of the expressions. This is especially suitable for the direct use of homogeneity
after the averaging of Eq. (\ref{eq:v_inc}), since terms like
$ \frac{\partial}{\partial
X_n} \langle...\rangle=0 $.
Unfortunately, this treatment can not be applied for terms in Eq. (\ref{eq:v_inc}) that involve
advection by either ${\bf b}$ or ${\bf H}$. However, we are able to derive the complementary terms
from the evolution equation of the magnetic increment (\ref{eq:b_i}). To this end, we multiply
Eq. (\ref{eq:b_i}) by $b_j$ and again interchange the indices, which leads to
\begin{eqnarray}\nonumber
\fl &~& \frac{\partial}{\partial t} b_i b_j + \frac{\partial}{\partial r_n} v_n b_i b_j +
\frac{\partial}{\partial X_n} U_n b_i b_j  - b_n b_i \frac{\partial}{\partial r_n} v_j - b_n b_j
\frac{\partial}{\partial r_n} v_i - H_n b_i \frac{\partial}{\partial X_n} v_j - H_n b_j
\frac{\partial}{\partial X_n} v_i \\
\fl &~&=  \lambda b_j (\nabla^2_{\bf x} + \nabla^2_{{\bf x}'}) b_i
 +\lambda b_i (\nabla^2_{\bf x} + \nabla^2_{{\bf x}'}) b_j .
 \label{eq:b_inc}
\end{eqnarray}
It can readily be seen that equation (\ref{eq:v_inc}) and (\ref{eq:b_inc}) are linked
together by terms in
(\ref{eq:v_i}) and (\ref{eq:b_i}) that are advected by either ${\bf b}$ or ${\bf H}$,
since we are not able to write the balance equation for the kinetic and
magnetic increments in a closed form separately. This can be seen as a significant feature of
locally isotropic MHD turbulence.\\
Adding equation (\ref{eq:v_inc}) to (\ref{eq:b_inc}) and taking the averages, gives an
evolution equation for the symmetric tensor $\langle v_i v_j + b_i b_j \rangle$ in MHD turbulence
\begin{eqnarray}\nonumber
\fl &~& \frac{\partial }{\partial t} \langle  v_i v_j+ b_i b_j \rangle + \frac{\partial }{\partial
r_n}\langle v_n (v_i v_j+b_i b_j)  \rangle  -  \frac{\partial }{\partial r_n} \langle b_n( v_i b_j +
v_j b_i)  \rangle \\
\nonumber
\fl  &~&+\frac{\partial }{\partial X_n}\langle U_n (v_i v_j+b_i b_j)  \rangle -
\frac{\partial }{\partial X_n} \langle H_n( v_i b_j + v_j b_i)\rangle
+\langle v_i P_j + v_j P_i \rangle \\
\fl &=&  2   \nu \left( \nabla^2_{\bf r} + \frac{1}{4}
\nabla^2_{\bf X} \right) \langle v_i v_j \rangle -2\langle \epsilon_{i\,j}^{{\bf u}{\bf u}}
\rangle  +2 \lambda \left( \nabla^2_{\bf r} + \frac{1}{4} \nabla^2_{\bf X} \right)
\langle b_i b_j \rangle -2 \langle \epsilon_{i\,j}^{{\bf h}{\bf h}} \rangle,
\label{eq:vv}
\end{eqnarray}
where we have introduced the tensors of the local
energy dissipation rate (\ref{eq:eps_u},{\ref{eq:eps_h}).
The treatment of the viscous terms is described in \ref{app:viscous}. Under the assumption
of homogeneity, terms that stand behind the center derivative $\frac{\partial }{\partial X_n}$
can be neglected. Furthermore, the pressure term vanishes on the basis of homogeneity, which
has been discussed by Hill \cite{hill:1997}.\\
Summing over equal indices $i=j$ in equation (\ref{eq:vv}) leads to the averaged
equation of energy
balance of MHD turbulence in a briefer form
\begin{eqnarray} \nonumber
 \fl & &\frac{\partial}{\partial t}
 \left\langle\frac{ { v}^2\left( {\bf r}, t \right)  +{ b}^2\left( {\bf r}, t \right)}{2} \right\rangle
\\ \nonumber
\fl & &	+ \nabla_{{\bf r}} \cdot
	\left\langle {\bf v} \left( {\bf r} , t \right) \frac{v^2 \left({\bf r} , t \right)+b^2
\left({\bf r} , t \right)}{2} \right\rangle- \nabla_{ {\bf r}} \cdot \left\langle
 	{\bf b} \left( {\bf r} , t \right)
	{\bf v} \left( {\bf r} , t \right) \cdot
	{\bf b} \left( {\bf r} , t \right) \right\rangle \\
\fl &=& \nu       \nabla_{{\bf r}}^{2} \left\langle v^2 \left( {\bf r} ,t \right) \right\rangle
	+  \lambda \nabla_{{\bf r}}^{2} \left\langle b^2 \left( {\bf r}, t \right) \right\rangle
	-  2\langle \varepsilon^{{\bf u}{\bf u}} \left( {\bf x},t \right)+
	\varepsilon^{{\bf h}{\bf h}} \left( {\bf x},t
\right) \rangle+ Q({\bf r},t),
\label{eq:energy}
\end{eqnarray}
where
\begin{eqnarray}
 \varepsilon^{{\bf u}{\bf u}} ({\bf x},t)= \sum_{i=j} \epsilon_{i\,j}^{{\bf u}{\bf u}}
 =\frac{\nu}{2} \sum_{i,j}  \left( \frac{\partial u_i}{\partial x_j} +
\frac{\partial u_j}{\partial x_i} \right)^2,  \\
\varepsilon^{{\bf h}{\bf h}} ({\bf x},t)= \sum_{i=j} \epsilon_{i\,j}^{{\bf h}{\bf h}}
 =\frac{\lambda}{2} \sum_{i,j}  \left( \frac{\partial h_i}{\partial x_j} +
\frac{\partial h_j}{\partial x_i} \right)^2,
\end{eqnarray}
denote the corresponding local energy dissipation rates
and $Q({\bf r},t) = \langle {\bf
v}({\bf r},t)\cdot {\bf F}({\bf r},t) \rangle +\langle {\bf b}({\bf r},t)\cdot {\bf G}({\bf r},t)
\rangle$ takes into account a suitable forcing procedure.
\subsection{The 4/5 law in MHD turbulence}\label{app:45law}
In the following we introduce the tensors
\begin{eqnarray}
	 S^{{\bf v} {\bf v}} _{i\,j} \left({\bf r} , t \right)
	&=& \langle v_i v_j \rangle\\
	 S^{{\bf b} {\bf b}} _{i\,j} \left({\bf r} , t \right)
	&=& \langle b_i b_j \rangle\\
\label{eq:vvv}
	 S^{{\bf v} {\bf v} {\bf v}} _{i\,j\,n} \left({\bf r} , t \right)
	&=& \langle v_i v_j v_n \rangle \\
	 S^{{\bf b} {\bf b} {\bf v}} _{i\,j\,n} \left({\bf r} , t \right) - S^{{\bf v} {\bf b} {\bf
b}} _{i\,j\,n} \left({\bf r} , t \right)
	&=&\langle b_i b_j  v_n \rangle- \langle (v_i b_j+ v_j b_i)  b_n \rangle \label{eq:letzt}
\end{eqnarray}
 where the subscripts denote the corresponding increments.\\
The last tensor (\ref{eq:letzt}) decomposes in terms of the corresponding correlation functions
according to
\begin{eqnarray}\nonumber
 S^{{\bf b} {\bf b} {\bf v}} _{i\,j\,n} \left({\bf r} , t \right) - S^{{\bf v} {\bf b} {\bf
b}}_{i\,j\,n} \left({\bf r} , t \right)
 = &+&2 (  \langle h_j h_n u_i' \rangle +\langle h_i h_n u_j' \rangle - \langle h_i h_j u_n'
\rangle)  \\ \nonumber
 &-&2 (\langle (u_n h_j-u_j h_n) h_i' \rangle +\langle (u_n h_i -u_i h_n) h_j' \rangle)\\
&+&2 \langle (u_jh_i+ u_i h_j ) h_n' \rangle
\label{eq:decomp}
\end{eqnarray}
where terms like $\langle u_i h_j h_n \rangle - \langle u_i' h_j' h_n' \rangle $ vanish under the
assumption of homogeneity. It can readily be seen, that the structure function does not decompose
into the corresponding correlation functions like the structure function in the hydrodynamic case
(\ref{eq:vvv}) that we considered in the appendix \ref{app:Kol}.
This is crucial for the appearance of the
antisymmetric tensor
\begin{equation}
C^{{\bf u} {\bf h} {\bf h}}_{ \,j;n,i}(r,t)=
\langle (h_j u_n-u_j h_n) h_i' \rangle= C^{{\bf u} {\bf h} {\bf h}}_{ \,r;t\,t}(r,t) \left(\frac{r_j}{r}
\delta_{in} - \frac{r_n}{r} \delta_{ij}\right).
\label{eq:P(r,t)}
\end{equation}
 whose defining scalar $C^{{\bf u} {\bf h} {\bf h}}_{ \,r;t\,t}(r,t)$ is not touched by the incompressibility
condition. The second correlation function that enters in Eq. ({\ref{eq:decomp}) follows
the usual tensorial form (\ref{eq:Cijk}) and reads
\begin{equation}
C^{{\bf h} {\bf h} {\bf u}}_{ i\, j\, ,n}(r,t)= \langle h_i h_j u_n'\rangle.
\label{eq:S(r,t)}
\end{equation}
The averaged equation of energy balance in spherical coordinates can now be written in the form
\begin{eqnarray}
\fl	\frac{1}{2} \frac{\partial}{\partial t} \left( S^{{\bf v} {\bf v}} \left( {\bf r}, t
\right)+ S^{{\bf b} {\bf b}} \left( {\bf r}, t \right) \right)
	+ \frac{1}{2r^2}\frac{\partial}{\partial r} \left(
	  r^2 S^{{\bf v} {\bf v} {\bf v}} \left( r , t \right) +r^2 (S^{{\bf b} {\bf b} {\bf v}}
\left( r , t \right)
	- S^{{\bf v} {\bf b} {\bf b}} \left( r , t \right) ) \right)\nonumber \\
\fl	= \nu \frac{1}{r ^2} \frac{\partial}{\partial r} \left( r ^2 \frac{\partial}{\partial r}
S^{{\bf v} {\bf v}}\left(r , t) \right) \right)+\lambda \frac{1}{r ^2} \frac{\partial}{\partial r}
\left( r ^2 \frac{\partial}{\partial r} S^{{\bf b} {\bf b}}\left(r , t) \right) \right) - 2 \langle
\varepsilon^{{\bf u}{\bf u}} + \varepsilon^{{\bf h}{\bf h}} \rangle+ Q(r,t),
\label{eq:energy_spherical}
\end{eqnarray}
where we have inserted the following functions, which can be expressed in terms of longitudinal
structure functions and correlation functions as derived in the appendix \ref{app:kar3d} and \ref{app:Kol}.
\begin{eqnarray}
	S^{{\bf v} {\bf v}} \left( r , t \right)
        &=&\ \langle v^2({r},t)\rangle=  \frac{1}{r^2}\frac{\partial}{\partial r} \left( r^3
S_{r\,r}^{{\bf v} {\bf v} } (r,t)  \right),\\
        \frac{\partial}{\partial r} S^{{\bf v} {\bf v}} \left( r , t \right)&=& \frac{1}{r^3}
\frac{\partial}{\partial r}\left(r^4 \frac{\partial}{\partial r} S^{{\bf v} {\bf v}}_{r\, r}(r,t)
\right),\\ \nonumber
S^{{\bf b} {\bf b}} \left( r , t \right)
        &=&\ \langle b^2({r},t)\rangle=  \frac{1}{r^2}\frac{\partial}{\partial r} \left( r^3
S_{r\,r}^{{\bf b} {\bf b} } (r,t)  \right),\\
        \frac{\partial}{\partial r} S^{{\bf b} {\bf b}} \left( r , t \right)&=& \frac{1}{r^3}
\frac{\partial}{\partial r}\left(r^4 \frac{\partial}{\partial r} S^{{\bf b} {\bf }}_{r\, r}(r,t)
\right),\\ \nonumber
       S^{{\bf v} {\bf v} {\bf v}} \left( r , t \right)&=& \langle v_r(r,t) {\bf v}(r,t)^2 \rangle
        = \frac{1}{3r^3}\frac{\partial}{\partial r}\left( r^4 S^{{\bf v} {\bf v} {\bf b}}_{r\, r\,
r\,}(r,t) \right),\\ \nonumber
       S^{{\bf b} {\bf b} {\bf v}} \left( r , t \right)
	- S^{{\bf v} {\bf b} {\bf b}} \left( r , t \right)&=& \langle v_r(r,t) {\bf b}(r,t)^2
\rangle - 2 \langle b_r(r,t) {\bf v}(r,t) \cdot {\bf b}(r,t)\rangle \\
       &=& -\frac{4}{r^3} \frac{\partial}{\partial r} \left(r^4 C^{{\bf h}{\bf h}{\bf
u}}_{t\,\,t\,\,r}(r,t) \right) -8 C^{{\bf u} {\bf h} {\bf h}}_{ \,r;t\,t}(r,t).
 \end{eqnarray}
For stationary turbulence the fields are driven by a time independent source term $Q(r,t)=Q(r)$, and
we obtain
\begin{eqnarray} \nonumber
\fl \qquad \qquad&~&\frac{1}{r^2} \frac{\partial}{\partial r} \left( \frac{1}{r} \frac{\partial}{\partial r}
	\left( r^4 \left \lbrace
	\frac{1}{6} S^{{\bf v} {\bf v} {\bf v}}_{r\, r\, r} \left( r \right)
	- 2         C^{{\bf h} {\bf h} {\bf u}}_{t\, \, t\, \, r} \left( r \right)
	- \nu \frac{\partial}{\partial r} S^{{\bf v} {\bf v}}_{r \,r}(r)
         -\lambda \frac{\partial}{\partial r} S^{{\bf b} {\bf b}}_{r\, r}(r)
	 \right \rbrace \right) \right)\\
 \fl      &~& -  \frac{4}{ r^2} \left( \frac{\partial}{\partial r} r ^2 C^{{\bf u} {\bf h} {\bf h}}_{ \,r;t\,t}(r)
\right)
	= - 2  \langle \varepsilon^{{\bf u}{\bf u}} +\varepsilon^{{\bf h}{\bf h}} \rangle +Q(r).
\end{eqnarray}
Two integrations with respect to $r$ yield
\begin{eqnarray}
\fl	S^{{\bf v} {\bf v} {\bf v}}_{ r\, r\, r} \left( r \right)
	- 12 C^{{\bf h} {\bf h} {\bf u}}_{ t\, \, t\, \,r} \left( r \right)-	\frac{24}{r^4}
\int_{0}^{r} \textrm{d} r' ~r'^3 C^{{\bf u} {\bf h} {\bf h}}_{ \,r;t\,t}(r')
	= &-& \frac{4}{5} \langle \varepsilon^{{\bf u}{\bf u}} +\varepsilon^{{\bf h}{\bf h}} \rangle r \\
\nonumber
\fl  &+& 6 \nu \frac{\partial}{\partial r} S^{{\bf v} {\bf v}}_{r\, r} \left( r \right)
	 + 6 \lambda \frac{\partial}{\partial r} S^{{\bf b} {\bf b}}_{r\, r} \left( r
\right)+q(r)
\label{eq:K41}
\end{eqnarray}
where the source $q(r)$ is given by
\begin{equation}
 q(r)= \frac{6}{r^4} \int_{0}^{r} \textrm{d} r' r' \int_0^{r'} \textrm{d}r{''}r{''}^2 Q(r'').
\end{equation}
Eq. (\ref{eq:K41}) is the generalization of the 4/5-law from hydrodynamic turbulence in the
presence of a magnetic field. The 4/5-law in hydrodynamic turbulence is an exact relation between
the second and third order longitudinal velocity structure function and the energy dissipation rate
$ \langle \varepsilon^{{\bf u}{\bf u}}\rangle$. However, in the case of MHD turbulence this relation is not
closed, since the source term from $C^{{\bf u} {\bf h} {\bf h}}_{ \,r;t\,t}(r)$ does not vanish in the inertial
range. In the following, we want to discuss the implications of Eq. (\ref{eq:K41}).
\subsubsection{Dissipation range}\label{app:45law_diss} ~\\~\\
The third order longitudinal velocity structure
function $S^{{\bf v} {\bf v} {\bf v}}_{ r\, r\, r} \left( r \right)$ scales as $\sim r^3$ for small
$r$, whereas the mixed correlation functions only scale as $r$, which first has been established by
Chandrasekhar \cite{chandra:1951}. Furthermore, he was able to show that the mixed correlation
functions for small $r$ are related by
\begin{eqnarray}
C^{{\bf h} {\bf h} {\bf u}}_{ t\, \, t\, \,r} \left( r \right)&=& -2 C_0 r, \\
C^{{\bf u} {\bf h} {\bf h}}_{ \,r;t\,t}(r)&=& 5 C_0 r,
\label{eq:C0}
\end{eqnarray}
where $C_0$ can be seen as the contribution to the magnetic
energy from the stretching of the lines of force by the velocity field. The source term can
thus be expressed in terms of the defining scalar $C^{{\bf h} {\bf h} {\bf u}}_{ t\, \, t\, \,r}
\left( r \right)$, namely $C^{{\bf u} {\bf h} {\bf h}}_{ \,r;t\,t}(r)= -\frac{5}{2}C^{{\bf h} {\bf h} {\bf u}}_{
t\, \, t\, \,r}(r)$.\\
In the dissipation range ($r \ll \textrm{min}(\eta^{{\bf u}{\bf u}}, \eta^{{\bf h}{\bf h}})$)
the velocity structure function of third order $S^{{\bf v} {\bf
v} {\bf v}}_{ r\, r\, r} \left( r \right)$ can thus be neglected. Furthermore, by inserting
(\ref{eq:C0}) into (\ref{eq:K41}), one can readily see that the two mixed correlation terms exactly cancel
each other, which implies that energy is only lost due to dissipative effects. In the dissipation
range Eq. (\ref{eq:K41}) thus reads
\begin{equation}
 6 \nu \frac{\partial}{\partial r} S^{{\bf v} {\bf v}}_{r\, r} \left( r \right)
	 + 6 \lambda \frac{\partial}{\partial r} S^{{\bf b} {\bf b}}_{r\, r} \left( r \right)=
\frac{4}{5} \left\langle \varepsilon^{{\bf u}{\bf u}} +\varepsilon^{{\bf h}{\bf h}} \right\rangle r,
\end{equation}
where the forcing is assumed to take place on larger scales, which implies $q(r)=0$. An integration with
respect to $r$ yields
\begin{equation}
 S^{{\bf v} {\bf v}}_{r\, r} \left( r \right)+ \frac{1}{\textrm{Pm}}S^{{\bf b} {\bf b}}_{r\, r}
\left( r
\right) = \frac{ \left\langle \varepsilon^{{\bf u}{\bf u}} +\varepsilon^{{\bf h}{\bf h}} \right\rangle}{15} r^2,
\end{equation}
where $\textrm{Pm} = \frac{\nu }{\lambda}$ is the magnetic Prandtl number.
\subsubsection{Inertial range}\label{app:45law_inert}
In the inertial range the viscous and forcing terms
can be neglected and the 4/5-law (\ref{eq:K41}) reads
\begin{eqnarray}
	S^{{\bf v} {\bf v} {\bf v}}_{ r\, r\, r} \left( r \right)
	- 12 C^{{\bf h} {\bf h} {\bf u}}_{ t\, \, t\, \,r} \left( r \right)-	\frac{24}{r^4}
\int_{0}^{r} \textrm{d} r' ~r'^3 C^{{\bf u} {\bf h} {\bf h}}_{ \,r;t\,t}(r')
	= - \frac{4}{5} \langle \varepsilon^{{\bf u}{\bf u}} +\varepsilon^{{\bf h}{\bf h}} \rangle r
\label{eq:K41_inert}
\end{eqnarray}
In contrast to the dissipation range, there exists no general relation between
$C^{{\bf h} {\bf h} {\bf u}}_{ t\, \, t\, \,r} \left( r \right)$ and
$C^{{\bf u} {\bf h} {\bf h}}_{ \,r;t\,t}(r')$ in the inertial range. The implications
for the scaling of each of the correlations on the r.h.s of (\ref{eq:K41_inert}) are
therefore far from obvious.
\section{Next-order structure function equation in MHD turbulence}\label{app:next-order}
In this appendix we derive an evolution equation that contains the fourth-order structure functions
in MHD turbulence in analogy to the next-order equations given by Hill and Boratav
\cite{hill-boratav:2001}. In multiplying (\ref{eq:v}) and (\ref{eq:b}) by the corresponding increments
we obtain a first evolution equation for the triple velocity structure function $\langle v_i v_j v_k
\rangle$, which also involves the dynamics of three terms similar to $\langle v_i b_j b_k \rangle$
(indices interchanged). These three terms have to be added in order to arrive at closed expressions
in much the same way as it has been done for the symmetric tensor of second order
$\langle v_iv_j +b_i b_k\rangle$ in \ref{app:energy-bal}.
The tensor $\big \langle v_i v_j v_k + v_i b_j b_k + b_i v_j b_k + b_i b_j v_k
\big \rangle $ can therefore be considered as the symmetric tensor of third order in MHD turbulence.
We obtain
\begin{eqnarray}\label{sym}\nonumber
\fl \frac{\partial}{\partial t} \big \langle v_i v_j v_k + v_i b_j b_k + b_i v_j b_k + b_i b_j v_k
\big \rangle \\ \nonumber
\fl +   \frac{\partial}{\partial r_n}
\langle v_n(v_i v_j v_k + v_i b_j b_k + b_i v_j b_k + b_i b_j v_k) \rangle -
\frac{\partial}{\partial r_n}  \langle b_n(b_i b_j b_k + v_i v_j b_k + b_i v_j v_k + v_i b_j v_k) \rangle \\
\nonumber
\fl +   \frac{\partial}{\partial X_n} \Big[
\langle U_n(v_i v_j v_k + v_i b_j b_k + b_i v_j b_k + b_i b_j v_k) \rangle -
 \langle H_n(b_i b_j b_k + v_i v_j b_k + b_i v_j v_k + v_i b_j v_k) \rangle \Big]\\
\nonumber
\fl = -  \left \langle(v_i v_j + b_i b_j)  P_k +   ( v_i v_k+ b_i b_k) P_j+ (v_j v_k + b_j b_k)  P_i
\right\rangle  \\ \nonumber
\fl + \nu \left \langle v_i v_j (\nabla_x^2 +\nabla_{x'}^2) v_k + v_j v_k (\nabla_x^2
+\nabla_{x'}^2)v_i + v_i v_k (\nabla_x^2 +\nabla_{x'}^2)v_j \right \rangle \\ \nonumber
\fl + \nu \left \langle b_i b_j (\nabla_x^2 +\nabla_{x'}^2) v_k + b_j b_k (\nabla_x^2
+\nabla_{x'}^2)v_i + b_i b_k (\nabla_x^2 +\nabla_{x'}^2)v_j \right \rangle \\ \nonumber
\fl + \lambda \left \langle v_i b_j (\nabla_x^2 +\nabla_{x'}^2) b_k +  v_k b_j (\nabla_x^2
+\nabla_{x'}^2)b_i + v_k b_i (\nabla_x^2 +\nabla_{x'}^2)b_j \right \rangle \\
\fl +\lambda \left \langle v_j b_i (\nabla_x^2 +\nabla_{x'}^2) b_k +  v_j  b_k(\nabla_x^2
+\nabla_{x'}^2)b_i +  v_i b_k (\nabla_x^2 +\nabla_{x'}^2)b_j \right \rangle
\end{eqnarray}
We introduce the following tensors
\begin{eqnarray}
 S_{i\,j\,k}^{{\bf v}{\bf v}{\bf v}}({\bf r},t) &=& \langle v_i v_j v_k \rangle \\
 S_{i\,j\,k}^{{\bf v}{\bf b}{\bf b}}({\bf r},t) &=& \langle v_i b_j b_k + b_i v_j b_k + b_i b_j v_k
\rangle \\
 S_{i\,j\,k\,n}^{{\bf v}{\bf v}{\bf v}{\bf v}}({\bf r},t) &=& \langle v_i v_j v_k v_n \rangle \\
 S_{i\,j\,k\,n}^{{\bf b}{\bf b}{\bf b}{\bf b}}({\bf r},t) &=& \langle b_i b_j b_k b_n \rangle \\
 S_{i\,j;k\,n}^{{\bf v}{\bf v}{\bf b}{\bf b}}({\bf r},t) &=& \langle v_i v_j b_k b_n - b_i b_j v_k
v_n \rangle\\
 T_{ijk} ({\bf r},t) &=& \left \langle(v_i v_j + b_i b_j)  P_k +   ( v_i v_k+ b_i b_k) P_j+ (v_j v_k
+ b_j b_k)  P_i  \right\rangle
\end{eqnarray}
For the viscous terms, we restrict ourselves to the case where the magnetic Prandtl number
Pm$=\frac{\nu}{\lambda}$ is unity, which simplifies the treatment. This restriction seems somehow
arbitrary, but for later times we are only interested in the inertial range behavior of the fourth
order structure functions, where it was shown for the hydrodynamic case \cite{hill-boratav:2001}
that the
viscous terms should have no contribution. We require homogeneity in order to neglect terms in
(\ref{sym}), where $\frac{\partial}{\partial X_n}$ acts on an ensemble average and
arrive at the following equation
\begin{eqnarray}\label{sym2}
\fl\qquad &~&\frac{\partial}{\partial t}\left(  S_{i\,j\,k}^{{\bf v}{\bf v}{\bf v}}({\bf r},t)+
S_{i\,j\,k}^{{\bf v}{\bf b}{\bf b}}({\bf r},t)\right )+\frac{\partial }{\partial r_n} \left(
S_{i\,j\,k\,n}^{{\bf v}{\bf v}{\bf v}{\bf v}}({\bf r},t)- S_{i\,j\,k\,n}^{{\bf b}{\bf b}{\bf b}{\bf
b}}({\bf r},t) \right)\\ \nonumber
\fl &~&- \frac{\partial }{\partial r_n} \left(  S_{i\,j;k\,n}^{{\bf v}{\bf v}{\bf b}{\bf b}}({\bf r},t)
+ S_{j\,k;i\,n}^{{\bf v}{\bf v}{\bf b}{\bf b}}({\bf r},t)+ S_{i\,k;j\,n}^{{\bf v}{\bf v}{\bf b}{\bf
b}}({\bf r},t) \right) \\ \nonumber
\fl &=& -T_{ijk} ({\bf r},t) + 2\nu \left[ \nabla_{\bf r}^2 \left(  S_{i\,j\,k}^{{\bf v}{\bf v}{\bf
v}}({\bf r},t)+ S_{i\,j\,k}^{{\bf v}{\bf b}{\bf b}}({\bf r},t)\right)- Z_{i\,j\,k}^{{\bf v}{\bf
v}{\bf v}}({\bf r},t)- Z_{i\,j\,k}^{{\bf v}{\bf b}{\bf b}}({\bf r},t) \right]
\end{eqnarray}
where we have introduced
\begin{eqnarray}\nonumber
 \fl \qquad Z_{i\,j\,k}^{{\bf v}{\bf v}{\bf v}}({\bf r},t)&=& \langle v_i \varepsilon_{j\,k}^{{\bf u}{\bf u}}
+v_j \varepsilon_{k\,i}^{{\bf u}{\bf u}}+ v_k \varepsilon_{i\,j}^{{\bf u}{\bf u}} \rangle\\
\nonumber
\fl \qquad Z_{i\,j\,k}^{{\bf v}{\bf b}{\bf b}}({\bf r},t)&=&\langle v_i \varepsilon_{j\,k}^{{\bf h}{\bf h}}
+v_j \varepsilon_{k\,i}^{{\bf u}{\bf u}}+ v_k \varepsilon_{i\,j}^{{\bf u}{\bf u}}+ b_i(
\varepsilon_{j\,k}^{{\bf u}{\bf h}} + \varepsilon_{k\,j}^{{\bf u}{\bf h}})+ b_j(
\varepsilon_{k\,i}^{{\bf u}{\bf h}} + \varepsilon_{i\,k}^{{\bf u}{\bf h}})+b_k(
\varepsilon_{i\,j}^{{\bf u}{\bf h}} + \varepsilon_{j\,i}^{{\bf u}{\bf h}})\rangle
\end{eqnarray}
and
\begin{eqnarray}
\varepsilon_{i\,j}^{{\bf u}{\bf u}}&=& \left(\frac{\partial u_i}{\partial x_l}
\right)\left(\frac{\partial u_j}{\partial x_l} \right)+\left(\frac{\partial u_i'}{\partial x_l'}
\right)\left(\frac{\partial u_j'}{\partial x_l'} \right)\\
\varepsilon_{i\,j}^{{\bf u}{\bf h}}&=& \left(\frac{\partial u_i}{\partial x_l}
\right)\left(\frac{\partial h_j}{\partial x_l} \right)+\left(\frac{\partial u_i'}{\partial x_l'}
\right)\left(\frac{\partial h_j'}{\partial x_l'} \right)
\end{eqnarray}
The treatment of the viscous terms follows the same procedure as described in \ref{app:viscous}.\\
In the inertial range (\ref{sym2}) the viscous terms can be neglected and we require statistical
stationarity, which yields
\begin{eqnarray}\nonumber
 &~&\quad \frac{\partial }{\partial r_n} \left(  S_{i\,j\,k\,n}^{{\bf v}{\bf v}{\bf v}{\bf v}}({\bf r},t)-
S_{i\,j\,k\,n}^{{\bf b}{\bf b}{\bf b}{\bf b}}({\bf r},t) \right) \\
&~&-\frac{\partial}{\partial r_n} \left(  S_{i\,j;k\,n}^{{\bf v}{\bf v}{\bf b}{\bf
b}}({\bf r},t) + S_{j\,k;i\,n}^{{\bf v}{\bf v}{\bf b}{\bf b}}({\bf r},t)+ S_{i\,k;j\,n}^{{\bf v}{\bf
v}{\bf b}{\bf b}}({\bf r},t) \right)  = T_{ijk}({\bf r},t)
\end{eqnarray}

\section{The viscous term}\label{app:viscous}
As an example of the treatment of the viscous terms in (\ref{eq:vv}) we focus on the viscous
velocity contributions. The calculation is the same for the diffusive magnetic field contributions.
The viscous terms in (\ref{eq:vv}) read
\begin{equation}
   v_j (\nabla^2_{\bf x} + \nabla^2_{{\bf x}'}) v_i
 +  v_i (\nabla^2_{\bf x} + \nabla^2_{{\bf x}'}) v_j.
 \label{eq:v_jv_i}
\end{equation}
We rewrite the Laplacian in ${\bf x}$- and ${\bf x'}$-space according to
\begin{equation}
 \nabla^2_{\bf x} = \frac{\partial}{\partial x_n}\frac{\partial}{\partial x_n}       \qquad
\textrm{and} \qquad \nabla^2_{{\bf x}'} = \frac{\partial}{\partial x_n'}\frac{\partial}{\partial
x_n'},
\end{equation}
and make use of the identity
\begin{equation}
\frac{\partial}{\partial x_n}\frac{\partial}{\partial x_n} (fg) = f \frac{\partial}{\partial
x_n}\frac{\partial}{\partial x_n} g + 2 \left(\frac{\partial f}{\partial x_n}\right)
\left(\frac{\partial g}{\partial x_n} \right) + g \frac{\partial}{\partial
x_n}\frac{\partial}{\partial x_n} f.
\end{equation}
Therefore, we can rewrite (\ref{eq:v_jv_i}) as
\begin{eqnarray}
\fl &~& v_j \left( \frac{\partial}{\partial x_n}\frac{\partial}{\partial x_n} + \frac{\partial}{\partial
x_n'}\frac{\partial}{\partial x_n'}  \right) v_i +v_i \left( \frac{\partial}{\partial
x_n}\frac{\partial}{\partial x_n} + \frac{\partial}{\partial x_n'}\frac{\partial}{\partial x_n'}
\right) v_j \\ \nonumber
\fl &=& \left( \frac{\partial}{\partial x_n}\frac{\partial}{\partial x_n} + \frac{\partial}{\partial
x_n'}\frac{\partial}{\partial x_n'} \right) v_i v_j - 2 \left[ \left(\frac{\partial v_i}{\partial
x_n}\right) \left(\frac{\partial v_j}{\partial x_n} \right) +  \left(\frac{\partial v_i}{\partial
x_n'}\right) \left(\frac{\partial v_j}{\partial x_n'} \right) \right].
\end{eqnarray}
Note that $ \left(\frac{\partial v_i}{\partial x_n}\right) \left(\frac{\partial v_j}{\partial x_n}
\right)=  \left(\frac{\partial u_i}{\partial x_n}\right) \left(\frac{\partial u_j}{\partial x_n}
\right)$ and  $ \left(\frac{\partial v_i}{\partial x_n'}\right) \left(\frac{\partial v_j}{\partial
x_n'} \right)=  \left(\frac{\partial u_i'}{\partial x_n'}\right) \left(\frac{\partial u_j'}{\partial
x_n'} \right)$ and that
\begin{equation}\label{x-r}
 \left( \frac{\partial}{\partial x_n}\frac{\partial}{\partial x_n} + \frac{\partial}{\partial
x_n'}\frac{\partial}{\partial x_n'} \right) = 2 \left(  \frac{\partial}{\partial
r_n}\frac{\partial}{\partial r_n} + \frac{1}{4}  \frac{\partial}{\partial
X_n}\frac{\partial}{\partial X_n} \right),
\end{equation}
where we made use of (\ref{eq:transform}).
If we rewrite the equations again with the Laplacian in ${\bf r}$- and ${\bf X}$-space, we
get
\begin{equation}
  v_j (\nabla^2_{\bf x} + \nabla^2_{{\bf x}'}) v_i
 + v_i (\nabla^2_{\bf x} + \nabla^2_{{\bf x}'}) v_j=  2 \left [ \left( \nabla^2_{\bf r} + \frac{1}{4}
\nabla^2_{\bf X} \right) v_i v_j - \epsilon_{i\,j}^{{\bf u}{\bf u}} \right],
\end{equation}
with
\begin{eqnarray}
\epsilon_{i\,j}^{{\bf u}{\bf u}}= \left(\frac{\partial u_i}{\partial x_n}
\right)\left(\frac{\partial u_j}{\partial x_n} \right)+\left(\frac{\partial u_i'}{\partial x_n'}
\right)\left(\frac{\partial u_j'}{\partial x_n'} \right) .
\end{eqnarray}

\vspace{1cm}

\bibliographystyle{unsrt}
\bibliography{mhd_long_trans.bib}

\end{document}